\documentclass[longauth]{aa}
\usepackage[switch]{lineno}
\usepackage{graphicx}
\usepackage{subfig}
\usepackage{txfonts,float}
\usepackage{booktabs}
\usepackage{natbib}
\usepackage[utf8]{inputenc}
\usepackage{multirow}
\bibpunct{(}{)}{;}{a}{}{,} 

\usepackage{longtable}

\newcommand{\swift}{\textit{Swift}}

\newcommand{\agile}{\textit{AGILE}}
\newcommand{\chandra}{\textit{Chandra}}
\newcommand{\suzaku}{\textit{Suzaku}}

\usepackage[usenames,dvipsnames]{color}
\usepackage{colortbl,xcolor}

\begin{document}

\title{A deep search for the host galaxies of GRBs with no detected
optical afterglow\thanks{Based on observations
collected at the Very Large Telescope of the European Southern
Observatory,  Paranal, Chile (ESO programmes 381.A-0647, 383.A-0399,
384.A-0414; PI: S. Klose; 081.D-0739, PI: A. Rossi, and 086.A-0533, PI:
T. Kr\"uhler), GROND (PI: J. Greiner), and  the Kitt Peak National
Observatory (Program ID 2008B-0070; PI:  A. C. Updike). Other
observations are obtained from the ESO/ST-ECF Science Archive Facility.}}

\author{
A.~Rossi\inst{1},
S.~Klose\inst{1},
P.~Ferrero\inst{2,3},
J.~Greiner\inst{4},
L.~A.~Arnold\inst{5}, 
E.~Gonsalves\inst{6},
D.~H.~Hartmann\inst{7},
A.~C.~Updike\inst{8},
D.~A.~Kann\inst{1}, 
T.~Kr\"uhler\inst{4,9,10},  
E.~Palazzi\inst{11},    
S.~Savaglio\inst{4},
S.~Schulze\inst{12},    
P.~M.~J.~Afonso\inst{13},
L.~Amati\inst{11},         
A.~J.~Castro-Tirado\inst{14},
C.~Clemens\inst{4},
R.~Filgas\inst{4,15},
J.~Gorosabel\inst{14},
L.~K.~Hunt\inst{16},
A.~K\"upc\"u~Yolda\c{s}\inst{17},
N.~Masetti\inst{11},              
M.~Nardini\inst{18},
A.~Nicuesa~Guelbenzu\inst{1},
F.~Olivares~E.\inst{4},
E.~Pian\inst{11,19,20},                 
A.~Rau\inst{4},
P.~Schady\inst{4},
S.~Schmidl\inst{1},
A.~Yolda\c{s}\inst{17},
\and
A.~de~Ugarte~Postigo\inst{10,14}       
}

\offprints{A. Rossi, rossi@tls-tautenburg.de}

\institute{Th\"uringer Landessternwarte Tautenburg, Sternwarte 5, 07778 Tautenburg, Germany 
\and
   Instituto de Astrof\'{\i}sica de Canarias (IAC), E--38200 La Laguna, Tenerife, Spain 
\and
   Departamento de Astrof\'{\i}sica, Universidad de La Laguna (ULL), E--38205 La Laguna, Tenerife, Spain 
\and
   Max-Planck-Institut f\"ur Extraterrestrische Physik, Giessenbachstra{\ss}e, 85748 Garching, Germany 
\and
   University of Rochester, Department of Physics and Astronomy, Rochester, NY 14627-0171, USA 
\and
   Florida Institute of Technology, Melbourne, FL 32901, USA  
\and
   Department of Physics and Astronomy, Clemson University, Clemson, SC 29634, USA 
\and 
   Department of Physics and Astronomy, Dickinson College, Carlisle, PA 17013, USA
\and
   Universe Cluster, Technische Universit\"at M\"unchen, Boltzmannstra{\ss}e 2, 85748, Garching, Germany 
\and
   Dark Cosmology Centre, Niels Bohr Institute, Univ. of Copenhagen, Juliane Maries Vej 30, 2100 K{\o}bnhaven, Denmark 
\and 
   INAF-IASF Bologna, Area della Ricerca CNR, via Gobetti 101, I--40129 Bologna, Italy 
\and
   Centre for Astrophysics and Cosmology, Science Institute, University of Iceland, Dunhagi 5, 107 Reykjav\'ik, Iceland 
\and
   American River College, Physics Dpt., 4700 College Oak Drive, Sacramento, CA 95841, USA 
\and
   Instituto de Astrof\'isica de Andaluc\'ia (IAA-CSIC), Glorieta de la Astronom\'ia s/n, 18.008 Granada, Spain 
\and
   Institute of Experimental and Applied Physics, Czech Technical University in Prague, Horsk\'a 3a/22, 12800 Prague, Czech Republic 
\and
   INAF-Osservatorio Astrofisico di Arcetri, Largo Fermi 5, I-50125 Firenze, Italy 
\and
   Institute of Astronomy, University of Cambridge, Madingley Road CB3 0HA, Cambridge, UK 
\and
   Universit\'a degli Studi di Milano-Bicocca, Piazza della Scienza 2, 20126, Milano, Italy 
\and
   Scuola Normale Superiore di Pisa - Piazza dei Cavalieri, 7 - 56126 Pisa, Italy 
\and
   INAF-Osservatorio Astronomico di Trieste, Via G.B. Tiepolo 11, 34143 Trieste 
}   
   
\date{Received May 6, 2011; accepted XXXX}

\authorrunning{Rossi et al.}
\titlerunning{Dark GRB host galaxies}

\abstract
{
Gamma-ray bursts (GRBs) can provide information about star formation at high
redshifts. Even in the absence of a bright  optical/near-infrared/radio
afterglow, the high detection rate of X-ray afterglows by \swift/XRT
and its localization precision of 2--3 arcsec facilitates the identification
and the study of GRB host galaxies. \\[-2.8mm]}
{We focus on the search for the host galaxies of 17 bursts with
arcsec-sized XRT error circles but no detected long-wavelength
afterglow, in spite of their deep and rapid follow-up observations. Three of these
events can also be classified as truly dark bursts, i.e., the observed upper 
limit on the optical flux of the afterglow was less than expected based on the
measured X-ray flux.  Our goals are to
identify the GRB host galaxy candidates and characterize their
phenomenological parameters.\\[-2.8mm]}
{Our study is based on deep $R_C$ and $K_s$-band observations performed with
FORS1, FORS2, VIMOS, ISAAC, and HAWK-I at the ESO/VLT, partly supported by
observations with the seven-channel imager GROND at the 2.2-m telescope
on La Silla, and supplemented by observations with NEWFIRM at the 4-m
telescope on Kitt Peak. To be 
conservative, we searched for host galaxy candidates 
within an area of twice the radius of each associated 90\% c.l. \swift/XRT error circle.\\[-2.8mm]}
{For 15 of the 17 bursts, we find at least one galaxy within the searching area,
and in the remaining two cases only a deep upper limit to $R_C$ and $K_s$ can be
provided. In seven cases, we discover extremely red objects
in the error circles, at least four of which 
might be dust-enshrouded galaxies. The most remarkable
case is the host of \object{GRB 080207} which  has a color of $(R_C-K_s)_{\rm
AB}\sim4.7$ mag, and is one of the reddest galaxies ever associated with a
GRB. As a by-product of our study we identify the optical afterglow of 
\object{GRB 070517}.\\[-2.8mm]}
{Only a minority of optically dim afterglows are due to Lyman dropout 
($\lesssim1/3$). Extinction by dust in the host
galaxies might explain all other events. Thereby, a seemingly
non-negligible fraction of these hosts are globally dust-enshrouded,
extremely red galaxies. This suggests that at least a fraction of GRB
afterglows trace a subpopulation of massive starburst galaxies,
which are markedly different from the main body of the GRB host
galaxy population, namely the blue, subluminous, compact galaxies.}


\maketitle

\section{Introduction}

\subsection{Optical afterglows}

By the end of 2010, about 900 gamma-ray bursts (GRBs) have been  localized at
the arcmin scale (see J. Greiner's www page\footnote{\tt
http://www.mpe.mpg.de/$\sim$jcg/grbgen.html}), most of them ($>$80\%) by the
\swift\ satellite (\citealt{Gehrels2004}). Nearly 600 events have a detected
X-ray afterglow, and nearly 400 have been detected in the optical and
near-infrared (NIR) bands, too. The observed brightness distribution of the
optical afterglows is broad and time-dependent, spanning at least 14
magnitudes within the first hour after the burst, and at least 10 magnitudes
at around 1 day, after correction for Galactic extinction
(\citealt{Kann2010a,Kann2011a}).

In principle, the observed brightness distribution reflects the luminosity
distribution of the afterglows (an intrinsic property), but it is affected by
physical processes that can block the optical light on its way to the observer
(external processes). The latter consists of two possible mechanisms,
extinction by dust in the GRB host galaxies (GRBHs) represented by the
parameter $A_V^{\rm host}$, and cosmological Lyman absorption owing to the high
redshift of the objects. If an afterglow is still detected in the optical/NIR bands, these
two processes can be recognized if a redshift ($z$) can be measured and a
broad-band spectral energy distribution (SED) of an afterglow 
constructed. The analysis of optically detected afterglows shows that (in the
$R$ band) Lyman absorption is rather the exception than the rule; only a small
fraction of bursts lie at $z \gtrsim 5$ (cf. \citealt{Haislip2006,Kawai2006,
Tagliaferri2005,Greiner2009,Tanvir2009,Salvaterra2009,PerezRamirez2010a,
Cucchiara2011}). In addition, it is found that while extinction by dust in
GRBHs along the lines of sight is usually rather small, a long tail of
possible extinction values is apparent in the data
(cf. \citealt{Kann2010a,Greiner2011,Kruhler2011A&A534}),  implying that at
least some optical afterglows are extinguished by dust in their host galaxies.

Even if no spectrum of the afterglow can be obtained, a precise optical
localization usually means that a search for an underlying host galaxy can be
undertaken and, if successful, its redshift can be measured even years after
the corresponding burst. The host extinction along the line of sight can then
be measured if a broad-band SED of the afterglow can be constructed. 
In principle, the influence of Lyman dropout and host extinction on the
observed SED can be then distinguished from each other (e.g.,
\citealt{Rossi2008AA491,Rossi2011a}).  However, the precise interplay between intrinsic
luminosity, redshift, and host extinction cannot be determined in an easy
way if no optical/NIR afterglow is detected at all. 

\subsection{Bursts with optically undetected afterglows}

The reason why a non-negligible fraction of bursts have no optically detected
afterglow despite a rapid localization in X-rays on the arcsec scale  has been
unclear for many years. While many of these non-detections are simply due to
the lack of rapid and deep optical follow-up observations, some events  (after
correction for Galactic extinction) are truly optically dark (e.g.,
\citealt{Fynbo2001a,DePasquale2003a,Castro-Tirado2007a,Rol2007,Hashimoto2010,Holland2010}). 

Theoretically, the shape of the SED of an afterglow is well-defined  and the
flux expected in the optical bands is determined by the  observed X-ray
flux. A comparison with observed optical upper limits can then tell us whether the
observations did not go deep enough or if an additional dimming of the
afterglow light is required
(\citealt{Rol2005a,Jakobsson2004a,VanderHorst2009a}).  Once the real dark
nature of a GRB is established, the question is which of the aforementioned
three physical mechanisms led to the non-detection  in the optical bands.
To tackle this problem, one can try to identify the most likely GRB host
candidate within the corresponding X-ray error circle and study  the
corresponding galaxy population. 

After a putative host has been identified, the observations can constrain 
not only the redshift but may also help to explain the optical dimness of the afterglow
 Several
studies of individual events have already found that dust extinction in the
corresponding GRBHs was the main reason for the optical dimness of some events
(e.g.,
\citealt{Piro2002a,Gorosabel2003a,Levan2006a,Berger2007a,KupcuYoldas2010AA515,
Kruhler2011A&A534}). This conclusion then  naturally leads to the question
of whether extinction by cosmic dust can explain the entire ensemble of optically
dark bursts or whether their high redshifts (seen as Lyman dropouts) is also an important 
factor. This question was finally answered when
\cite{Greiner2011}, based on a homogeneous data set of multi-color
follow-up observations of bursts, were able to show that extinction by dust in the
GRBHs is the main reason for the optical dimness of most dark events. This was
later confirmed by \cite{Melandri2012a} in an independent analysis of a
complete sample of bright \swift\ bursts.

\subsection{The present work}

 Even though the dominant role of extinction in explaining 
optically dim/dark events has long been established, we still wish to
identify and characterize  the host galaxies of all bursts with
no detected long-wavelength afterglow at all.
Therefore, in the case of \swift\ bursts, host galaxies have to be identified in  X-ray error
circles with sizes on the order of some arcsec, which is still observationally
challenging and usually requires the largest optical telescopes.  A first
study of this kind was published by \cite{Perley2009}, who reported on the
results of an imaging campaign at Keck Observatory. In their analysis, they
focused on a homogeneous sample of  29 \swift\ GRBs with rapid follow-up
observations by the robotic  Palomar 60 inch telescope (\citealt{Cenko2009a}),
among which seven were undetected by the P60 down to $R$=20-23 only 1 ks
after the burst.  They found that a significant fraction of the afterglows  in
their sample was affected by host extinction at moderate redshift and were able to 
constrain the fraction of high-$z$ \swift\ events to at most 7\% (at the
80\% c.l.\footnote{c.l. stands for confidence level}). In particular, on the basis of mainly optical observations they
concluded that the hosts  of dark bursts seem to be rather normal galaxies
in terms of their colors, suggesting that the obscuring dust is rather
local to the vicinity of the GRB progenitor or highly unevenly distributed
within the host galaxy.

Here, we report on the results of a search for the potential hosts of 
17 bursts with no detected optical/NIR afterglow. All bursts in our sample
have an observed duration in the \swift/BAT \citep{Barthelmy2005a} energy window of $T_{90} > 2$~s, i.e., they are classified as long GRBs
(\citealt{Kouveliotou1993a}). All events have optical upper limits well below
the average brightness of detected long-GRB afterglows
(\citealt{Kann2010a}).  Our goal is to identify the host-galaxy candidates
and to study the galaxy population of these events in order to 
ascertain more clearly the cause of the optical dimness of the corresponding
afterglows. 
In contrast to \cite{Perley2009}, we also make use of
deep NIR observations in order to identify and characterize the 
GRB host galaxy population.

This paper is organized as follows. In section~\ref{Selection}, we describe
the sample selection and the data-reduction procedures. In section
\ref{Targets}, we provide a detailed overview of the objects found in the
corresponding \swift/XRT \citep{Gehrels2004,Burrows2005a} error circles. Section~\ref{disc} then compiles 
information about these objects and characterizes host-galaxy candidates
and subsamples based on different selection criteria.  Finally, a summary is
given in section~\ref{Sum}. 

Throughout this work, we adopt the convention that the flux density of the
afterglow can be described as $F_{\nu}(t) \propto t^{-\alpha} \nu^{-\beta}$
and we use a $\Lambda$CDM world model with $\rm \Omega_M$ = 0.27, $\rm
\Omega_\Lambda$ = 0.73, and $\rm H_0$ = 71 km/s/Mpc  (\citealt{Spergel2003}).

\section{Target selection and observations \label{Selection} }

\subsection{The GRB sample}

 In the years from 2005 to 2008, there were about 100  bursts
at a declination $\le25$ deg (i.e., fields easily observable
from either ESO Paranal or La Silla) with detected X-ray afterglows
but no detected optical afterglows\footnote{See footnote 1}. From
this sample, we selected 17 fields with the following properties: (1) the X-ray
error circle radius is smaller than six arcsec, (2) rapid  (within one day), 
deep, but unsuccessful follow-up observations performed by
various optical telescopes, and (3) Galactic visual extinction along the
line of sight of less than 1 mag. In addition, when we selected
these targets  (usually  several months before they were observed) no
corresponding studies had been reported in the literature\footnote{ 
We have been unable to investigate several other bursts that also fulfil 
our selection criteria, owing to the limited amount of granted telescope time.}.


\begin{table*}[htbp!]
\caption[]{Characterizations of the GRB fields of our sample.}
\renewcommand{\tabcolsep}{8.5pt}
\begin{center}
\begin{tabular}{rlclccc}
\toprule
\# & GRB    & $T_{90}$& R.A., Dec. (J2000) XRT & Error   & Galactic coordinates $(l,b)$ & $E(B-V)$ \\
   &        & seconds &                       & arcsec   & degrees          & mag       \\[1mm]   
\midrule
1 &   050717  &   85    &14:17:24.48, $-$50:32:00.7 & 1.5      &  316.61, 10.04   & 0.24   \\
2 &   050922B &  150.9  &00:23:13.37, $-$05:36:16.7 & 1.7      &  104.35, $-$67.45& 0.04   \\
3 &   060211A &  126.3  &03:53:32.65, +21:29:19.0   & 1.4      &  169.74, $-$24.40& 0.19   \\[-2.5mm]
  &         &         &                           &          &                  &         \\
4 &   060805A &  5.3    &14:43:43.47, +12:35:11.2   & 1.6      &  9.53, 59.97     & 0.02   \\
5 &   060919  &  9.1    &18:27:41.80, $-$51:00:52.1 & 1.7      &  343.87, $-$17.50& 0.07   \\
6 &   060923B &  8.6    &15:52:46.70, $-$30:54:13.7 & 1.8      &  342.74, 17.61   & 0.15   \\[-2.5mm]
  &         &         &                           &          &                  &         \\
7 &   061102  &  45.6   &09:53:37.84, $-$17:01:26.0 & 2.9      &  253.43, 28.29   & 0.04   \\
8 &   070429A &  163.3  &19:50:48.92, $-$32:24:17.8 & 2.1      &  8.06, $-$25.90  & 0.17   \\
9 &   070517 &  7.6    &18:30:28.93, $-$62:17:51.7 & 2.1      & 332.76,$-$21.47  & 0.15   \\[-2.5mm]
  &         &         &                           &          &                  &         \\
10&   080207  &  340    &13:50:02.93, +07:30:07.9   & 1.4      &  340.92, 65.95   & 0.02   \\
11&  080218B &  6.2    &11:51:49.65, $-$53:05:48.5 & 1.6      &  293.94, 8.73    & 0.17   \\[-2.5mm]
  &         &         &                           &          &                  &         \\
12&   080602  &  74     &01:16:42.17, $-$09:13:55.9 & 1.7      &  142.56, $-$71.13& 0.03   \\
13&   080727A &  4.9    &13:53:33.81, $-$18:32:40.5 & 1.6      &  322.88, 41.91   & 0.07   \\
14&   080915A &  14     &01:11:47.63, $-$76:01:13.1 & 3.7      &  301.30, $-$41.04& 0.05   \\
  &         &         &                           &          &                  &         \\[-2.5mm]
15&   081012  &  29     &02:00:48.17, $-$17:38:17.2 & 1.8      &  185.87, $-$71.40& 0.02   \\
16&   081105  &$\sim10$ &00:15:48.50,   +03:28:15.5 & 4.8      &  105.87, $-$58.22& 0.03   \\
17&   081204  &$\sim20$ &23:19:09.13, $-$60:13:31.7 & 5.3      &  321.96, $-$53.36& 0.03   \\[-2.5mm]
                                                             &                  &         \\
\bottomrule
\end{tabular}
\label{tab:obs_field}
\end{center} 
\tablefoot{(1) The \swift/XRT positions for GRB 061102 and 070517 are from N.
Butlers's webpage ({\it
http://astro.berkeley.edu/$^\sim$nat/swift/xrt\_pos.html})
\citep{Butler2007}.  The XRT position for GRB 080915A, 081105, and 
081204 are from \cite{Oates2008GCNrep168}, \cite{Beardmore2008GCN8487}, and
\cite{Mangano2008GCN8620}, respectively. All other XRT data are from {\it
http://www.swift.ac.uk/xrt\_positions/index.php} 
(\citealt{Evans12250,Evans12273}). (2) The burst duration,
$T_{90}$, was mostly taken from {\it
http://heasarc.gsfc.nasa.gov/docs/swift/archive/grbtable/}.  For GRB 081105,
the reference is \cite{Cummings08_GCN8484}, for  GRB 080727A it is
\cite{McLean2008}, and for  GRB 081204 it is \cite{Gotz08_GCN8614}. (3)
 The Galactic reddening along the line of sight,
$E(B-V)$, was obtained from the NASA Extragalactic Database Coordinate
Transformation and Extinction calculator at {\it
http://nedwww.ipac.caltech.edu/forms/calculator.html}.}
\end{table*}

\begin{figure}[htbp]
\begin{center}
\includegraphics[width=0.48\textwidth]{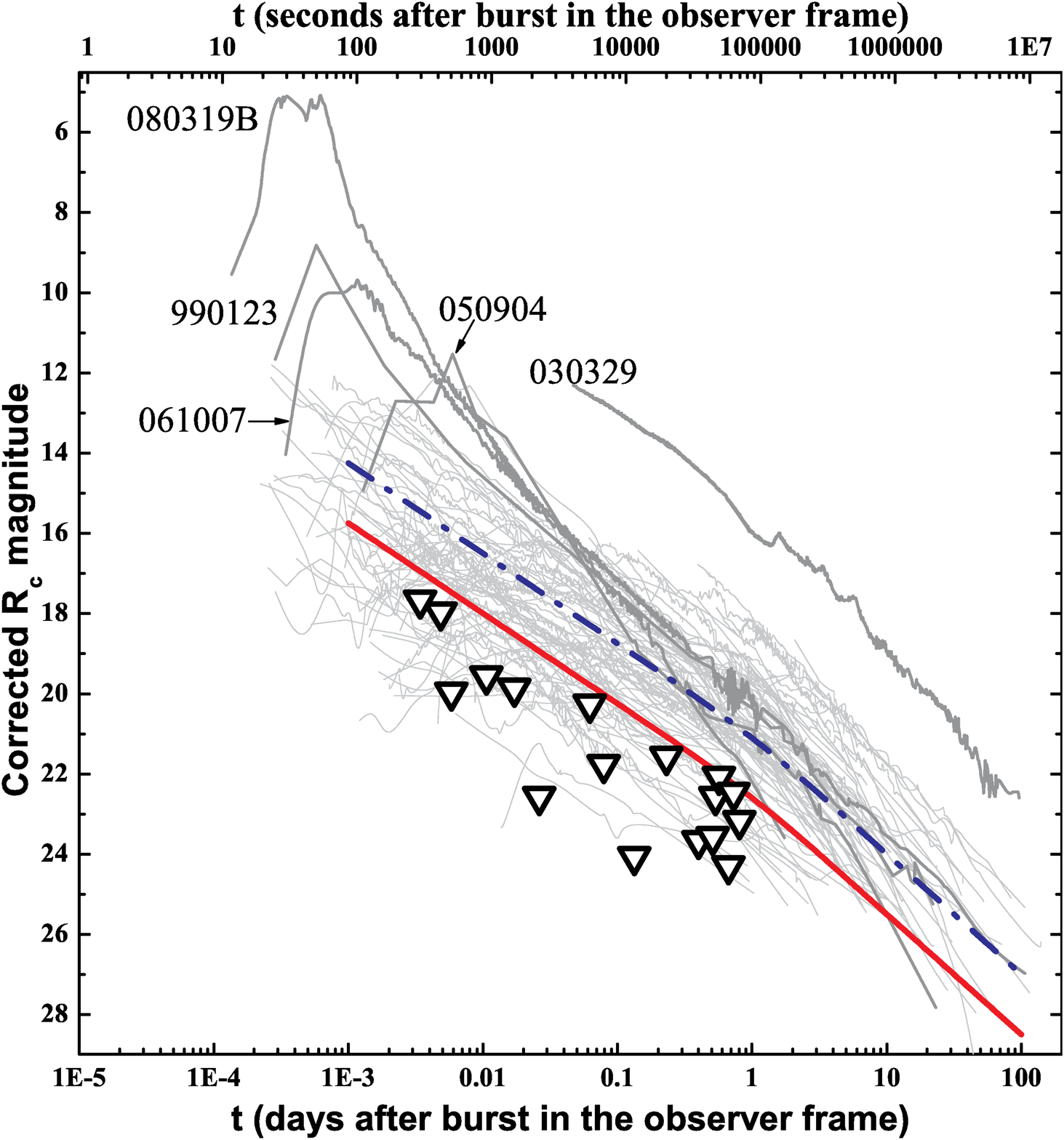}
\caption{ 
$R_C$-band light curves of all (long) afterglows in the sample of
\citet{Kann2010a, Kann2011a};  some extraordinary bright events
are indicated. All data have been corrected for Galactic
extinction. Triangles indicate equivalent $R_C$-band upper limits of the
afterglows in  our sample (Table~\ref{tab:box}).  The blue dashed/dotted line
approximately indicates the mean of the afterglow brightness distribution.
 The red straight line, 1.5~mag below the blue line, indicates the border
line of all targets in our study. }
\label{fig:DarkGRBs}
\end{center}
\end{figure}

All 17 bursts have upper limits to their optical afterglow magnitudes that 
lie at least 1.5 mag below the mean value of the afterglow brightness
distribution (Fig.~\ref{fig:DarkGRBs}). The observed GRB fields are
summarized in Table~\ref{tab:obs_field} and further details of the
corresponding world-wide observing campaigns are given in 
Appendix~\ref{sec:addnotes}.

Deep follow-up observations of 14 of these 17 X-ray error circles were
performed with VLT/FORS1, FORS2, VIMOS, ISAAC, and 
HAWK-I\footnote{See \citep{Appenzeller1998a}, \citep{LeFevre2003a}, \citep{Moorwood1998a} 
and \citep{Kissler-Patig2008a} for more informations about FORS, VIMOS, ISAAC, and 
HAWK-I, respectively.} in the years 2008 to
2010, months to years after the corresponding burst
(Table~\ref{tab:obs_log}). Limiting 3$\sigma$ AB magnitudes were typically
$R_C$=26.5 and $K_s$=23.5. In the case of GRBs 050717, 060211A, and 060805A,
multi-band imaging was performed using GROND on La Silla \citep{Greiner2008a} and, in the case of
GRBs 050922B and 060211A, data were obtained using the  near-infrared imager
NEWFIRM mounted at the 4-m Mayall telescope at Kitt Peak National
Observatory \citep{Autry2003a}. In the case of GRB 081204, a late $J$-band observation was
executed using NTT/SOFI on La Silla \citep{Moorwood1998b}.

\subsection{Optical/NIR data analysis \label{OAGana} }

VLT, GROND, and NEWFIRM data were reduced using IRAF
tasks\footnote{http://iraf.noao.edu} and analyzed by performing point-spread
function (PSF) and aperture photometry using DAOPHOT and APPHOT
\citep{Tody1993}. The procedure is mainly based on the pipeline written to
reduce GROND data (\citealt{KupcuYoldas2008AIPC,Kruhler2008a}). Aperture
photometry, if not otherwise specified, was performed by using an aperture
diameter of twice the full width half maximum (FWHM) of the stellar
PSF. The ISAAC, HAWK-I, and GROND NIR fields were calibrated  using 2MASS field
stars. The VLT optical data were calibrated using standard star 
fields limited to the Vega photometric system, while  the calibration performed for the optical
$g'r'i'z'$ images of GROND employed SDSS stars (Table~\ref{tab:obs_log}).

 We used the following transformations between AB and Vega magnitudes:  
 (1) for FORS1, FORS2, and VIMOS, $R_{\rm AB}=R_{\rm Vega} + 0.23$~mag
(\citealt{Klose2004}) and (2) for ISAAC, HAWK-I, and NEWFIRM $K_{\rm AB}=K_{s,\rm
  Vega} + 1.86$~mag (\citealt{Klose2004}).  For GROND, the Vega-to-AB
conversion is  $J_{\rm AB}= J_{\rm Vega} + 0.93$~mag, 
$H_{\rm AB} = H_{\rm Vega} + 1.39$~mag, as well
as $K_{\rm AB} = K_{s,\rm Vega} + 1.80$~mag, except for observations after an
intervention on the instrument on March 2008,  for which $K_{\rm AB} =
K_{s,\rm Vega} + 1.86$~mag.


\begin{figure}[htbp!]
\includegraphics[width=8.9cm]{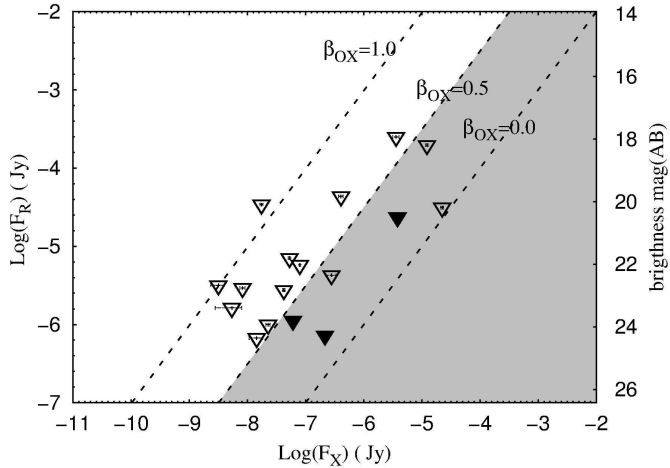}
\caption{{ 
Application of the J04 criterion:} Observed upper limits in the $R_C$ band
relative to the measured flux density at 1.73 keV (the logarithmic mean of
the \swift/XRT window, $0.3-10$~keV) for the 17 bursts in our sample. 
When no $R_C$-band data were available, we used the observed
 spectral slope $\beta_{\rm OX}$ to shift the flux density from the
native filter  to the $R$ band (Table~\ref{tab:box}). The bursts
falling in the  gray area fulfil the J04 criterion. The three bursts
that can be assumed to be securely classified dark bursts according to
J04 as well as V09  (see Fig.~\ref{fig:boxbx})  are marked with a
filled black triangle (see Sect.~\ref{sec:darksample}).}
\label{fig:box}
\end{figure}

\begin{figure}[htbp!]
\includegraphics[width=8.9cm]{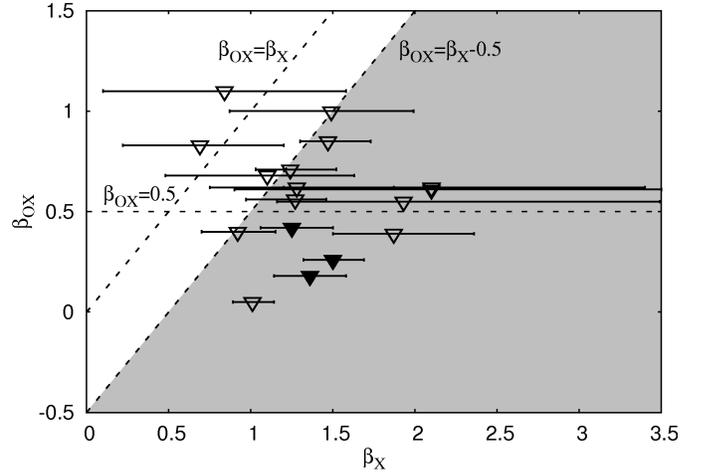}
\caption{
{Application of the V09 criterion:}  Deduced upper limits to the
spectral slope $\beta_{\rm OX}$ relative to the measured spectral slope of
the afterglow in the X-ray band.  We use the same symbols as used in
Fig.~\ref{fig:box}. The bursts falling in the gray area fulfill the V09
criterion. Here $\beta_{\rm OX} = 0.5$ is highlighted in order to compare
with the J04 criterion.}
\label{fig:boxbx}
\end{figure}

\subsection{Adding X-ray data: which bursts are truly optically 
dark? \label{sec:darksample}}

According to \citet[][ hereafter J04]{Jakobsson2004a}, a GRB with a detected
X-ray afterglow is considered dark if the spectral slope between the optical
and the X-ray regimes obey the relation $\beta_{\rm OX} < 0.5$, while
according to \citet[][ hereafter V09]{VanderHorst2009a} a burst is optically
dark if $\beta_{\rm OX} < \beta_{\rm X} - 0.5$ (see also
\citealt{Rol2005a}). Both definitions are suited to identify additional dimming of
the optical flux relative to the observed X-ray flux, assuming standard
afterglow theory (e.g., \citealt{Sari1998a}).

To determine $\beta_{\rm X}$ and $\beta_{\rm OX}$, we used the data
from the \swift/XRT GRB light curve and spectrum repository \citep{Evans2007a,
Evans2009a}. 
Optical upper limits and X-ray data were typically not
obtained at the same time. We therefore fit the X-ray light curves in order
to interpolate the X-ray flux that was contemporaneous with the corresponding optical/NIR
upper limits. Thereby, no calibration issues occurred for the optical upper
limits for the afterglows of GRBs 070429A, 070517, 080207, 080218B, 080727A,
080915A, 081012, 081105, and 081204, since here the calibration  was performed
based on our data sets. In the other cases, optical upper limits  were taken
from GCN circulars and, therefore,
can be affected by systematic errors (e.g., \citealt{Rossi2011a}). Assuming a conservative
systematic error of 0.5 mag, this would translate into a $\sim0.07$ systematic
error in the upper limit of  $\beta_{\rm OX}$ and an error of 0.5 mag in
the extrapolated $R_C$-band upper limit. Fortunately, in no case did this
uncertainty affect our potential classification of a burst as dark. 

For the optical/NIR bands, we proceeded as follows.  To compare the
different bursts, we shifted upper limits to a common band, the $R_C$
band. This required us to know the spectral slope $\beta_{\rm OX}$, and a
conservative upper limit on it was obtained by taking into account all
afterglow upper limits in time and filter. Thereby, if possible, we used only
optical upper limits that were taken at times reasonably separated from the
prompt GRB phase.  Special care was taken when the
deepest afterglow upper limits had only been obtained in bands shortwards of
$R_C$. 
We then chose to use the reddest filter, since the blue part of the
spectrum is more affected by both Lyman dropout and uncertainties in extinction
than the red part. Among all computed values, we 
 finally chose those that implied the
lowest $\beta_{\rm OX}$ values. The resulting $R_C$-band upper limits are
shown in Fig.~\ref{fig:DarkGRBs} and they were used to depict the J04
criterion in Fig.~\ref{fig:box}. 

For the X-rays, we used the flux density at 1.73 keV
(corrected for Galactic absorption), which is the logarithmic mean of the
\swift/XRT window ($0.3-10$~keV). Where the X-ray data are concerned, we gave
priority to time intervals where the light curves were smoothly decaying
(which differ from burst to burst) and during which $\beta_{\rm X}$ was
constant (within the errors).  

Table~\ref{tab:box} summarizes our results where $\Delta =
\beta_X-\beta_{\rm OX}-0.5$.  To be conservative, we used its minimum value
$\Delta_{\rm min}$, based on the 90\% c.l. error of  $\beta_{\rm X}$.
If $\Delta_{\rm min} > 0$,  a burst is classified as dark according to
V09. Five events, namely GRBs 050922B, 070429A, 080207, 080218B, and 080602,
are dark according to both definitions, while GRB 050717 is dark
according only to J04 and GRBs 080915A and 081204 are
only dark according to V09 (Figs.~\ref{fig:box}, \ref{fig:boxbx}). Following the
above discussion, we assume that a GRB is dark if it fulfills the criterion
of J04 as well as V09. 


\begin{table*}[htbp!]
\caption{Summary of the darkness properties of our sample.}
\renewcommand{\tabcolsep}{5.5pt}
\begin{center}
\begin{tabular}{rlrccccrcrr}
\toprule
\# & GRB    & Time (s)   &UL   & Filter  & UL$_R$ & Ref.          &$\beta_{\rm OX}$ & $\beta_{\rm X}$ & $\Delta_{\rm min}$ & Comments \\
(1)& (2)    & (3)       &(4)  & (5)     & (6)  & (7)           &(8)             & (9)             & (10)& (11) \\
\midrule
 1 & 050717  & 420       & 19.0 & $v $      & 18.2 & [1]        	         &$<0.40$      &$0.92^{+0.23}_{-0.22}$    & $-0.20$  & a; b \\ 
 2 & 050922B & 49000     & 22.5 & $r^\prime$& 22.3 & [2]        	         &$<0.39$      &$1.87^{+0.49}_{-0.37}$    &    0.61  & c \\   
 3 & 060211A & 19980     & 22.0 & $R_C$     & 21.8 & [3]         	         &$<0.71$      &$1.24^{+0.28}_{-0.21}$    & $-0.18$  & --\\[1mm]    
 4 & 060805A & 63000     & 22.9 & $r^\prime$& 22.7 & [4]         	         &$<1.00$      &$1.49^{+0.50}_{-0.62}$    & $-0.63$  & --\\    
 5 & 060919  & 918       & 20.2 & $v$       & 19.8 & [5]         	         &$<0.68$      &$1.10^{+0.53}_{-0.62}$    & $-0.70$  & a \\    
 6 & 060923B & 295       & 18.5 & $v$       & 17.9 & [6]         	         &$<0.62$      &$1.28^{+0.59}_{-0.53}$    & $-0.37$  & b \\[1mm]   
 7 & 061102  & 1480      & 20.5 & $v$       & 20.1 & [7]         	         &$<1.10$      &$0.84^{+0.74}_{-0.74}$    & $-1.50$  & d \\   
 8 & 070429A & 44064     & 24.0 & $R_C$     & 23.8 & [8]         	         &$<0.42$      &$1.25^{+0.25}_{-0.19}$    &    0.14  & $\bullet$; g \\ 
 9 & 070517 & 57600     & 24.5 & $i'$    & 24.3  &  [9]           	         &$<0.56$      &1.27$^{+0.19}_{-0.30}$    & $-$0.09  & f \\[1mm]   
10 & 080207  & 5364      & 20.3 & $R_C$     & 20.5 & [10]        	         &$<0.26$      &$1.50^{+0.19}_{-0.18}$    &    0.56  & $\bullet$ \\    
11 & 080218B & 11520     & 24.7 & $r^\prime$& 24.3 & Table~\ref{tab:darkULs}      &$<0.18$       &$1.36^{+0.22}_{-0.22}$    &    0.46  & $\bullet$ \\    
12 & 080602  & 504       & 20.3 & $v$       & 20.2 & [11]           	         &$<0.05$	 &$1.01^{+0.13}_{-0.12}$    &	 0.34  & e \\[1mm]   
13 & 080727A & 2268      & 19.8 & $K$       & 22.8 & [12]          	         &$<0.85$	 &$1.47^{+0.26}_{-0.17}$    & $-0.05$  & --  \\	
14 & 080915A  & 6840      & 22.1 & $I_C$     & 22.1 & [13]          	         &$<0.62$	 &$2.10^{+1.30}_{-0.90}$    &	 0.08  & d \\   
15 & 081012  & 69660     & 23.5 & $r^\prime$& 23.4 & Table~\ref{tab:darkULs}      &$<0.83$       &$0.69^{+0.51}_{-0.47}$    & $-1.11$  & -- \\[1mm]    
16 & 081105  & 46224     & 23.0 & $r^\prime$& 22.8 & Table~\ref{tab:darkULs}      &$<0.61$       &$2.10^{+1.70}_{-1.20}$    & $-0.21$  & d \\   
17 & 081204  & 34560     & 24.1 & $r^\prime$& 23.9 & Table~\ref{tab:darkULs}      &$<0.55$       &$1.93^{+1.56}_{-0.77}$    &    0.11  & d \\
\bottomrule 
\end{tabular}
\label{tab:box}
\end{center}
\linespread{1.0}
\tablefoot{
{\it Columns:} (3 to 5) Time after the burst and reported upper limits
(UL) of the afterglow (observed magnitudes); $r^\prime$-band
magnitudes are given in the AB system, all others in the Vega/UVOT system.  (6)
Deduced UL in the $R_C$ band (AB system) after correcting for Galactic
extinction and shifting from the native filter wavelength 
(column 5) to the $R_C$ band
using the upper limit to $\beta_{\rm OX}$. (8) If $\beta_{\rm OX} < 0.5$,
then a burst is dark according to J04.  (10) The minimum value (based on the
90\% confidence error of  $\beta_{\rm X}$) of the quantity $\Delta
= \beta_X-\beta_{\rm OX}-0.5$.  If $\Delta_{\rm min} > 0$, then a burst is dark according
to V09. 
(11)
(a) During the time of the observed UL the SED in the X-ray band is not constant. 
(b) The observed UL lies close to the end of the prompt GRB phase.
(c) No X-ray data exist for the time when the UL was obtained.
(d) Very faint X-ray flux; no well-defined X-ray light curve.
(e) Flat X-ray light curve during the time when the UL was obtained.
(f) The optical afterglow was detected; see Sect.~\ref{notesind}.
(g) No UL is reported in the corresponding GCN (\citealt{Price2007GCN6371}).
We used $R_C=24.0$ based on the original data, which are available in the Gemini archive.
A bullet ($\bullet$) indicates that the burst is truly dark 
according to J04 and V09 (Sect.~\ref{sec:darksample}).
{\it Column} (7; references): 
1. \cite{Blustin2005GCN3638};
2. \cite{GuziyGCN4025}; 
3. \cite{Sharapov2006GCN4927}; 
4. \cite{Rol2006GCN5406}; 
5. \cite{Breeveld2006GCN5580}; 
6. \cite{Holland2006GCN5603}; 
7. \cite{Holland2006GCN5784}; 
8. \cite{Price2007GCN6371}; 
9. \cite{Fox6420}; 
10. \cite{Andreev08_GCN7333}; 
11. \cite{Beardmore2008GCNR145}; 
12. \cite{Levan08_GCN8048}; 
13. \cite{Rossi2008GCN8268}.}
\end{table*}


We note that when the optical upper limits were obtained some
shortcomings might have limited the validity of this approach: (1) an X-ray light curve
that is rather flat might be indicative of an additional X-ray
component, while the aforementioned procedure assumes a single radiation
component (this affects \object{GRB 080602}), (2) a substantial time gap in the
X-ray data base (\object{GRB 050922B}), (3) an evolving X-ray spectral slope (\object{GRB
050717}), and (4) a large error ($>1$) in the X-ray spectral slope (\object{GRB
080915A},  \object{GRB 081105}, and \object{081204}). Taking all this into account, only three
events in our sample can be  securely classified as dark bursts 
(\object{GRB 070429A}, \object{GRB 080207}, and \object{GRB 080218B}). 
All other bursts, except for \object{GRB 070517}, may
still be truly dark bursts but the available data are insufficient to
claim this with certainty. 


\linespread{1.1}

\begin{table*}[htbp!]
\vspace{1em}
\enlargethispage{15.6cm}
\caption{Summary of the photometry of all objects found in the XRT error circles
based on the VLT observations.}
\vspace{1em}
\renewcommand{\tabcolsep}{3.5pt}
\begin{center}
\begin{tabular*}{1.0\textwidth}{@{\extracolsep{\fill}}rlccccccc}
\toprule
\# & GRB  &  Object  &  R.A., Dec. (J2000)    &  $R_{\rm AB}$  &  $K_{\rm AB}$   &  UL$_R$  &  UL$_K$  &  XRTpos \\
\midrule                        
2  &  050922B &   --    &  no candidates              &  $>$26.5           &  $>$22.8          &  26.5  &  22.8 &    \\[2.8mm]
5  &  060919  &   A     &  18:27:41.78, $-$51:00:51.0 &  26.14 $\pm$ 0.24  &  $>$23.4          &  26.5  &  23.4 &  1 \\[2.8mm]
6  &  060923B &   A\tablefootmark{a} &  15:52:46.49, $-$30:54:12.3 &  23.10 $\pm$ 0.11  &  21.76 $\pm$ 0.09 &  26.6  &  24.3 &  2 \\
   &          &   B\tablefootmark{a} &  15:52:46.61, $-$30:54:10.3 &  21.67 $\pm$ 0.02  &  18.95 $\pm$ 0.03 &  26.6  &  24.3 &  2 \\
   &          &   C\tablefootmark{a} &  15:52:46.56, $-$30:54:14.6 &  24.49 $\pm$ 0.04  &  22.87 $\pm$ 0.15 &  26.6  &  24.3 &  1 \\
   &          &   D\tablefootmark{a} &  15:52:46.63, $-$30:54:16.4 &  25.74 $\pm$ 0.12  &  21.63 $\pm$ 0.06 &  26.6  &  24.3 &  2 \\
   &          &   E\tablefootmark{a} &  15:52:46.66, $-$30:54:12.9 &  blended with A    &  blended with A   &  26.6  &  24.3 &  1 \\[3.0mm]
7  &  061102  &   A     &  09:53:37.93, $-$17:01:22.7 &  24.10 $\pm$ 0.06  &  $>$22.8           &  26.9 &  22.8 &  2 \\
   &          &   B     &  09:53:37.89, $-$17:01:30.8 &  23.96 $\pm$ 0.06  &  $>$22.8           &  26.9 &  22.8 &  2 \\[3.0mm]
8  &  070429A &   A     &  19:50:48.78, $-$32:24:13.6 &  25.01 $\pm$ 0.20  &  22.57 $\pm$ 0.25  &  26.5 &  23.8 &  3 \\
   &          &   B\tablefootmark{a} &  19:50:48.78, $-$32:24:18.1 &  24.14 $\pm$ 0.09  &  22.39 $\pm$ 0.21  &  26.5 &  23.8 &  1 \\
   &          &   C\tablefootmark{a} &  19:50:48.90, $-$32:24:17.4 &  24.32 $\pm$ 0.08  &  21.89 $\pm$ 0.14  &  26.5 &  23.8 &  1 \\
   &          &   D     &  19:50:49.10, $-$32:24:17.3 &  $>$26.5           &  23.01 $\pm$ 0.40  &  26.5 &  23.8 &  2 \\[3.0mm]
9  &  070517 &   A     &  18:30:29.08, $-$62:17:53.0 &  25.39 $\pm$ 0.21  &  $>$23.4           &  26.6 &  23.4 &  1 \\[3.0mm]
10 &  080207  &   A     &  13:50:03.03, $+$07:30:09.3 &  25.15 $\pm$ 0.17  &  23.02 $\pm$ 0.39  &  26.9 &  23.6 &  2 \\
   &          &   B     &  13:50:02.97, $+$07:30:07.2 &  26.49 $\pm$ 0.37  &  21.77 $\pm$ 0.14  &  26.9 &  23.6 &  1 \\[3.0mm] 
11 &  080218B &   A     &  11:51:49.69, $-$53:05:49.1 &  26.23 $\pm$ 0.13  &  21.74 $\pm$ 0.10  &  27.3 &  24.0 &  1 \\
   &          &   B     &  11:51:50.00, $-$53:05:47.4 &  24.62 $\pm$ 0.04  &  22.74 $\pm$ 0.24  &  27.3 &  24.0 &  3 \\[3.0mm]
12 &  080602  &   A     &  01:16:42.15, $-$09:13:55.0 &  22.95 $\pm$ 0.02  &  22.55 $\pm$ 0.05  &  26.9 &  23.5 &  1 \\
   &          &   B     &  01:16:42.12, $-$09:13:57.5 &  24.00 $\pm$ 0.06  &  $>$23.5           &  26.9 &  23.5 &  2 \\
   &          &   C     &  01:16:42.14, $-$09:13:53.4 &  $>$26.9           &  22.49 $\pm$ 0.14  &  26.9 &  23.5 &  2 \\[3.0mm]
13 &  080727A &   --    &   no candidates             &  $>$26.3           &  $>$23.0           &  26.3 &  23.0 &   \\[3.0mm] 
14 &  080915A &   A     &  01:11:47.80, $-$76:01:13.9 &  21.63 $\pm$ 0.01  &  20.42 $\pm$ 0.02  &  26.3 &  23.4 &  1 \\
   &          &   B     &  01:11:45.27, $-$76:01:10.4 &  21.28 $\pm$ 0.01  &  19.19 $\pm$ 0.01  &  26.3 &  23.4 &  3 \\
   &          &   C\tablefootmark{a} &  01:11:47.47, $-$76:01:10.0 &  24.71 $\pm$ 0.07  &  $>$23.4           &  26.3 &  23.4 &  1 \\
   &          &   D\tablefootmark{a} &  01:11:46.98, $-$76:01:09.5 &  24.57 $\pm$ 0.08  &  $>$23.4           &  26.3 &  23.4 &  2 \\
   &          &   E\tablefootmark{a} &  01:11:47.16, $-$76:01:15.1 &  25.44 $\pm$ 0.15  &  $>$23.4           &  26.3 &  23.4 &  1 \\[3.0mm] 
15 &  081012  &   A     &  02:00:48.18, $-$17:38:15.2 &  25.16 $\pm$ 0.17  &  $>$23.9           &  26.7 &  23.9 &  2 \\[3.0mm]
16 &  081105  &   A     &  00:15:48.42, $+$03:28:11.6 &  23.73 $\pm$ 0.08  &  22.78 $\pm$ 0.18  &  26.1 &  24.5 &  1 \\
   &          &   B     &  00:15:48.30, $+$03:28:13.8 &  24.34 $\pm$ 0.13  &  22.13 $\pm$ 0.14  &  26.1 &  24.5 &  1 \\
   &          &   C     &  00:15:48.46, $+$03:28:10.7 &  $>$25.3           &  21.74 $\pm$ 0.13  &  26.1 &  24.5 &  1 \\[3.0mm]
17 &  081204  &   A\tablefootmark{b} &  23:19:09.39, $-$60:13:31.5 &  23.21 $\pm$ 0.04  &  22.37 $\pm$ 0.16  &  26.4 &  24.3 &  1 \\
   &          &   B\tablefootmark{a} &  23:19:09.13, $-$60:13:30.2 &  23.54 $\pm$ 0.06  &  21.59 $\pm$ 0.08  &  26.4 &  24.3 &  1 \\
   &          &   C     &  23:19:08.99, $-$60:13:23.4 &  23.16 $\pm$ 0.07  &  22.06 $\pm$ 0.11  &  26.4 &  24.3 &  2 \\
   &          &   D     &  23:19:08.89, $-$60:13:37.6 &  24.19 $\pm$ 0.10  &  22.16 $\pm$ 0.15  &  26.4 &  24.3 &  2 \\
   &          &   E     &  23:19:09.10, $-$60:13:39.4 &  24.32 $\pm$ 0.12  &  $>$24.3           &  26.4 &  24.3 &  2 \\
   &          &   F\tablefootmark{a} &  23:19:09.24, $-$60:13:29.4 &  24.65 $\pm$ 0.50  &  21.53 $\pm$ 0.07  &  26.4 &  24.3 &  1 \\
   &          &   G     &  23:19:08.30, $-$60:13:39.0 & blended with a star&  21.57 $\pm$ 0.15  &  26.4 &  24.3 &  2 \\[3.0mm]
\bottomrule
\end{tabular*}
\label{tab:PhotomVLT}
\end{center} 
\enlargethispage{0cm}
\linespread{1.0}
\tablefoot{
Column: (2) GRB 050717, GRB 060211A, and GRB 060805A are the
only bursts in our sample for which we  do not have VLT data (see
Table~\ref{tab:PhotomGROND}).  (3) Objects identified in the XRT error
circles (see Sect.~\ref{notesind}).  (5) Observed magnitudes.  UL stands
for the 3$\sigma$ upper limit. The last column defines the distance
of the object from the center of the 90\% XRT error circle of radius $r_0$
 (see Table~\ref{tab:obs_field}). A value $n$ means that the source lies
within  $[(n-1)\,r_0, n\,r_0]$. {\it Special notes about the photometry:} All
magnitudes are based on ($2\times$ FWHM) aperture photometry, except for 
cases where the object was affected by near-by objects.  In the
latter case, we used either \tablefoottext{a}{PSF photometry} or
 \tablefoottext{b}{$1\times$ FWHM aperture photometry}. In particular,
we gave preference to the latter in the case of elongated objects. }
\end{table*}
\vspace{1em}
 
\linespread{1.0}

\begin{table*}[htbp]
\caption{Summary of the photometry 
of all objects found in the XRT error circles based on observations with GROND.}
\renewcommand{\tabcolsep}{2.5pt}
\begin{center}
\begin{tabular}{rlcccccccccc}
\toprule
\#& GRB &Obj. &R.A., Dec. (J2000) &$g'_{\rm AB}$&$r'_{\rm AB}$&$i'_{\rm AB}$&$z'_{\rm AB}$&$J_{\rm AB}$& $H_{\rm AB}$&$K_{\rm AB}$& XRTpos\\
\midrule                                                
1  &  050717  &  A    &  14:17:24.56, $-$50:31:58.7    &  $>25.4$   &  23.65(10)  &  23.01(11) &  23.40(30) &  $>$22.6  &  $>$21.9   &  $>$21.1  &  2 \\
   &          &  B\tablefootmark{b}&  14:17:24.58, $-$50:32:01.6    &  $>25.4$   &  24.50(40)  &  23.50(40) &  22.80(30) &  $>$22.6  &  $>$21.9   &  $>$21.1  &  1 \\[1mm] 
3  &  060211A &  A    &  03:53:32.66, $+$21:29:19.8    &  $>25.2$   &  24.51(20)  &  $>24.8$   &  $>24.4$   &  $>$23.4\tablefootmark{n}  &  $>$21.6   &  $>$21.6\tablefootmark{n}  &  1 \\
   &          &  B    &  03:53:32.43, $+$21:29:16.3    &  23.60(08) &  23.09(06)  &  22.76(09) &  23.31(10) &  $<$23.1\tablefootmark{n}  &  $>$21.6   &  21.50(20)\tablefootmark{n}&   3 \\
   &          &  C    &  03:53:32.57, $+$21:29:18.0    &  $>25.2$   &  $>24.9$    &  $>24.8$   &  $>24.4$   &  23.10(30)\tablefootmark{n} &  $>$21.6   &  $>$21.6\tablefootmark{n}  &  1 \\[1mm]
   &          &  D    &  03:53:32.69, $+$21:29:20.9    &  $>25.2$   &  $>24.9$    &  $>24.8$   &  $>24.4$   &  23.40(40)\tablefootmark{n} &  $>$21.6   &  $>$21.6\tablefootmark{n}  &  2 \\[1mm]
4  &  060805A &  A    &  14:43:43.49, $+$12:35:12.5    &  $>$25.5   &  25.4(40)   &  $>24.6$   &  $>24.2$   &  $>$22.9  &  $>$21.8   &  $>$21.1  &  1 \\
   &          &  B    &  14:43:43.39, $+$12:35:10.1    &  23.42(16) &  23.68(12)  &  $>24.6$   &  $>24.2$   &  $>$22.9  &  $>$21.8   &  $>$21.1  &  2 \\[1mm]
10 &  080207  &  A-B  &  see Table~\ref{tab:PhotomVLT} &  $>$25.4   &  $>$24.9    &  $>$23.9   &  $>$23.8   &  $>$22.0  &  $>$20.8   &  $>$20.1  &  1,2 \\[1mm]
11 &  080218B &  A    &  see Table~\ref{tab:PhotomVLT} &  $>$25.5   &  $>$24.9    &  $>$24.2   &  $>$24.1   &  $>$22.8  &  $>$21.4   &  $>$21.2  &  1 \\ 
   &          &  B    &  see Table~\ref{tab:PhotomVLT} &  25.10(30) &  24.30(30)  &  ---       &  23.27(10) &  $>$22.8  &  $>$21.4   &  $>$21.2  &  3 \\[1mm]
12  &  080602 &  A    &  see Table~\ref{tab:PhotomVLT} &  22.96(10) &  22.93(08)  &  22.86(13) &  22.60(14) &  $>$21.4  &  $ >$21.0  &  $>$20.6  &  1 \\
    &         &  B    &  see Table~\ref{tab:PhotomVLT} &  $>$25.3   &  23.73(12)  &  23.90(29) &  22.97(17) &  $>$21.4  &  $ >$21.0  &  $>$20.6  &  2 \\ 
    &         &  C    &  see Table~\ref{tab:PhotomVLT} &  $>$25.3   &  $>$25.5    &  $>$24.9   &  $>$24.6   &  $>$21.4  &  $ >$21.0  &  $>$20.6  &  2 \\[1mm]
14  &  080915A&  A    &  see Table~\ref{tab:PhotomVLT} &  22.05(10) &  21.27(10)  &  21.21(10) &  20.80(10) &  20.53(02)&  20.37(03) &  20.39(15)&  1 \\
    &         &  B    &  see Table~\ref{tab:PhotomVLT} &  23.30(30) &  21.14(05)  &  20.88(06) &  20.28(12) &  19.74(06)&  19.33(06) &  19.00(08)&  3 \\ 
    &         &  C-E  &  see Table~\ref{tab:PhotomVLT} &  $>$24.0   &  $>$24.3    &  $>$24.1   &  $>$23.9   &  $>$22.1  &  $>$21.3   &  $>$21.3  &  1,2 \\[1mm]
15  &  081012 &  A    &  see Table~\ref{tab:PhotomVLT} &  $>$23.8   &  $>$23.8    &  $>$23.4   &  $>$23.2   &  $>$21.8  &  $>$21.3   &  $>$21.0  &  2 \\[1mm]
16  &  081105 &  A-C  &  see Table~\ref{tab:PhotomVLT} &  $>$24.0   &  $>$23.9    &  $>$23.3   &  $>$22.9   &  $>$21.4  &  $>$20.7   &  $>$20.3  &  1 \\[1mm]
17  &  081204 &  A\tablefootmark{b}&  see Table~\ref{tab:PhotomVLT} &  23.69(10) &  23.74(08)  &  23.25(11) &  23.09(15) &  22.50(16)\tablefootmark{s}  &  $>$21.5   &  $>$20.9  &  1 \\
    &         &  B\tablefootmark{b}&  see Table~\ref{tab:PhotomVLT} &  24.18(30) &  23.74(08)  &  23.53(14) &  23.29(17) &  22.15(15)\tablefootmark{s}  &  $>$21.5   &  $>$20.9  &  1 \\
    &         &  C    &  see Table~\ref{tab:PhotomVLT} &  $>$25.0   &  23.96(17)  &  23.62(28) &  $>$24.0   &  22.55(19)\tablefootmark{s} &  $>$21.5   &  $>$20.9  &  2 \\ 
    &         &  D-G  &  see Table~\ref{tab:PhotomVLT} &  $>$25.0   &  $>$25.0    &  $>$24.7   &  $>$24.0   &  $>$22.0\tablefootmark{s}  &  $>$21.5   &  $>$20.9  &  1,2 \\[1mm]
\bottomrule
\end{tabular}
\label{tab:PhotomGROND}
\end{center} 
\tablefoot{
Columns (4-10): Observed magnitudes. Magnitude errors are given in
units of 10 mmag. 
 The letters $n$ and $s$ are used to mark those magnitudes
resulting from  \tablefoottext{n}{NEWFIRM} and \tablefoottext{s}{NTT/SOFI} imaging, respectively.
For GRB 080218B/object B the $i'$-band data are
affected by a ghost image from a bright star. Upper limits are 3$\sigma$. 
{\it Special notes about the photometry:} see Table~\ref{tab:PhotomVLT}.}
\end{table*}


\section{Results \label{Targets}}

\subsection{General}\label{sec:noteresults}

In the following, we report the results of our  deep late-time
observations for each GRB field. They are summarized in Tables
\ref{tab:PhotomVLT}, \ref{tab:PhotomGROND}, and \ref{tab:size}. 

If not stated otherwise, in the following $R_C,K_s$-band  magnitudes, and
colors are given in the AB magnitude system, in order to allow for a direct
comparison with data of confirmed GRB host galaxies  compiled by Savaglio
et al. (\citeyear{Savaglio2009a}; in the following SBG09), i.e., host
galaxies that are identified via optical afterglow detections obtained with
sub-arcsec accuracy. 
All $(R_C-K_s)$ colors were corrected for Galactic extinction,
estimated using the extinction  maps published by \cite{Schlegel1998}.
Extinction corrections for the GROND filters are $A(g^\prime)=
1.253 \,A_V,\; A(r^\prime)= 0.799 \,A_V,\; A(i^\prime)= 0.615 \,A_V,\;
A(z^\prime)= 0.454 \,A_V,\; A(J) = 0.292 \,A_V,\; A(H) = 0.184 \,A_V,$ and $A(K) =
0.136 \,A_V$, while for all other instruments we assumed the standard values of
$A(R_c)= 0.748 \,A_V$ and $A(K)  = 0.112 \,A_V$ (\citealt{Rieke1985}).
We always set $A_V=3.1 \ E(B-V).$

\subsection{Selecting host galaxy candidates \label{candidates}}

Even in the case of arcsec-sized error boxes, it is usually difficult  to
determine the most likely GRB host galaxy candidate.  The approach we used here to
identify a putative host is identical to the approach adopted 15 years ago,
when no afterglows were known at all and at best only arcmin-sized error boxes
obtained via satellite triangulation were available (e.g., \citealt{Vrba1995,
Klose1996,Schaefer1998,Vrba1999}). The main observational difference is the
size of the XRT error circles provided by \swift/XRT  that can go down to 1-2
arcsec, allowing meaningful searches for host galaxies. 

 We analyzed all
objects present in an XRT error circle and studied their properties
following different criteria in order to establish the best GRB
host candidate. The first criterion is the magnitude-probability criterion.

Following \cite{Bloom2002a} and \cite{Perley2009},  we calculated, for every
object, the probability $p$ of finding a galaxy of any type of the given
(extinction-corrected) $R_C$-band magnitude $m$ in a region of radius $r$,
where $r$ is the radius of the associated error circle.  It is  
\begin{equation}
p(m) = 1\,-\,\exp(-\pi\, r^2 \,\sigma(\leq m))\,, 
\label{p}
\end{equation}
where $\sigma(\leq m)$ is  the surface density of galaxies with magnitudes
$\leq m$ (equation 3 in \citealt{Bloom2002a}).
If the object  we have found is located within the 90\% 
c.l. XRT error
circle of radius $r_0$, we set $r=r_0$, if it is placed within  
$[(n-1)\,r_0, n\,r_0]$, then we set $r=n\, r_0$.  The input  for
$\sigma(\leq m)$ is the relation derived by  \cite{Hogg1997}, which is based
on galaxy counts down to about $R_{\rm Vega}=26.5$.  We consider galaxies of
all types as very likely GRB host galaxy candidates if the chance probability
$p$ of finding such an object of the given $R_C$-band magnitude in the
corresponding error circle is $\leq$10\% (within 1$\sigma$).  

Other criteria rely on the phenomenological appearance of the
galaxies, particularly their color.  Spiral galaxies can have a
$(R-K)_{\rm AB}$ color as red as about 3.5 mag before the Lyman dropout at high $z$
comes into play (SBG09; their figure 3). Galaxies with a redder $(R-K)_{\rm
AB}$ color are therefore of special interest, since they can be either 
dust-enshrouded or Lyman-dropped-out galaxies. These galaxies are
usually called extremely red objects (EROs) and were first addressed in
the context of deep NIR surveys \citep{Elston1988}. Number counts for these
galaxies are now available (\citealt{Gonzales2009a,Hempel2011a,Kim2011a}) and
will be used in the following. Finding an ERO galaxy in an XRT error
circle  can be considered
as strong evidence that this object is related to the
burst under consideration.

The ERO galaxy population follows a bimodal
distribution and consists of both passively evolving  ellipticals and
dusty star-forming galaxies in the redshift interval $1\lesssim z \lesssim
2$ \citep{Doherty2005a,Fontanot2010a,Gonzalez-Perez2011a}. 
The ratio of both populations is still  a matter of debate
(see \citealt{Conselice2008MNRAS383,Gonzales2009a,
Kong2009ApJ702,Fontanot2010a}). Since long GRBs tend to be hosted by
 star-forming galaxies, the
ERO number counts provide a conservative upper limit to the
probability of finding a star-forming ERO galaxy in a \swift/XRT error circle.

Finally,  a close pair of galaxies inside an XRT error circle, 
i.e., a hint of interaction and thus triggered star-formation, is 
a good candidate to be the birthplace of a (long) GRB progenitor.

\subsection{Notes for individual targets \label{notesind}}

We now report on the observations and results for each
individual target. In several cases, we could make use of
late-time as well as early-time GROND or VLT data. 
However, the comparison between different observing 
epochs did not reveal any fading afterglows, except for the case of GRB 070517, 
where a comparison with published 
Gemini-S data led to the identification of the afterglow.

\paragraph{\bf \object{GRB 050717}}

The burst occurred at relatively low Galactic latitude ($b=10^\circ$), and the
field is relatively crowded with stars. The foreground Galactic reddening is
moderate, $E(B-V)=0.24$ mag, but the highest in our sample. The 90\% c.l. XRT
error circle has a radius of $r_0=1\farcs5$.

\begin{figure}[t!]
\includegraphics[width=0.48\textwidth]{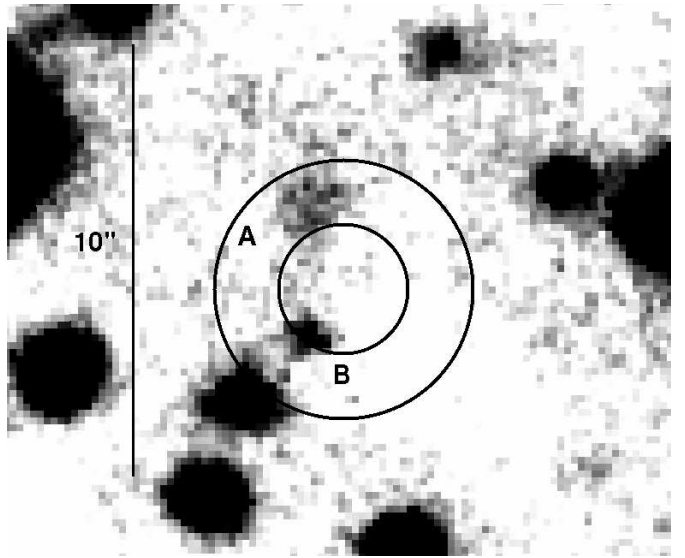}
\caption{GROND combined $r^\prime
i^\prime z^\prime$-band image of the field of GRB 050717. 
It shows the 90\% c.l. XRT error circle ($r_0=1\farcs5$), as well as a circle of 
radius $2r_0$. Here and in the following, east is left and north is up.}
\label{fig:050717}
\end{figure}

We observed the field with GROND two years after the burst.  Within 2$r_0$, two
objects (A and B) are visible in the combined $r^\prime i^\prime
z^\prime$-band image (Fig.~\ref{fig:050717}). Object A   ($r'_{\rm AB}$ =
23.6) lies outside 1$r_0$, appears fuzzy, and  has a size of about
$2\farcs1\,\times\,3\farcs9$.  On the basis of its visual appearance, this is a faint
galaxy.   Its outer parts extend into the 90\% c.l. error circle. The fainter
object B  ($r'_{\rm AB}$ = 24.5) lies  within  $1r_0$ close to the southern
boundary of the error circle.  Neither object is detected in $g^\prime$ and
they are also not seen in the NIR bands (Table~\ref{tab:PhotomGROND}).  Given the
non-detection in the NIR ($K_{\rm AB}> 21.1$),  for both objects only an
upper limit to $(R -K)_{\rm AB}$ can be given ( $<2.1$ mag and 
$<2.9$ mag,  respectively).\footnote{ Here and in the following, we
make the simplifying assumption  $(r'-K)_{\rm AB}= (R_C-K)_{\rm AB}$ and
write $(R-K)_{\rm AB}$ when we provide (extinction-corrected) colors based
on GROND data.} 

Assuming that A and B are galaxies, the probability $p$ of finding a galaxy of
the given $R_C$-band magnitude in a region of radius $2r_0$ and $1r_0$ is
about 0.05 and 0.03,  respectively.\footnote{ 
Here and in the following, we set
$R_C$(Vega) = $r'$(Vega) when calculating $p$-values based on Eq.~\ref{p}.} 
We therefore consider both objects as equally likely GRB host galaxy
candidates.

\paragraph{\bf \object{GRB 050922B}}

\begin{figure}[t!]
\includegraphics[width=0.48\textwidth]{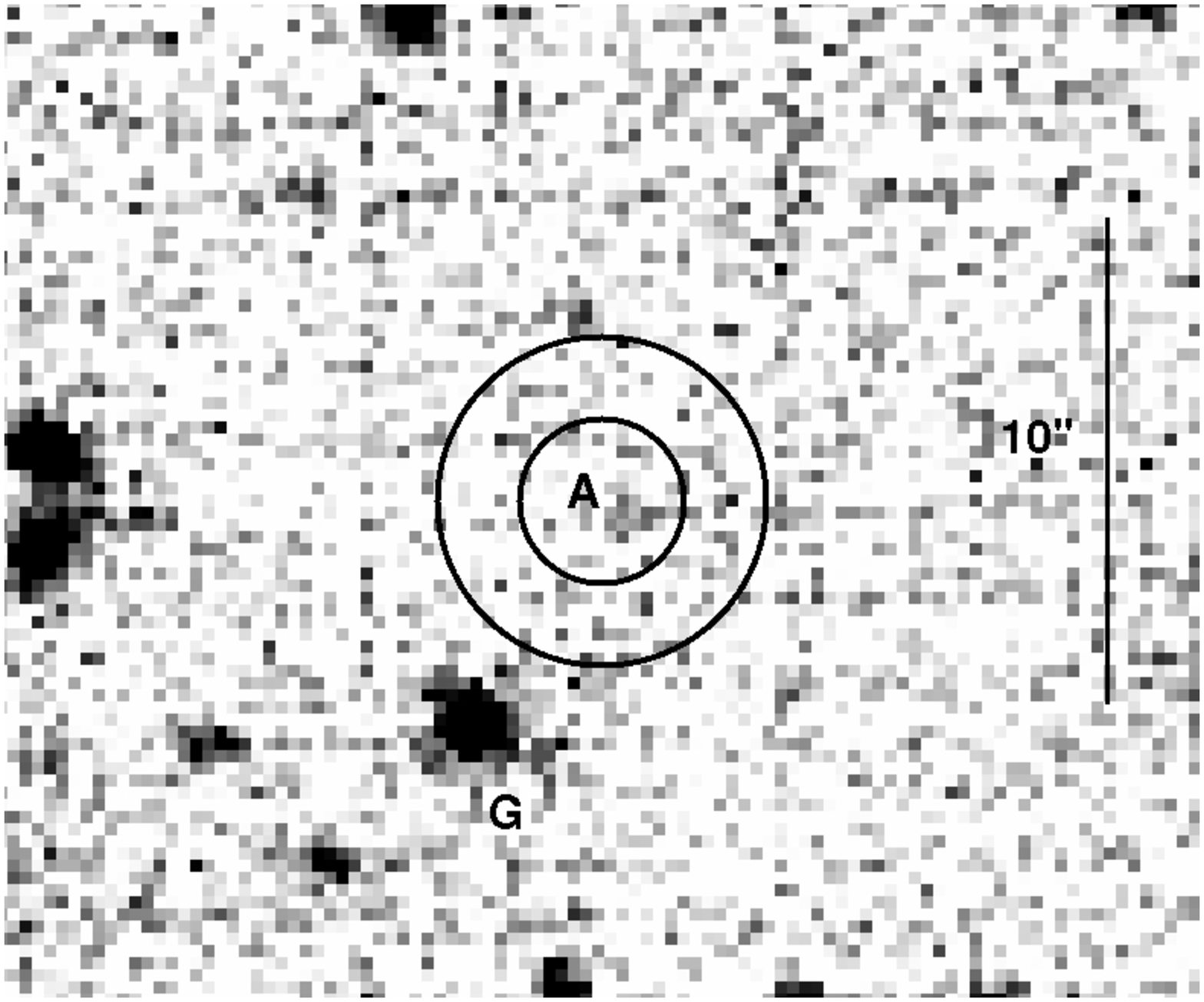}
\caption{FORS2 $R_C$-band image of the field of GRB 050922B. 
It shows the 90\% c.l. XRT error circle ($r_0=1\farcs7$), as well as a circle of 
radius $2r_0$.}
\label{fig:050922B}
\end{figure}

This burst occurred at high Galactic latitude ($b=-67^\circ$); the field is not
crowded with stars. The foreground Galactic reddening is very small, at $E(B-V) =
0.04$ mag. The 90\% XRT error circle has a radius of $r_0=1\farcs7$
(Fig.~\ref{fig:050922B}). The field was observed with NEWFIRM in the $K_s$-band
about three years after the burst. Additional data were obtained with FORS2 and
ISAAC one year later. No object is found in any band, neither within the 90\%
c.l.  XRT error circle nor within $2r_0$, down to deep $3\sigma$ upper limits
of $R_{\rm AB}>26.5$ and $K_{\rm AB}>22.8$.

\paragraph{\bf \object{GRB 060211A}}

The field of GRB 060211A lies at relatively low Galactic latitude
($b=-24^\circ$) but is not crowded with stars. The foreground Galactic
reddening is moderate, $E(B-V) = 0.19$ mag. The  90\% c.l. XRT error circle
has $r_0=1\farcs4$.

We observed the field 1.5 and 3 years after the burst with GROND and NEWFIRM
($J$ and $K$), respectively. In the 90\% c.l. error circle, we find one object
(A) in the GROND $r^\prime$ band, which looks slightly extended in the north-south
direction ($1\farcs1\,\times\,1\farcs2$).  The object is not visible in the
other GROND  bands. Complementary NEWFIRM observations also did not detect this
object  down to $J_{\rm AB}=23.4$ and $K_{\rm AB}=21.6$
 (Table~\ref{tab:PhotomGROND}). 
 Only an upper limit to the  $(R-K)_{\rm AB}$ color of object A can therefore 
 be given ($<2.5$ mag).

\begin{figure}[t!]
\begin{center}
\includegraphics[width=0.48\textwidth]{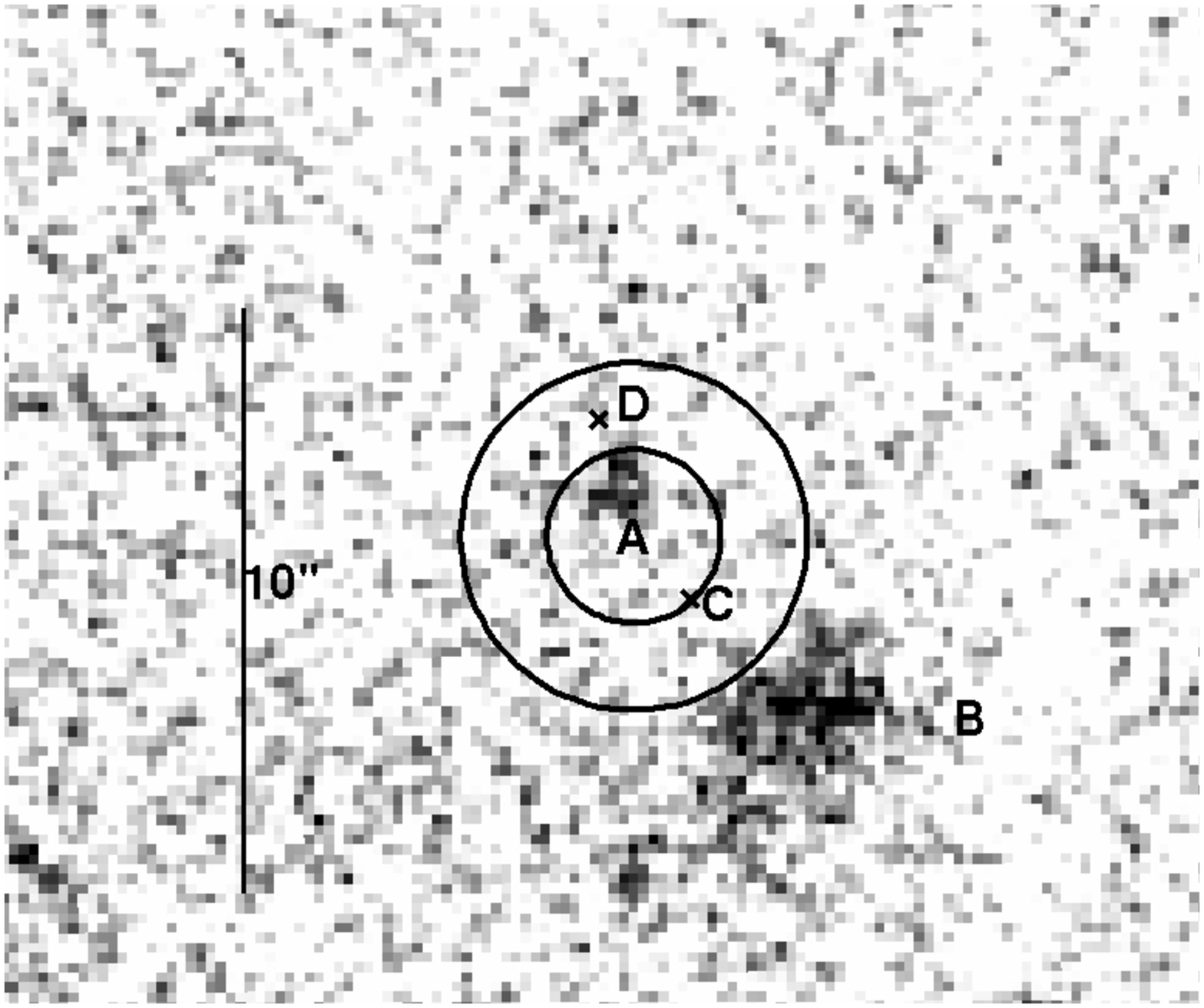}
\includegraphics[width=0.48\textwidth]{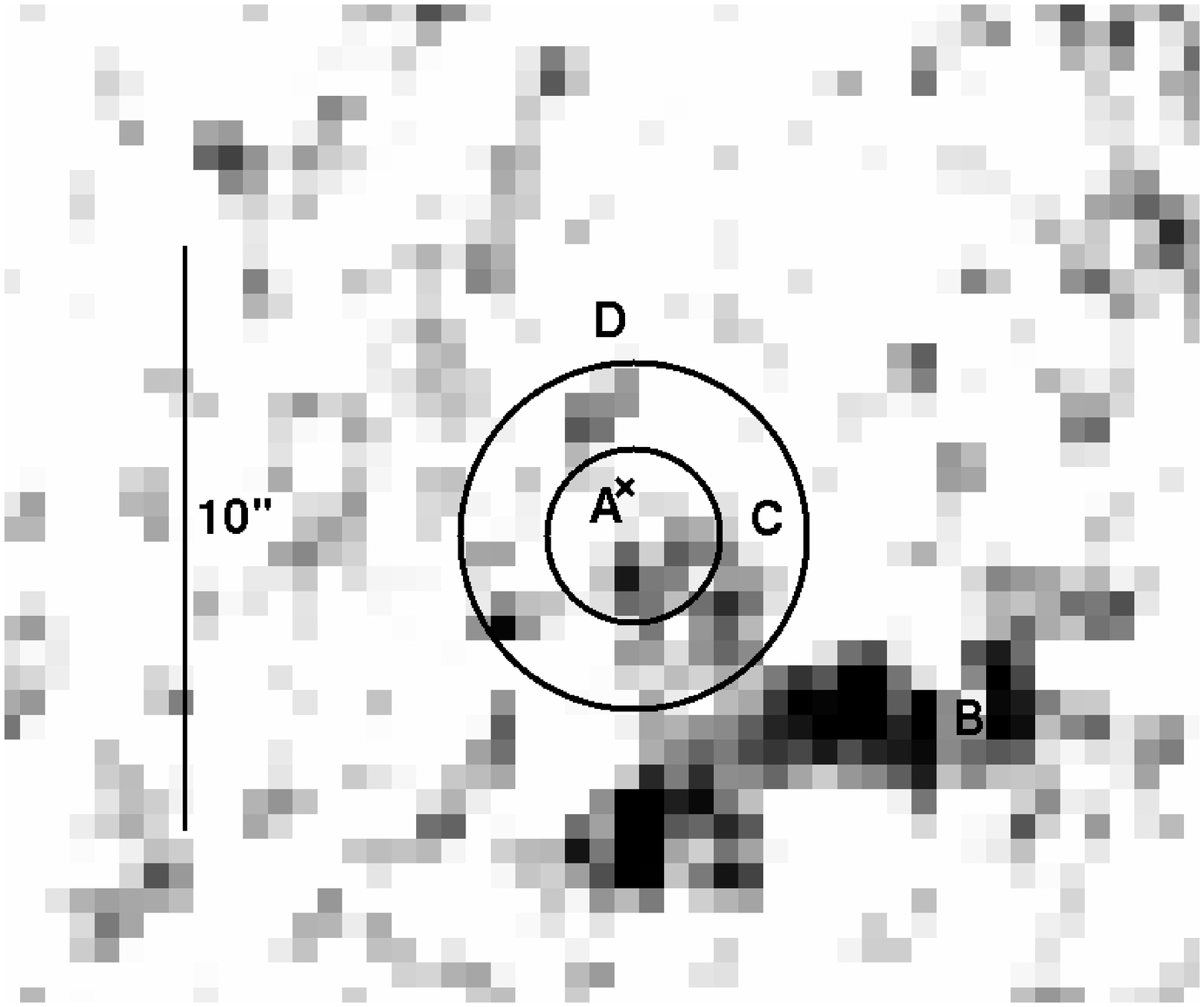}
\caption{ 
GROND $r'$-band (top) and NEWFIRM $J$-band image (bottom) 
of the field of GRB 060211A. 
It shows the 90\% c.l. XRT error circle ($r_0=1\farcs4$) as well as a circle of 
radius $2r_0$.  The cross in the GROND image indicates 
the position of objects C and D (only visible in $J$-band), 
while a cross in the NEWFIRM image indicates the position of object A
(detected only in the optical bands).}
\label{fig:060211J}
\end{center}
\end{figure}

In addition to A, an extended, fuzzy object (B)   is detected  with GROND
in $g^\prime r^\prime i^\prime z^\prime$, and located  about 4\farcs0 
($3r_0$)  south-west of the center of the  error circle 
(Fig.~\ref{fig:060211J}). This object is also seen in the NEWFIRM $J$-band
image, where it appears resolved into two or three sources.
In the $r^\prime$ band, its size is about
$3\farcs8\,\times\,2\farcs2$.

The NEWFIRM $J$-band image reveals an other two very faint sources, C and D.
Object C ($J_{\rm AB}\sim23.1$) is an extended object
lying within the 90\% c.l. error circle. This potential host galaxy is not
seen in any other band. Object D ($J_{\rm AB}\sim23.4$) lies outside the 90\%
c.l. error circle and is too faint for us to draw any conclusion about its morphology and
nature. The angular offset of D from the boundary of the 90\% c.l. error
circle is 0\farcs7. For a redshift of 0.5 or 1.0, this would correspond to  a
projected distance of 4.3 kpc and 5.6 kpc, respectively.  Compared to the
median projected angular offset of $1.3$~kpc found by \citet{Bloom2002a} for
a sample of 20 host galaxies of long bursts, this is a high but still
reasonable value for e.g. a Milky Way-like galaxy. In the case of object B,
the angular offset is 2\farcs5, corresponding to a projected distance of 15
kpc and 20 kpc, respectively, which most likely excludes object B as a host
galaxy  candidate. Unfortunately,  we cannot decide whether objects C and D are
potential ERO galaxies. An ERO  would have an $(r'-J)_{\rm AB}$ color of at
least 2 mag,  but our detection limit is insufficiently deep to check this
out. Assuming
that A is a single galaxy, the probability $p$ of finding a galaxy of the
measured $r'$-band magnitude within the 90\% c.l. error circle is about
0.03, while the corresponding value for object B is 0.08. We conclude that
in this case we cannot decide the most likely host galaxy candidate among 
A, C, or D.

\paragraph{\bf \object{GRB 060805A}}

The field lies at relatively high Galactic latitude ($b=60^\circ$).  It is not
crowded with stars but located close to a bright star ($R_C=13.5$) at RA,
Dec. (J2000) = 14:43:42.098, +12:35:20.63 (USNO-B1 catalog), which may affect
the background estimation.  The foreground Galactic reddening is small,
$E(B-V) = 0.02$ mag, among the smallest in our sample. The corresponding 90\%
c.l. XRT error circle has $r_0=1\farcs6$.

\begin{figure}[t!]
\includegraphics[width=0.48\textwidth]{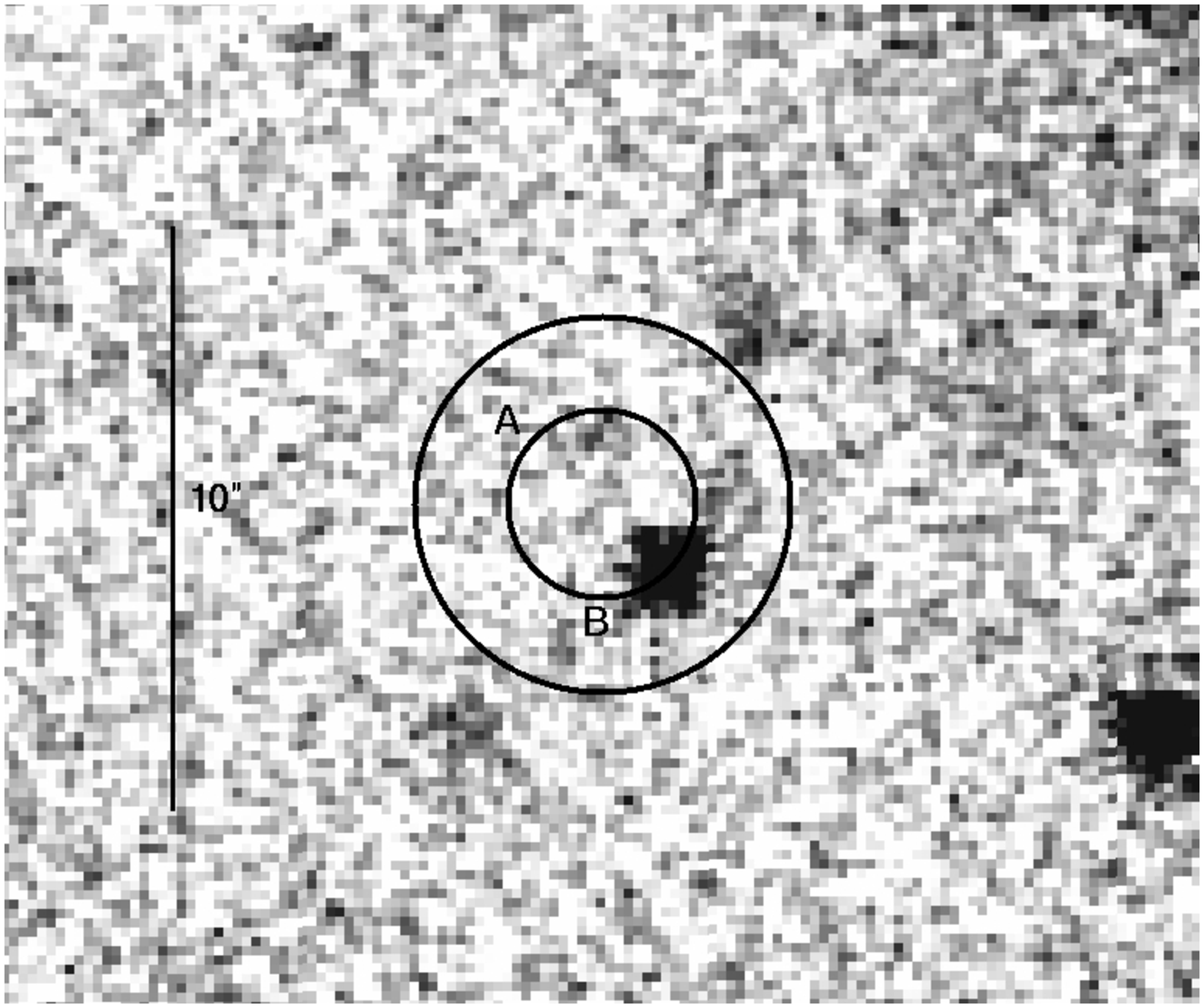}
\caption{GROND $r'$-band image of the field of GRB 060805A. 
It shows the 90\% c.l. XRT error circle ($r_0=1\farcs6$), as well as a circle of 
radius $2r_0$.}
\label{fig:060805A}
\end{figure}

The field was observed with GROND two years after the burst.  In the $r'$-band
image, we detect two sources (A, B; Fig.~\ref{fig:060805A}) within radii of $1r_0$
and $2r_0$ with magnitudes $r'=25.4$ and 23.7, respectively.  Both
objects appear extended. Object B, with a size of
$2\farcs7\,\times\,1\farcs3$, lies  about 2\farcs0 away from the center of the
XRT error circle but its outer regions extend into it.  In contrast to
object A, object B is also detected in the $g'$-band ($\sim23.4$) with a
$(g^\prime - r^\prime)_ {\rm AB}$  color consistent with a flat SED in this
wavelength region. This could imply that this galaxy is dominated by a young
stellar population. Both objects are not detected in the GROND  $i^\prime
  z^\prime JHK_s$ bands, where only deep upper limits could be derived
(Table~\ref{tab:PhotomGROND}).  The $(R -K)_{\rm AB}$ colors of A and B  are
$<4.3$ mag and $<2.5$ mag, respectively. 

Assuming that A and B are galaxies, the probability $p$ of finding a galaxy of
the measured $R_C$-band magnitude within $1r_0$ and $2r_0$, respectively, is
about 0.09 for object A and 0.09 for object B. We consider both, A and B, to be 
GRB host galaxy candidates.  The same conclusion was drawn by
\citet{Perley2009}, who observed this field in $g^\prime$ and $R_C$ using the
Keck telescopes. 

\paragraph{\bf \object{GRB 060919}}

The field of this burst lies at low Galactic latitude ($b=-17^\circ$) but
is not crowded with stars. The foreground Galactic reddening is small,
$E(B-V) = 0.07$ mag, and the 90\% c.l. XRT error circle is among the
smallest in our sample ($r_0=1\farcs7$).

\begin{figure}[t!]
\includegraphics[width=0.48\textwidth]{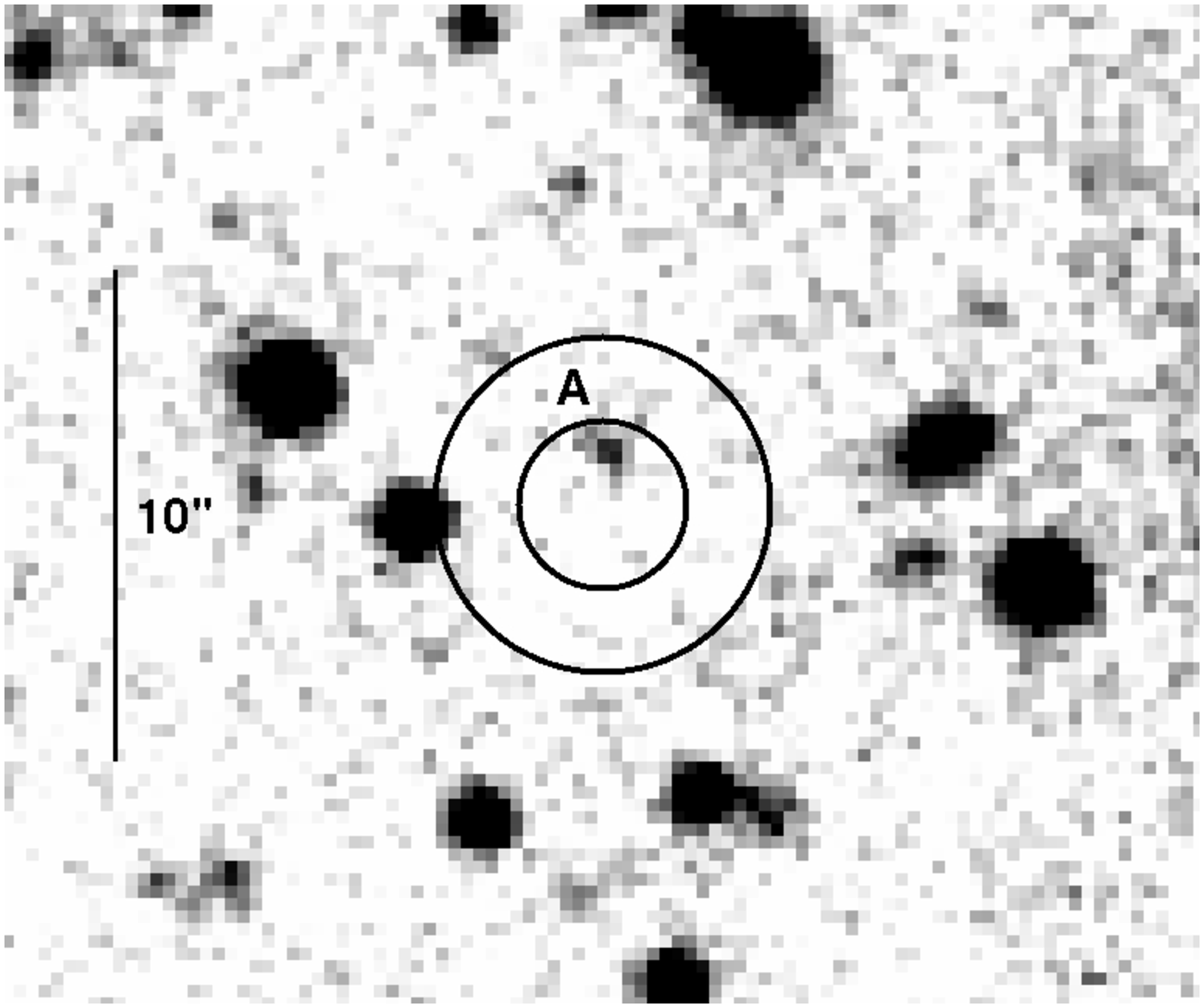}
\caption{FORS1 $R_C$-band image of the field of GRB 060919. 
It shows the 90\% c.l. XRT error circle ($r_0=1\farcs7$), as well as a circle of 
radius $2r_0$.}
\label{fig:060919A}
\end{figure}

The field was observed with FORS1 and ISAAC  about two years after the burst in
$R_C$ and $K_s$, respectively. We find only a single $R_C$-band source within
the 90\% c.l. error circle  (object A; Fig.~\ref{fig:060919A}),  with
$R_{\rm AB}=26.1$. No  other objects are visible even within $2r_0$. In the
$R_C$-band image, object A seems to be extended along the east-west direction
($1\farcs5\,\times\,1\farcs4$). It is undetected in the ISAAC image down to
deep flux limits ($K_{\rm AB}>23.4$). Its $(R-K)_{\rm AB}$ color is thus
$<2.6$ mag, well within the range of the colors of the known GRB host galaxy
population (SBG09). If this object is not the host, then we can provide 
upper limits for the GRB host galaxy of $R_{\rm AB}>26.5$ and $K_{\rm
AB}>23.4$. 

The probability of finding a galaxy of the measured $R_C$-band magnitude in a
region of radius $1r_0$ is 0.15. Given that object A is the only object we
detect within $2r_0$,  we suggest that it is the potential GRB host galaxy.  

\paragraph{\bf \object{GRB 060923B}}

The field lies at relatively low Galactic latitude ($b=18^\circ$) and is
relatively crowded with bright stars. The foreground Galactic reddening is
moderate, at $E(B-V) = 0.15$ mag. The corresponding 90\% c.l. XRT error circle has
$r_0=1\farcs8$.

\begin{figure}[t!]
\includegraphics[width=0.48\textwidth]{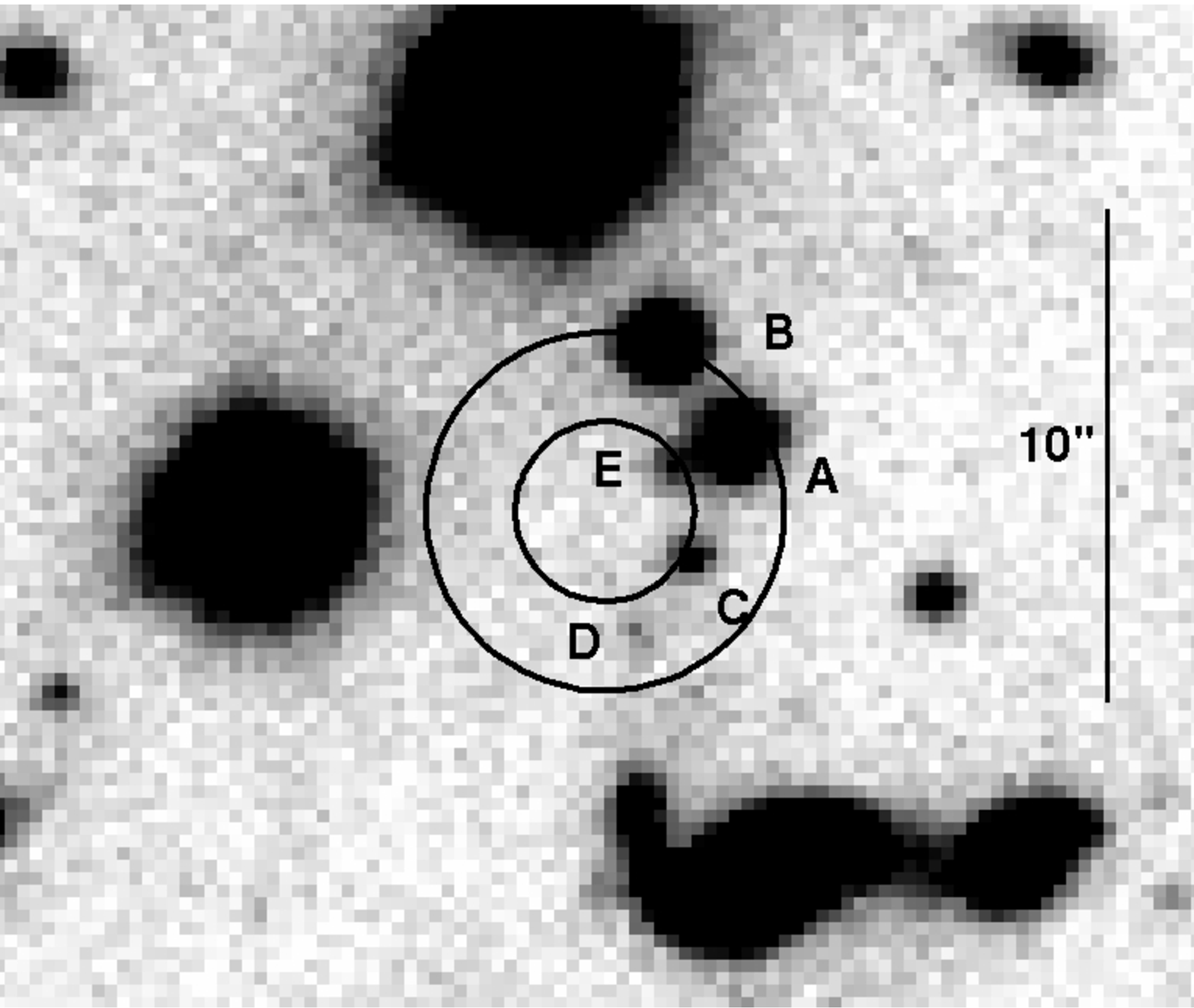}
\includegraphics[width=8.9cm]{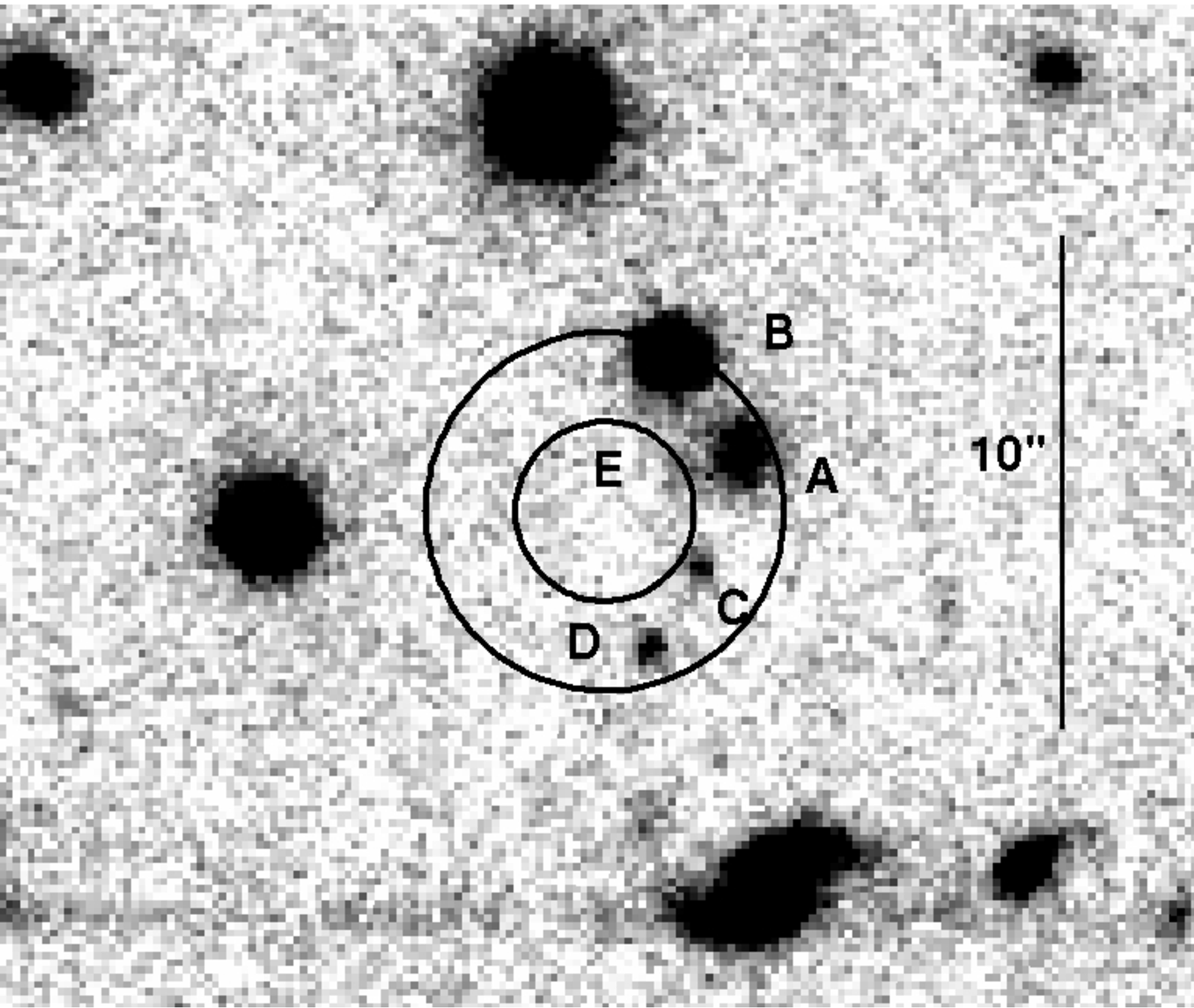}
\caption{FORS1 $R_C$-band (top) and 
ISAAC $K_s$-band image (bottom) of the field of GRB 060923B. Also shown
is the 90\% c.l. XRT error circle ($r_0=1\farcs8$), as well as a circle
of radius $2r_0$. The image reveals that object A is a galaxy. Object D is
an ERO.}
\label{fig:060923K}
\end{figure}

We observed the field with FORS1 and ISAAC about 1.5 years after the burst.
Our FORS1 $R_C$-band as well as our ISAAC $K_s$-band observations show 
two objects (E and C; Fig.~\ref{fig:060923K})  at the inner border of the
90\% c.l. error circle, while three more objects (A, B, and D) lie within 
$2r_0$.

Objects A ($R_{\rm AB}$= 23.1) and E are very close to each other,
making it difficult to  get a reliable $R$-band photometry, especially for
object E. Object B ($R_{\rm AB}$= 21.7) has a PSF that is point-like. In
the deep $K_s$-band image, object A shows an extended morphology
($2\farcs2\,\times\,2\farcs1$). Object C ($R_{\rm AB}=$
24.5) has a  point-like PSF but is probably too faint for  detecting the
faintest region of a galaxy. Object D appears slightly elongated in the
optical and the NIR images, but it is too faint for us to make any conclusion about its
morphology. Therefore, with high confidence only object A can be identified
as a galaxy.

The probability $p$ of finding a galaxy like A of the measured $R$-band
magnitude within a region of radius $2r_0$ is 0.06.  Objects C and E have
$p$-values of less than\footnote{Assuming for E a conservative
$R$=24.5$\pm$0.5 gives  $p=0.05\pm0.02$} 0.1, but it is difficult to conclude
anything about their nature. If E were a galaxy this would be extremely interesting,
because of its position close to galaxy A, which is indicative of a possible
interaction. On the other hand, object D ($R_{\rm AB}$=25.7) is an ERO with
$(R-K)_{\rm AB} \sim3.8$ mag, while A and C have moderately blue colors of 1.1 mag
and 1.3 mag, respectively. 

 Given the results mentioned above, we consider A, C, D,
and E as host galaxy candidates.

\paragraph{\bf \object{GRB 061102}}

The field lies at moderate Galactic latitude ($b=28^\circ$).  The foreground
Galactic reddening is small, $E(B-V) =0.04$ mag. The corresponding 90\% c.l. XRT
error circle has a radius of $r_0=2\farcs9$.

\begin{figure}[t!]
\includegraphics[width=0.48\textwidth]{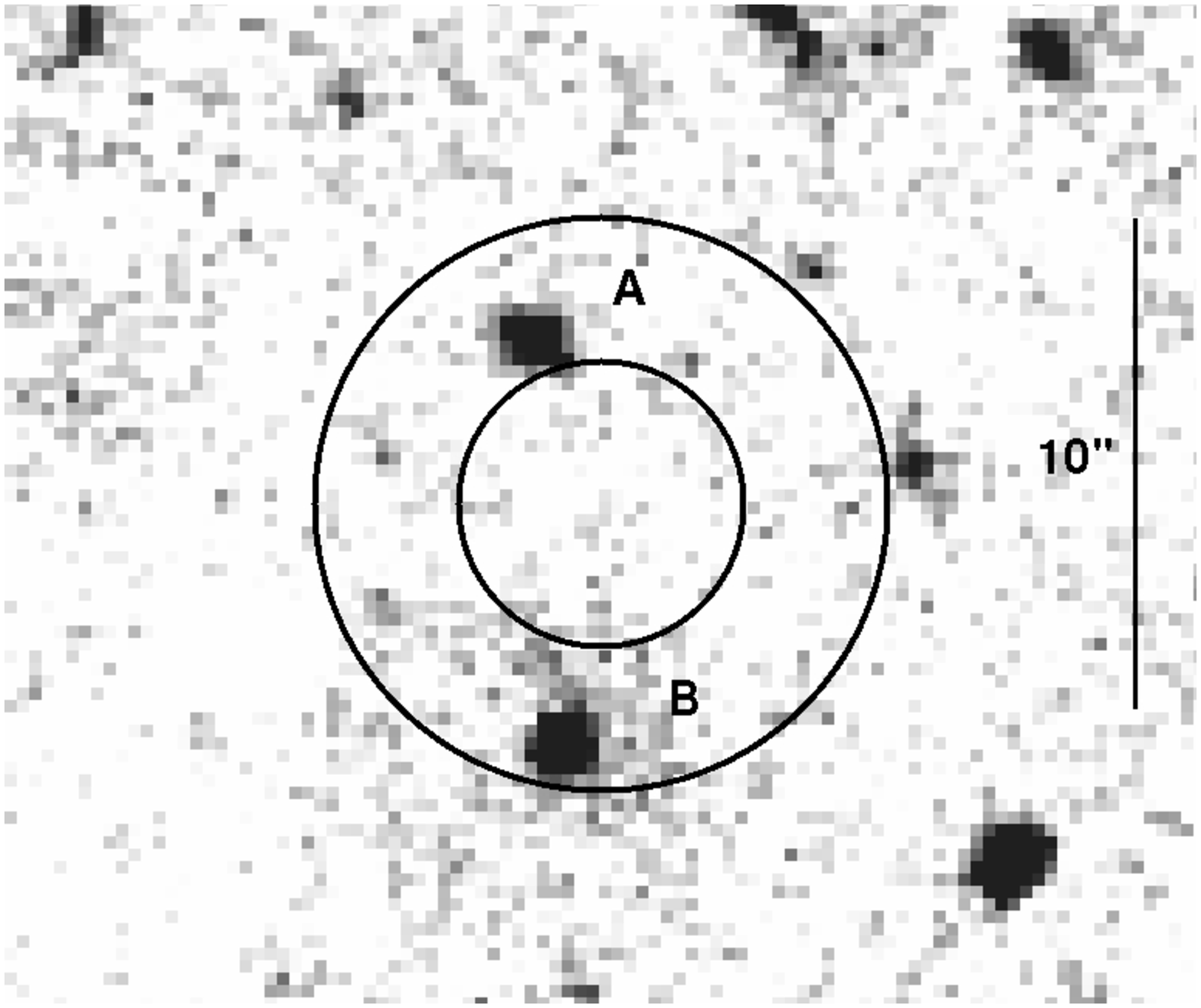}
\caption{FORS1 $R_C$-band image of the field of GRB 061102. 
It shows the 90\% c.l. XRT error circle ($r_0=2\farcs9$), as well as a circle of 
radius $2r_0$.}
\label{fig:061102A}
\end{figure}

We observed the field with FORS1 and ISAAC in $R_C$ and $K_s$, respectively,
about 1.5 years after the burst.  The VLT images show no object within the
90\% c.l. error circle down to $R_{\rm AB}=26.9$ and  $K_{\rm AB}=22.8$. Two
objects (A, B) are found within $2r_0$ 
(Fig.~\ref{fig:061102A})\footnote{ The enhanced 90\% c.l. XRT error circle 
(\citealt{Evans12250,Evans12273})
includes only object B. Its size is however 5\farcs7, which is 
twice as large as the error circle derived by 
\citet{Butler2007}.}. They are only
detected in $R_C$ but not in $K_s$.  Both objects are clearly extended
($2\farcs5\,\times\,1\farcs5$ and $1\farcs8\,\times\,1\farcs9$, respectively).
Their $(R-K)_{\rm AB}$ color ($\lesssim1.2$ mag and $\lesssim1.1$ mag,
respectively) matches the corresponding color of the GRB host galaxy
population  at a redshift of around $z=1$ (SBG09). In both cases,  the
probability of finding a galaxy of the given $R_C$-band magnitude inside a
region of $2r_0$ is 0.3. 

 Object A  touches the 90\% c.l. error circle, while  the brightness
center of object B lies 2\farcs0 away. However, B is surrounded by a
faint, asymmetric halo structure, which extends down to $1r_0$. We speculate
that this could be either a face-on spiral galaxy or the tidal tail of an
interacting system. If B lies at a redshift of, say, $z=0.3$ or 0.5, the
projected offset of the  afterglow from the center of this galaxy would be 9
kpc and 12 kpc, respectively.  This is a very large offset. It might be
smaller if the redshift were significantly 
lower, which would then point to a
rather  subluminous galaxy  (see also Sect.~\ref{redshiftsEstimate} and
Table~\ref{tab:Color}). 

If none of these sources is were the host galaxy, then the measured deep $R_C$ and
$K_s$-band upper limits would make the host galaxy of GRB 061102 one of the
faintest in our sample.

\paragraph{\bf \object{GRB 070429A}}

The field lies at moderate Galactic latitude ($b=-26^\circ$).  It is not
crowded with stars.  The foreground Galactic reddening is modest, at $E(B-V)
= 0.17$ mag. The corresponding 90\% c.l. XRT error circle has $r_0=2\farcs1$.

\begin{figure}[t!]
\begin{center}
\includegraphics[width=0.48\textwidth]{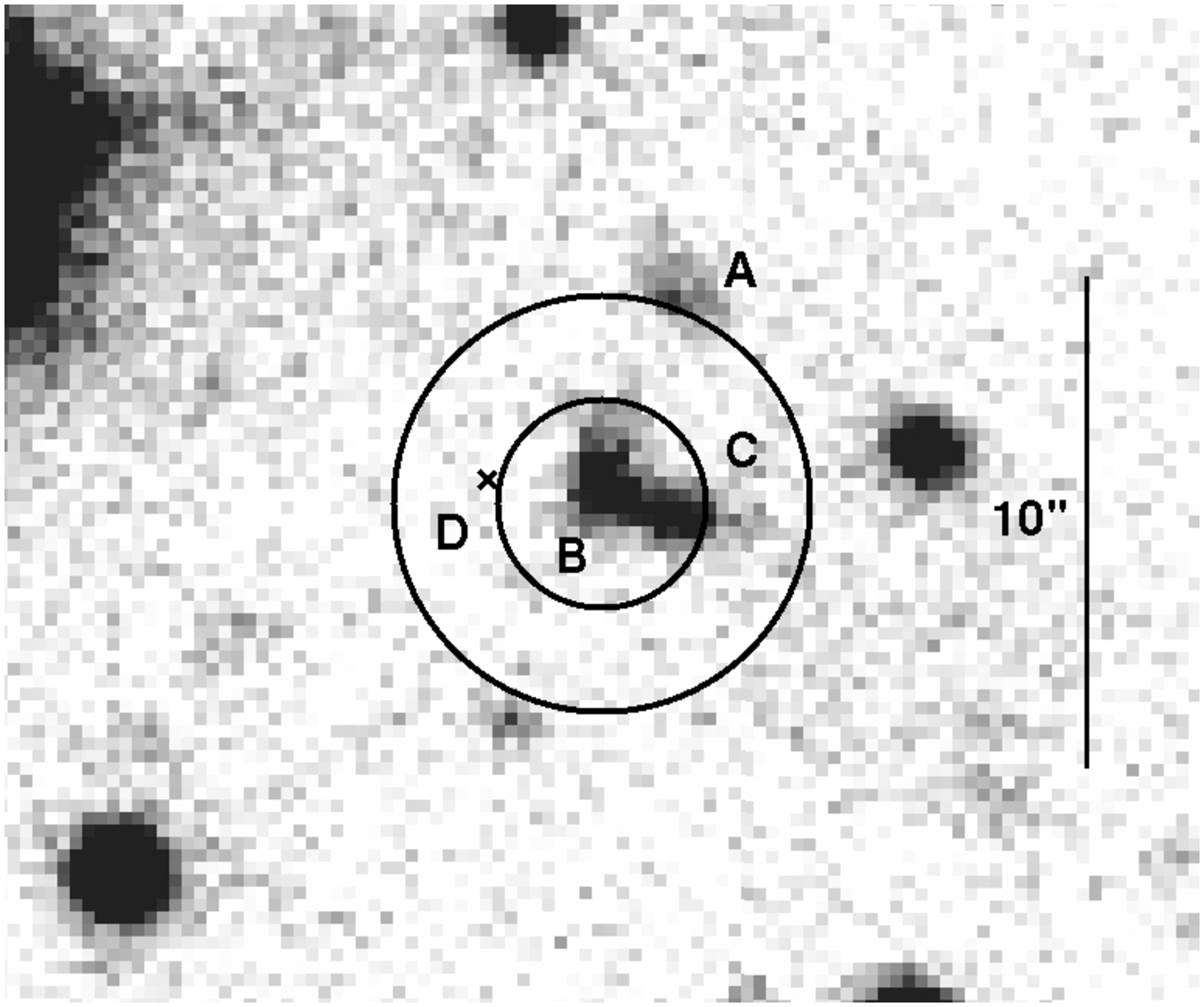}
\includegraphics[width=0.48\textwidth]{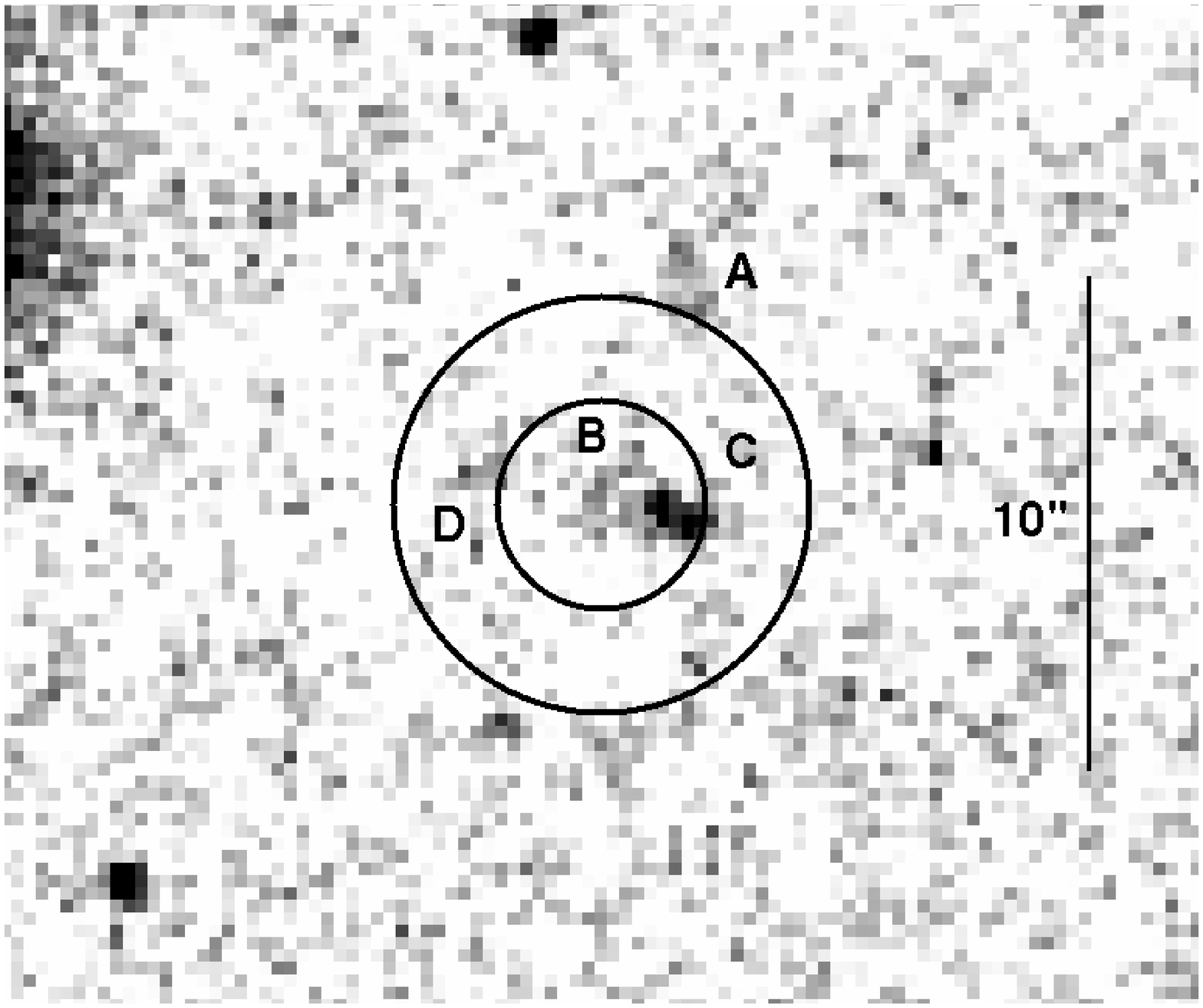}
\caption{ 
 FORS1 $R_C$-band image (top) and 
ISAAC $K_s$-band image (bottom) of the field of GRB 070429A. Also shown
is the 90\% c.l. XRT error circle ($r_0=2\farcs1$) as well as a circle
of radius $2r_0$. Object D is an ERO.
It is not visible in the $R_C$-band image, where it is indicated by a cross.}
\label{070429AK}
\end{center}
\end{figure}

We observed the field with FORS1 and ISAAC about one year after the burst.  In
the FORS1 $R_C$-band image (Fig.~\ref{070429AK}), we find one object (A) between
$2r_0$ and $3r_0$ and two other sources within the 90\% c.l. error circle 
(B,C), with
$R_C$-band magnitudes of $25.0 \pm 0.2, 24.1 \pm 0.1,$ and $24.3 \pm 0.1$,
respectively.  In the ISAAC $K_s$-band images, another object (D) is 
 marginally visible between $1r_0$ and $2r_0$
($K_s=23.0\pm0.4$; Fig.~\ref{070429AK}). All four objects are
extended (between 1\farcs7 and 3\farcs8 in their major axis). Objects B and C
may be an interacting pair because they have a fuzzy structure. The
individual $R_C,K_s$ magnitudes of objects A, B, and C 
(Table~\ref{tab:PhotomVLT}) and their
$(R-K)_{\rm AB}$ colors (Table~\ref{tab:size}) are compatible with the GRB
host population at a redshift $z<2$ (SBG09). Therefore, the observed colors of
objects A, B, and C  does not characterize any of them as very red.  
However, object D is very red ($(R-K)_{\rm AB}>3.2$ mag). Even though D
has a large 0.4 mag error in the $K_s$-band photometry, we consider it as a
potential ERO galaxy. Its center lies outside 1$r_0$, but its outskirts
reach into the 90\% c.l. error circle.

The probability-magnitude criterion gives the following numbers for the first
three galaxies (A-C): 0.54, 0.04, and 0.05, respectively.  Given that B and C
are located within the 90\% c.l. error circle, have low $p$-values, and are
probably an interacting galaxy system, we consider both galaxies as equally likely
host galaxy candidates.  In addition, D is also a host galaxy candidate
given its very red color.

\paragraph{\bf \object{GRB 070517} \label{ind070517} }

This burst is unique in our sample, because we could identify its afterglow by
comparing our late-time observations with the follow-up observations
reported by \citet{Fox6420}.

\begin{figure}[t!]
\includegraphics[width=0.48\textwidth]{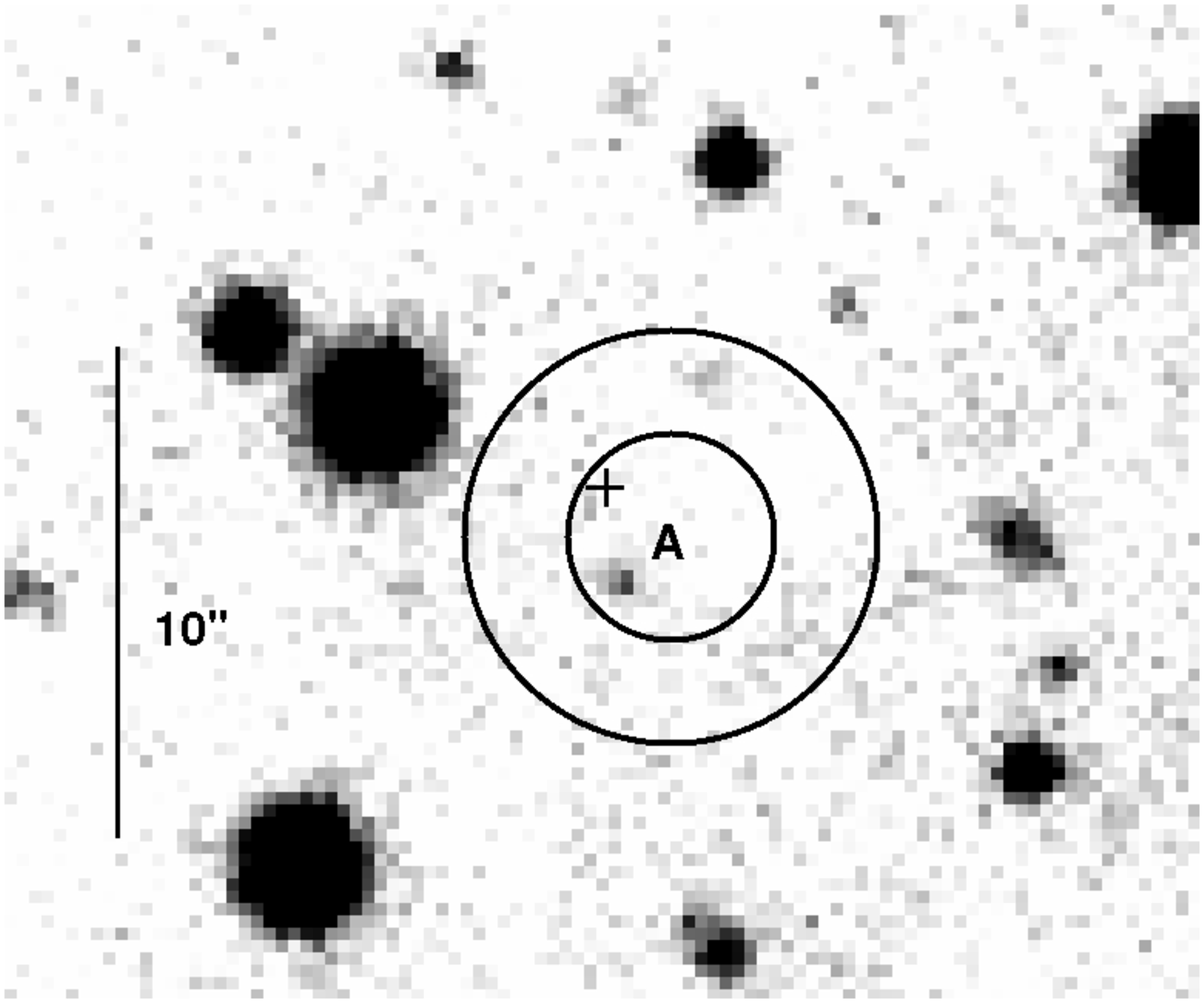}
\caption{FORS1 $R_C$-band image of the field of GRB 070517. 
It shows the 90\% c.l. XRT error circle ($r_0=2\farcs1$), as well as a circle of 
radius $2r_0$.}
\label{fig:070517}
\end{figure}

We observed the field with FORS1 and ISAAC about one year after the burst.
The field is at a relatively low Galactic latitude ($b=-21^\circ$) but is not very
crowded with stars. The foreground Galactic reddening is  modest, $E(B-V) =
0.15$ mag. The corresponding 90\% c.l. XRT error circle has $r_0=2\farcs1$ 
(Fig.~\ref{fig:070517}). 

In the $R_C$-band image, we detect only one object (A) within 
the 90\% c.l. XRT error circle with magnitudes $R_{\rm AB}=25.39\pm0.21$ 
and $K_{\rm AB} > 23.4$. No additional objects are apparent within $2r_0$. 
In particular, we do not detect the
$r^\prime=22.1$ afterglow candidate at RA, Dec. (J2000) = 18:30:29.12,
$-$62:17:50.7 (uncertainty of $<0\farcs75$ in each coordinate), which was
reported by \citet{Fox6420} based on Gemini-South observations about 16 h
after the burst (indicated by a cross in Fig.~\ref{fig:070517}). We conclude
that this was the GRB afterglow. 

The coordinates of object A agree with  the second object detected by
\cite{Fox6420} at RA, Dec. (J2000) = 18:30:29.08, $-$62:17:53.0 with
$i^\prime=24.5$.  On the basis of our images, we conclude that A is a galaxy.  Its
angular size is $1.6'' \times 1.3''$, which is about two times larger than the
stellar FWHM. If it is the GRB host galaxy, then its $(R-K)_{\rm AB}$ color
of $<1.7$ mag is compatible with the GRB host galaxy population at a
redshift around $z=1$ (SBG09). No underlying galaxy is found at the position
of the optical afterglow down to $R_{\rm AB}=26.6$ and $K_{\rm AB}=23.4$.

The angular distance between the afterglow and object A is
$1\farcs6\pm0\farcs3$. The probability $p$ of finding a galaxy of the given
$R_C$-band magnitude in a circle with this radius is  between 0.04 and
0.10. If A were the host galaxy of GRB 070517, then its angular distance would
translate into a projected distance of $12.8\pm2.4$ kpc, assuming a redshift
of $z=1$. This is ten times larger than the median projected angular offset of
$1.3$~kpc found by \citet{Bloom2002a} for a sample of 20 host galaxies of long
bursts, suggesting that object A is not the host. On the other hand, if we
require a projected angular distance of less than 10 kpc, then the upper limit
on the redshift of this galaxy is $z=0.4$. In this case, A would be a very
faint galaxy relative to the sample of SBG09.  Alternatively, the
true host galaxy could coincide with the optical afterglow position 
but be fainter than our detection limits. 
We conclude that we are unable to identify a good host galaxy candidate
for GRB 070517.

\paragraph{\bf \object{GRB 080207} \label{ind080207}}

The burst occurred at high Galactic latitude  ($b=66^\circ$), 
in a field that is not crowded with stars. 
The Galactic reddening is very small, at $E(B-V)=0.02$ mag. The
90\% c.l. XRT error circle is the smallest of our sample ($r_0=1\farcs4$).

\begin{figure}[t!]
\includegraphics[width=0.48\textwidth]{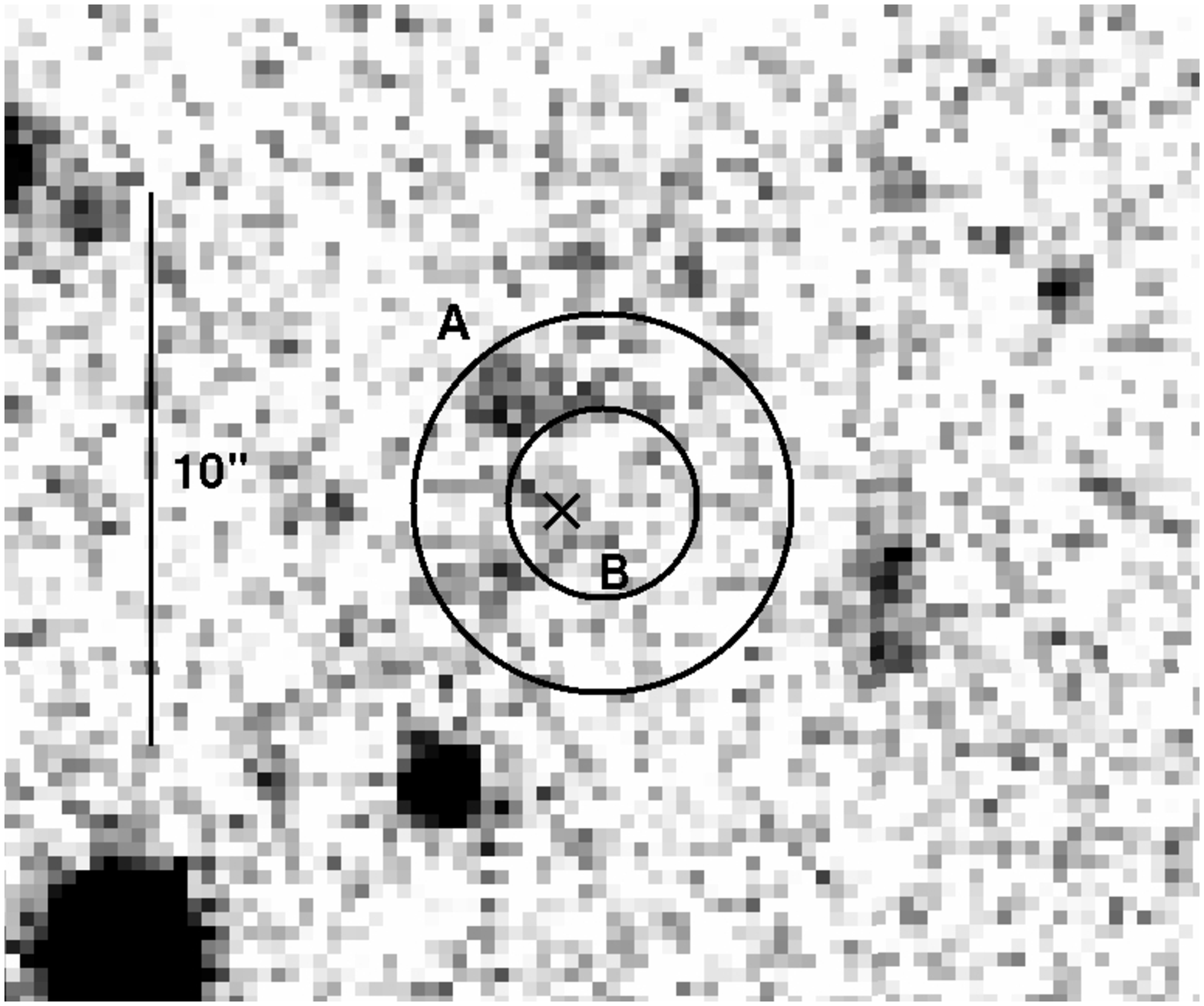}
\includegraphics[width=0.48\textwidth]{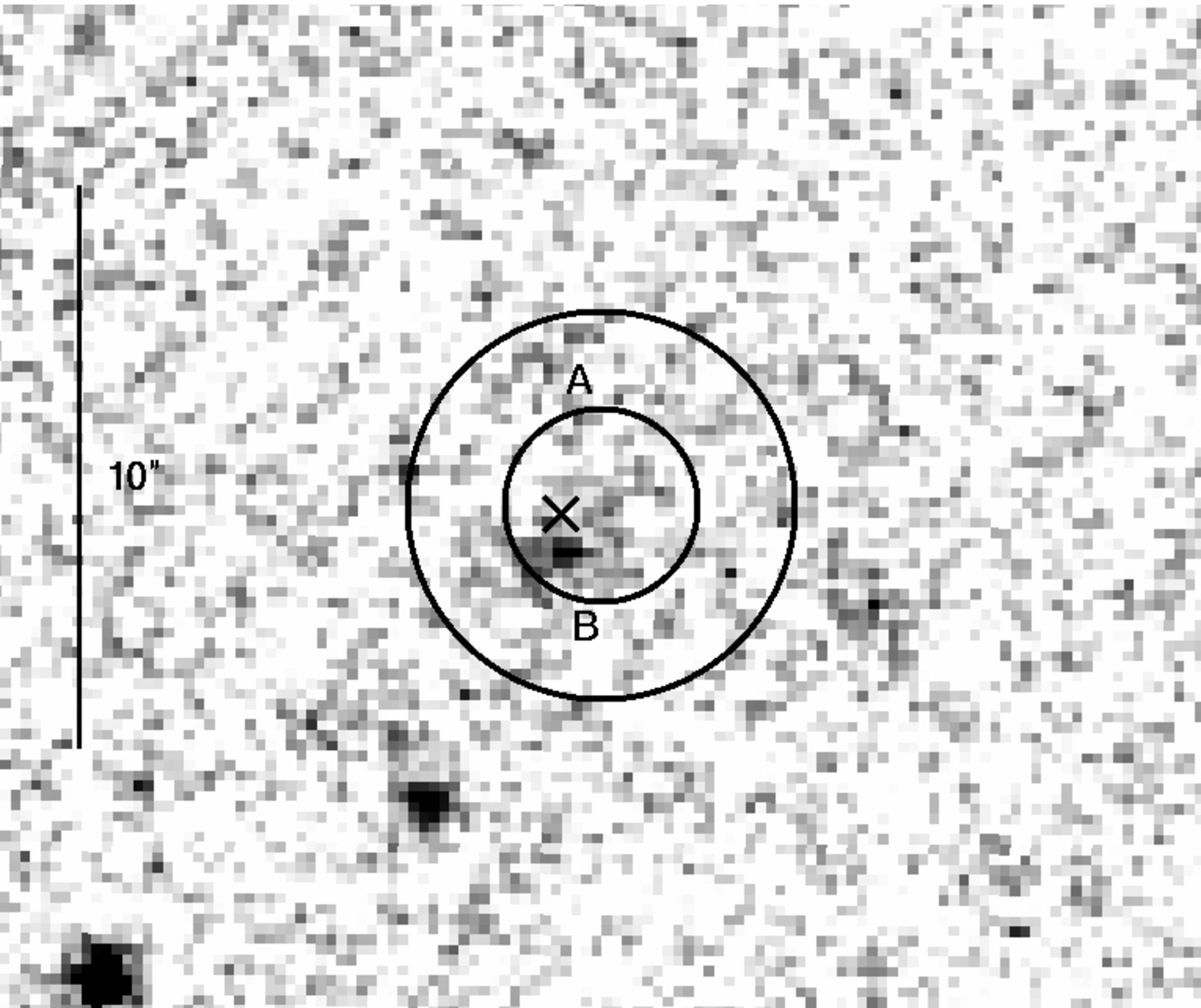}
\caption{ VIMOS $R_C$-band (top) and ISAAC $K_s$-band image
(bottom) of the field of GRB 080207, including the 90\%
c.l. XRT error circle ($r_0=1\farcs4$), as well as a circle of radius
$2r_0$. Object A is only visible in the VLT/VIMOS $R_C$-band image. 
Also indicated by  a cross is the position of the
\emph{Chandra} X-ray source.  Object B is an ERO.}
\label{080207K}
\end{figure}

We observed the field two years after the burst, with VLT/VIMOS in $R_C$ and
ISAAC in $K_s$. In addition, deep GROND imaging was performed at a mean time of ten
 hours after the burst, but no afterglow was detected
(Table~\ref{tab:darkULs}). Our deep VIMOS $R_C$-band image shows one fuzzy
object  of dimensions $2\farcs4\,\times\,1\farcs3$ 
at the northeast boundary of the 90\% c.l. error circle (A, Fig.~\ref{080207K}). This object is very faint in the
$K_s$-band. In addition, the ISAAC image shows another, elongated source (B;
$1\farcs6\,\times\,0\farcs9$) within the XRT error circle
that has a very faint $R_C$-band counterpart with $R_{\rm AB} \sim
26.5$. On the GROND images we do not detect these sources in any band, only upper limits can
be provided (Table~\ref{tab:PhotomGROND}). Object B is very red, $(R-K)_{\rm
  AB}\sim4.7$ mag. Its color and morphology defines it as an ERO galaxy. Given
its position within the 90\% c.l. error circle, we consider B as the most
likely GRB host candidate.

As this paper was being finalized, the  \emph{Chandra} source catalogue
(\citealt{Evans2010}) became public.  Inspection of the catalogue shows that a
X-ray observation of the field was performed 8 days after the burst and a
point source was detected (CXO J135002.9+073007) at coordinates RA, Dec.
(J2000) = 13:50:02.97, 07:30:07.8 $(\pm 0\farcs6$). The position of this
source is within $1\sigma$ consistent with the position of object B.
Therefore, we conclude that this is the host galaxy of GRB 080207. 
\cite{Hunt2011a} derived a photometric redshift of about 2.2
for this galaxy, which was subsequently confirmed spectroscopically
(z=2.086; \citealt{Kruhler2012a}).

\paragraph{\bf \object{GRB 080218B}}

The field is at a relatively low Galactic latitude ($b=9^\circ$), the lowest of
our sample. However, it is only moderately crowded with stars. The  Galactic
reddening along the line of sight  is modest, at $E(B-V)=0.17$ mag. The 90\% 
c.l. XRT error circle has a radius of $r_0=1\farcs6$.

\begin{figure}[t!]
\includegraphics[width=0.48\textwidth]{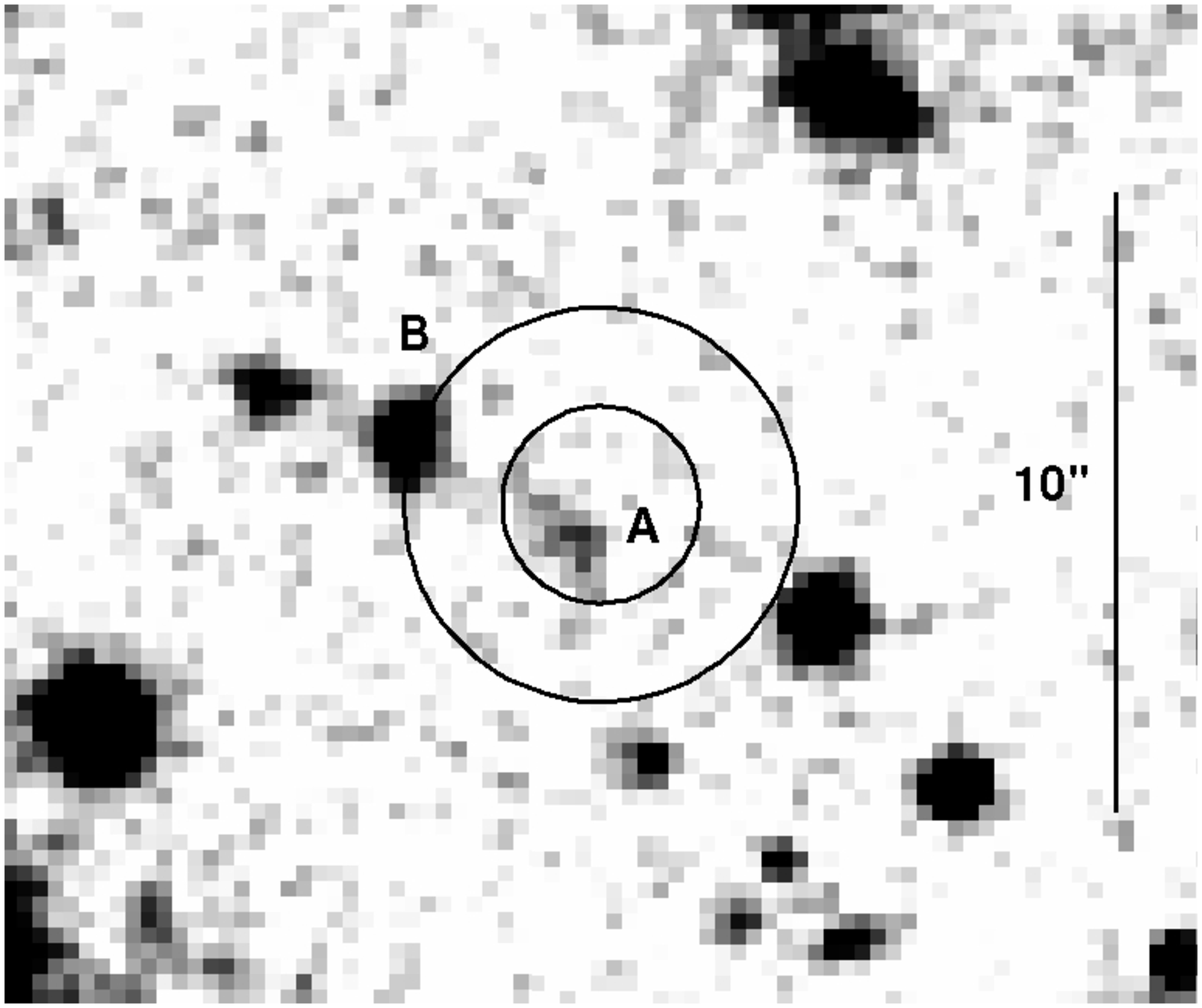}
\includegraphics[width=8.9cm]{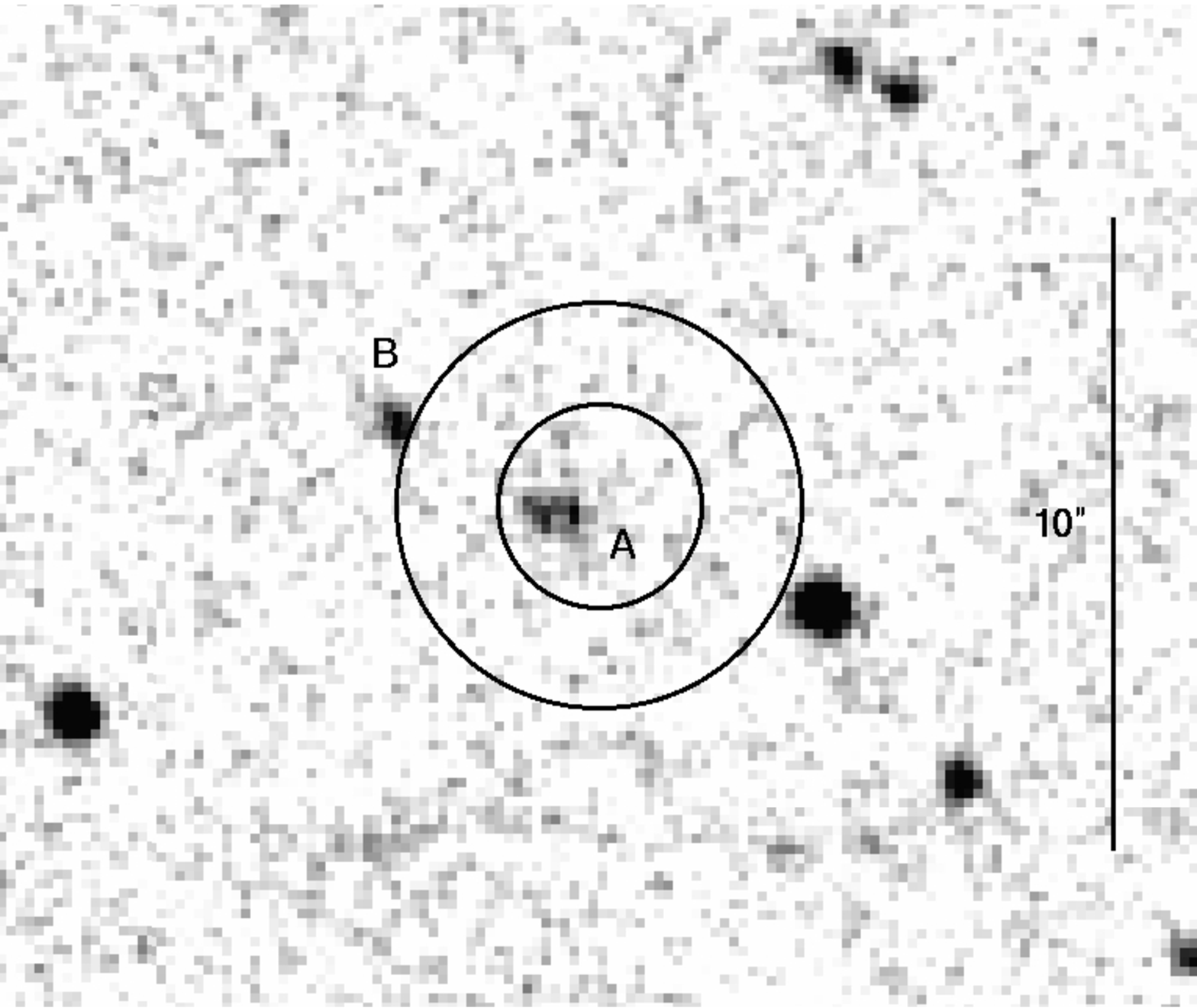}
\caption{FORS2 $R_C$-band image (top) and 
ISAAC $K_s$-band image (bottom) of the field of GRB 080218B,
including the 90\% c.l. XRT error circle ($r_0=1\farcs6$), as well as a circle
of radius $2r_0$.  Object A is an ERO.}
\label{080218K}
\end{figure}

We observed the field with FORS2 and ISAAC about one year after the burst.  In
addition, deep GROND imaging  was performed at a mean time of about 0.75~h 
after the burst, but no afterglow was detected
(Table~\ref{tab:darkULs}). Our deep FORS2 $R_C$-band image reveals one faint
($R_{\rm AB}=26.2$), extended object (A) within the 90\% c.l. error circle and
another object (B; $R_{\rm AB}=24.6$) inside  $3r_0$. Both objects are
also detected with ISAAC at magnitudes  $K_{\rm AB}=21.7$ and 22.7,
respectively (Fig.~\ref{080218K}). Object A is too faint to be detected by
GROND, while B is detected in $g^\prime r^\prime z^\prime$
(Table~\ref{tab:PhotomGROND}; it is not seen in $i^\prime$ owing to ghost images
in the field).

In the FORS image, object A is elongated in the southwest-northeast direction
($2\farcs5\,\times\,1\farcs1$).  It could be a spiral galaxy seen nearly
edge-on or a tight pair of galaxies. If it is a single galaxy, then its large
$(R-K)_{\rm AB}$ color (4.2 mag) defines it as an ERO galaxy. The probability of
finding a galaxy of this $R_C$-band magnitude within an area of radius $1r_0$
on the sky is about 0.1.  Given its extremely red color and its position inside
the 90\% c.l. error circle, we consider A as the most likely GRB host galaxy.

\paragraph{\bf \object{GRB 080602} \label{ind080602}}

The field is at high Galactic latitude ($b=-71^\circ$), among the highest in
our sample. The  Galactic reddening is very small, at $E(B-V)=0.03$ mag. The
90\% c.l. XRT error circle has $r_0=1\farcs7$.

\begin{figure}[t!]
\includegraphics[width=0.48\textwidth]{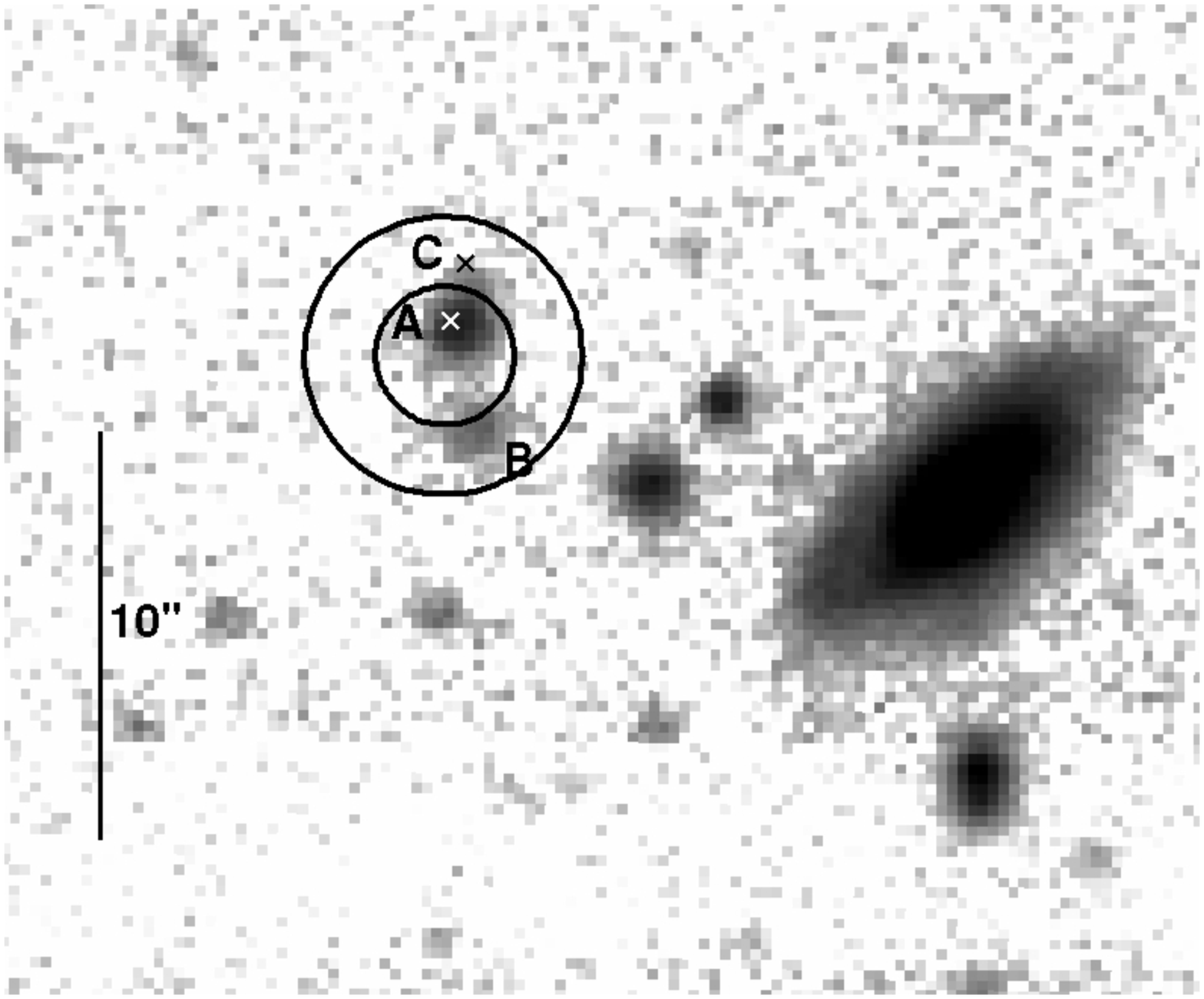}
\includegraphics[width=8.9cm]{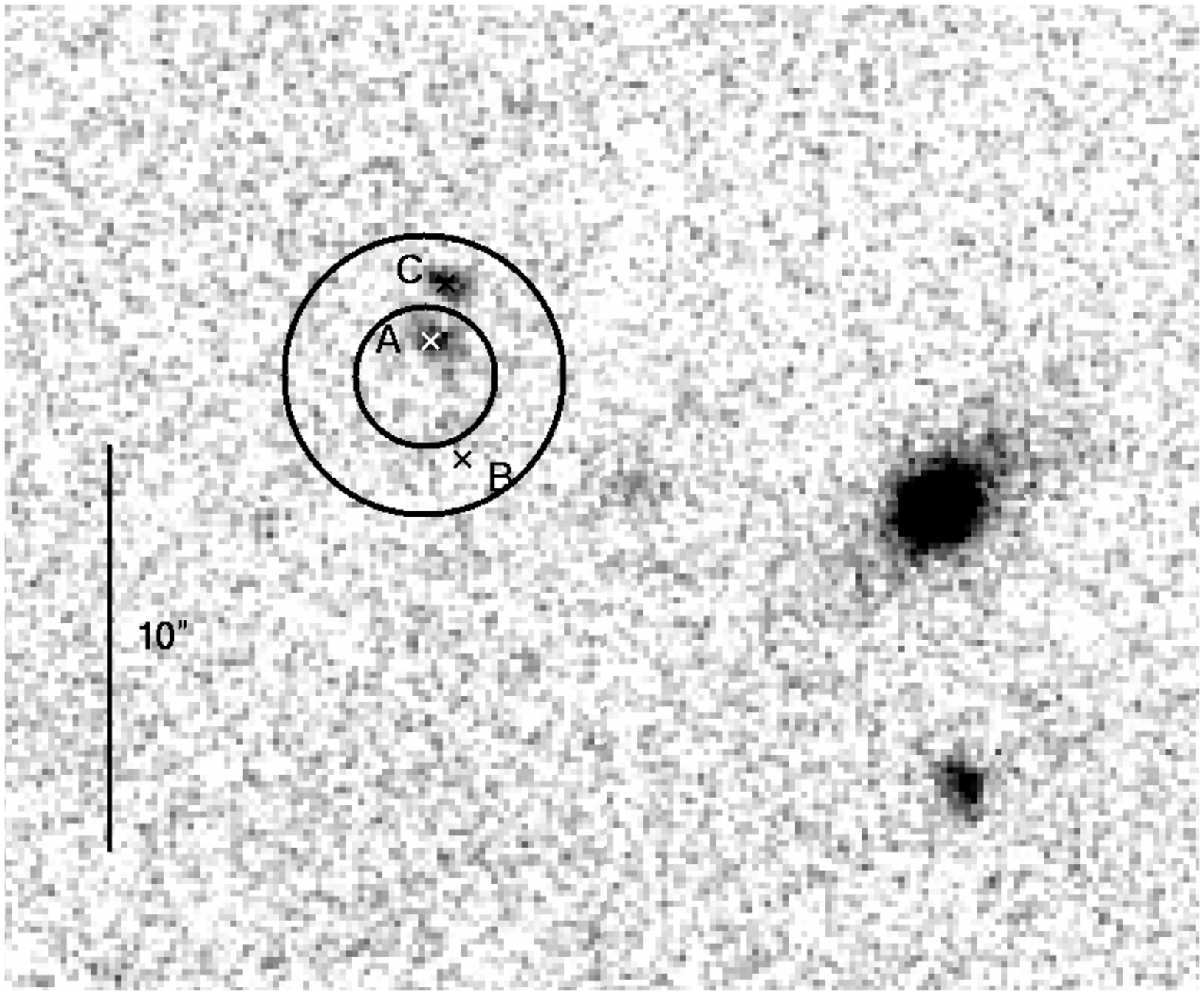}
\caption{FORS2 $R_C$-band (top) and
ISAAC $K_s$-band image (bottom) of the field of GRB 080602,
including the 90\% c.l. XRT error circle ($r_0=1\farcs7$), as well as a circle
of radius $2r_0$.  Object C is an ERO. The crosses indicate the positions of 
objects A and B in the FORS image.}
\label{080602}
\end{figure}

We retrieved VLT/FORS2 and ISAAC data obtained  about one year after the burst
from the ESO archive (program ID 081.A-0856; PI: P. Vreeswijk).  In addition,
deep GROND multi-color imaging was performed 1.5 years after the burst.  In
the FORS2 $R_C$-band image, we find one object (A; $R_{\rm AB}=22.9$) inside
the 90\% c.l. error circle. It is also detected in all GROND optical bands and
also seen in the ISAAC $K_s$-band image ($K_{\rm AB}=22.5$), where it  
seems to split into two objects, with the second one (C) being $1\farcs3$ north
of A (Fig.~\ref{080602}).  Even though in the FORS image, object A looks
extended in the northern direction,  C has no direct optical counterpart: The
angular distance between A and C on the ISAAC image is larger by about 0\farcs5
than the distance between the brightness center of A and its fainter
northern blob on the FORS image. Therefore, we consider C as a separate
object. Its $(R-K)_{\rm AB}$ color ($>$4.3 mag) defines it as an ERO.

\begin{figure}[htbp]
\includegraphics[width=8.9cm]{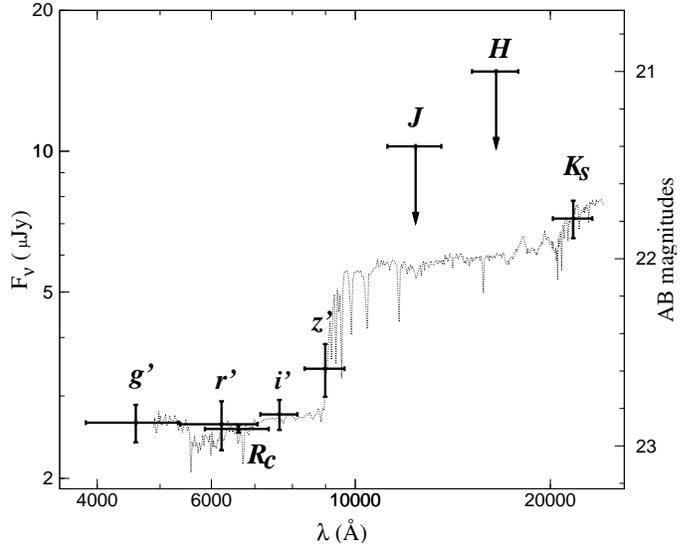}
\caption{
\emph{Hyperz} best-fit solution of the broad-band SED of object A  in the XRT
error circle of GRB 080602 (Tables \ref{tab:PhotomVLT}
and \ref{tab:PhotomGROND}).  From left to right: GROND $g^\prime$, GROND $r^\prime$, FORS2 $R_C$, 
GROND $i^\prime$, GROND $z^\prime$, GROND $J$, GROND $H$, and ISAAC $K_s$. 
The best fit corresponds to a spiral galaxy at a 
redshift of $z=1.40^{+0.30}_{-0.15}$ with no intrinsic extinction 
($\chi^2$/d.o.f $= 0.074$).}
\label{080602sed}
\end{figure}

At the southern boundary of the 90\% c.l. error circle lies another object (B;
size $2\farcs0\,\times\,1\farcs8$), which is possibly another
galaxy.  In both $R_C$ and $K_s$-band images, object A looks fuzzy and
extended ($2\farcs6\,\times\,2\farcs0$), while the nature of C is less
obvious. Assuming that A (including its northern blob) is a single
galaxy, a fit of its SED with \emph{Hyperz}  \citep{Bolzonella2000} gives good
solutions for both a spiral galaxy with no intrinsic extinction at a redshift of
$z=1.40^{+0.30}_{-0.15}$ ($\chi^2$/d.o.f $=0.074$), and a starburst
galaxy at a redshift of $z=2.10^{+0.20}_{-0.35}$ with a moderate Milky Way (MW)
extinction of $A_V=0.4$ mag ($\chi^2$/d.o.f $= 0.050$;
Fig.~\ref{080602sed}). 
This twofold solution is due to the SED being fit equally well by a $2175$\AA \ absorption feature or a 
$4000$\AA \ Balmer jump in the $z^\prime$-band.
We caution, however, that while the first solution implies an absolute magnitude
$M_B\sim -23.0$, 
in the case of the $z\sim2.1$ solution we obtain $M_B\sim -24.0$,
which is very unlikely when compared to the luminosity function found in the
Las Campanas redshift survey \citep{Lin1996}. Therefore, we consider
$z=1.4^{+0.30}_{-0.15}$ as the most likely redshift estimation.

Objects A and B have colors $(R-K)_{\rm AB}=0.3$ mag and $<0.4$ mag,
respectively, which is well within the  range of the observed colors for GRB
host galaxies (SBG09). In the case of object A, the probability of finding a
galaxy of the given $R_C$-band magnitude inside a circular area of radius
$1r_0$ is 0.01, while for B the corresponding value is 0.13 (within $2r_0$).
However, the probability of finding an ERO (object C) within the same area is
much smaller (see Sect.~\ref{EROsII}). Therefore, we consider  object C as
well as its  (possibly interacting) partner A as  the most likely 
host galaxy candidates.

\paragraph{\bf \object{GRB 080727A}}

The field lies at moderate Galactic latitude ($b=42^\circ$) and is not   
crowded by stars. The  Galactic reddening is very small, at $E(B-V)=0.07$ mag. The
90\% c.l. XRT error circle has $r_0=1\farcs6$.

We observed the field with ISAAC about 1.5 years after the burst.  The deep
FORS1 $R_C$-band image was taken from the ESO archive (program ID 081.A-0856;
PI: P. Vreeswijk; FWHM of $0\farcs8$).   No GRB host galaxy is detected
within $2r_0$, down to $R_{\rm AB}=26.3$ and $K_{\rm AB}=23.0$.  This is the
second case (besides GRB 050922B) in our sample where only deep upper
limits can be provided for the GRB host galaxy within $2r_0$.  A moderately
bright, nearly edge-on galaxy ($R_{\rm AB}=23.4$;
size $4\farcs5\,\times\,2\farcs0$) lies $10''$ west of the  center of the
XRT error circle. This object lies too far away  from the XRT 
error circle to be physically related to the GRB. 

\begin{figure}[t!]
\includegraphics[width=0.48\textwidth]{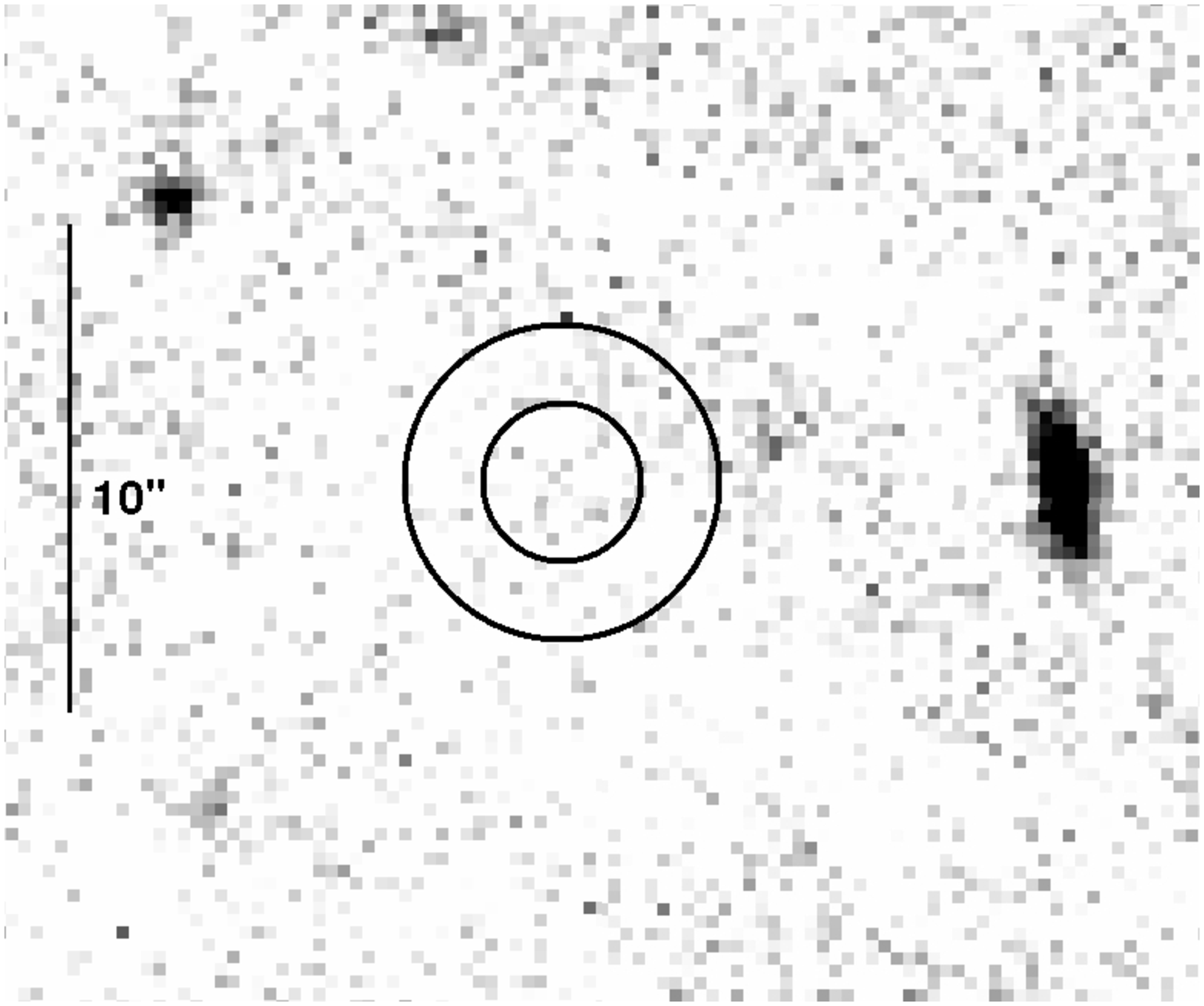}
\caption{FORS1 $R_C$-band image of the field of GRB 080727A. 
It shows the 90\% c.l. XRT error circle ($r_0=1\farcs6$), as well as a circle of 
radius $2r_0$.}
\label{080727}
\end{figure}

\paragraph{\bf \object{GRB 080915A}}

The field does not lies at low Galactic latitude ($b=-41^\circ$), but it is
relatively crowded with stars.  The  Galactic reddening is very small,
 at $E(B-V)=0.05$ mag. The 90\% c.l.  XRT error circle is of median size
($r_0=3\farcs7$).

\begin{figure}[t]
\includegraphics[width=0.48\textwidth]{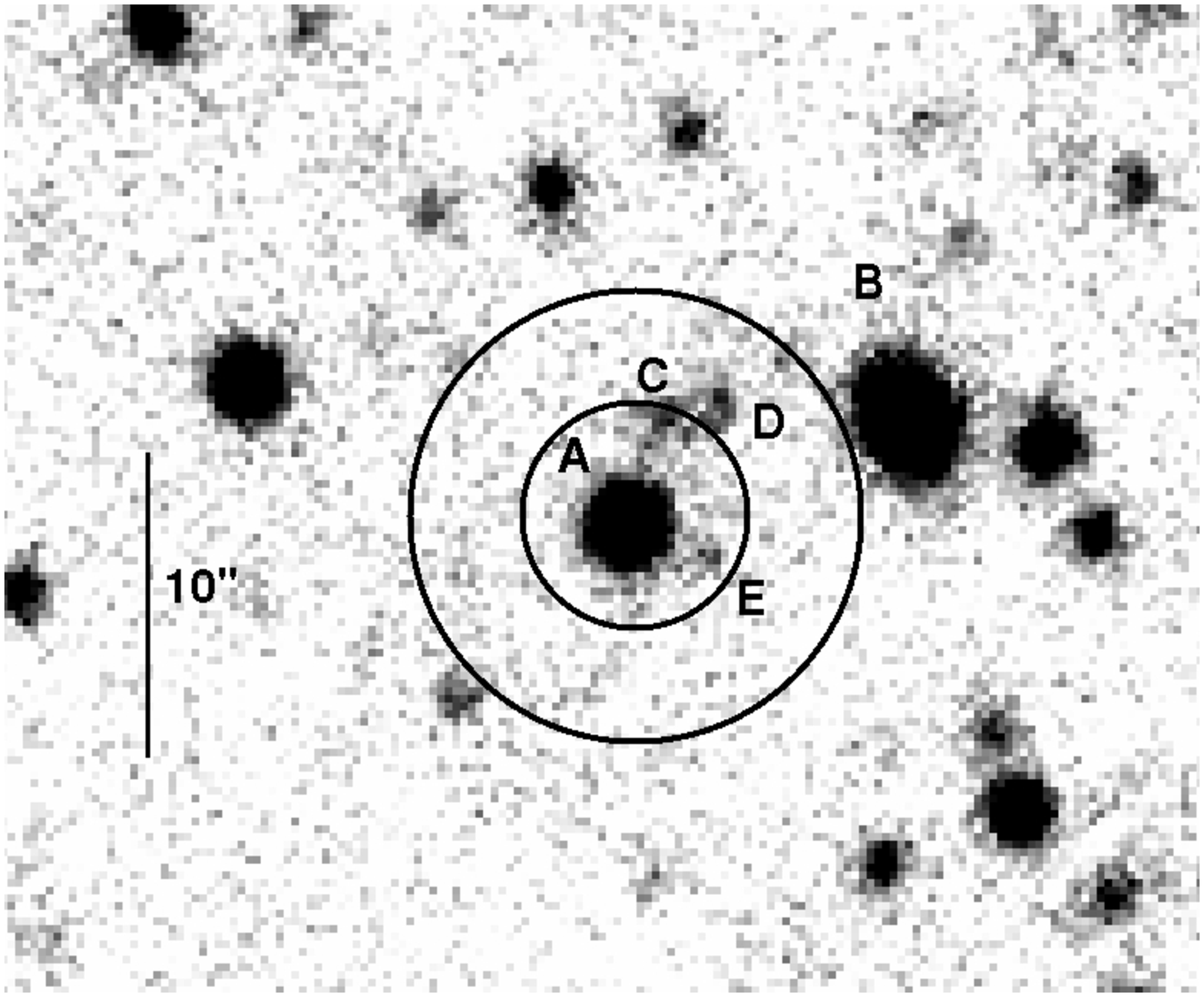}
\includegraphics[width=0.48\textwidth]{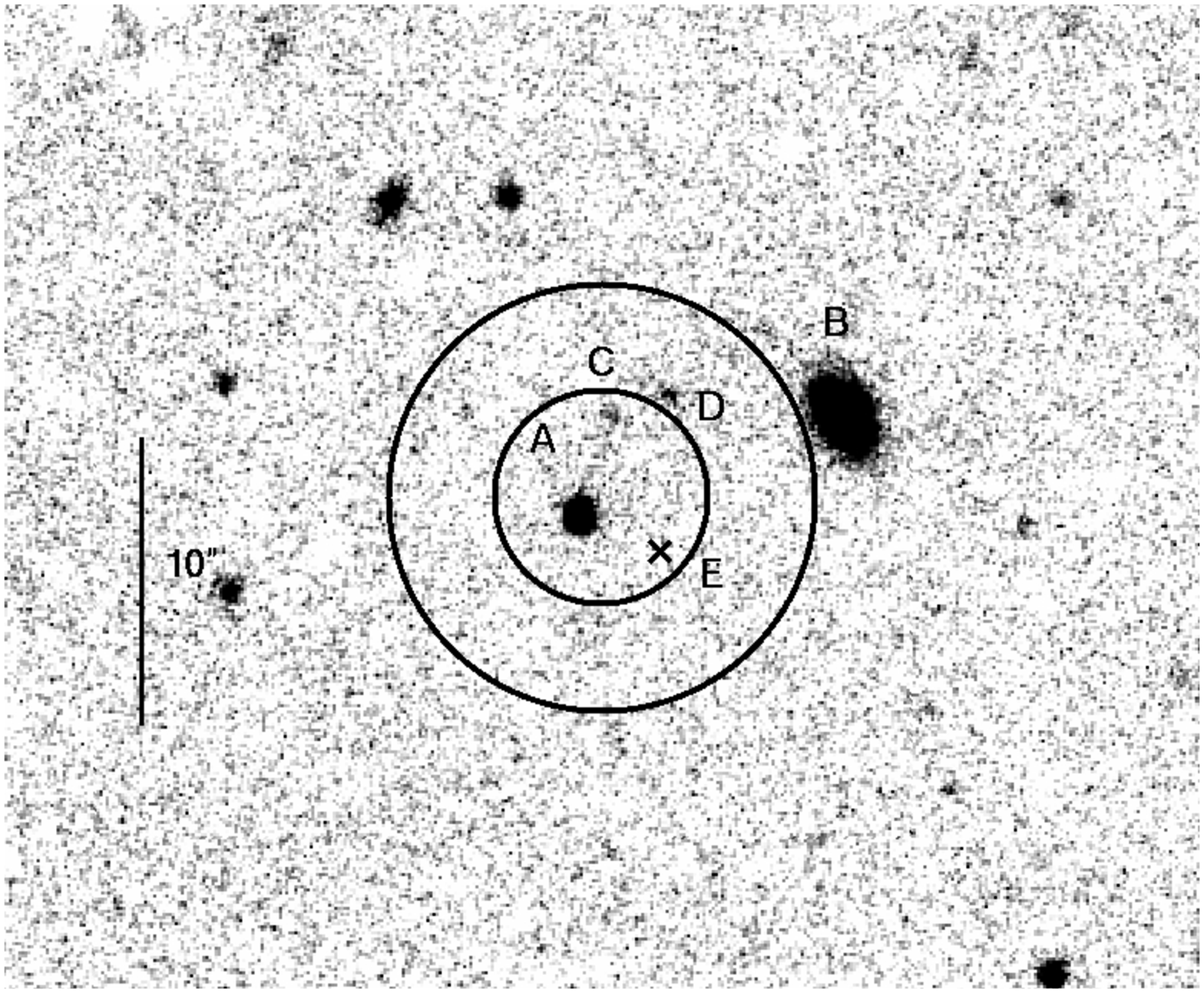}
\caption{ Deep VLT/FORS1 $R_C$-band (top) 
and HAWKI $K_s$-band image (bottom) of the XRT error circle of GRB
080915A taken 28 h after the burst. Also shown is the 90\% c.l. XRT error
circle ($r_0=3\farcs7$), as well as a circle of radius $2r_0$. 
 In the  $K_s$-band image, object E
is not visible and, therefore, is indicated by a cross. 
Note that all objects visible in the ISAAC image have a counterpart 
in the FORS1 image taken 11 days later.}
\label{080915K}
\end{figure}

The field was observed  in the $K_s$-band  with HAWK-I in target of
opportunity  mode starting 28 hours after the burst, lasting for 14 minutes.
No candidate NIR afterglow was found within $2r_0$ down to $K_{\rm AB}=23.4$.
The HAWK-I  observations reveal two  bright objects, one (A) within the 90\%
c.l. error  circle  and one (B) just outside $2r_0$ with AB magnitudes
$K_{\rm AB}=20.42\pm0.02$ and $19.19\pm0.01$, respectively (Fig.~\ref{080915K}). These
objects were also detected with GROND in all bands during the same night
(Table~\ref{tab:PhotomGROND}).  Additional $R_C$-band data were obtained with
FORS1 12 days after the burst (FWHM of $1\farcs4$).   The FORS image shows
objects A ($R_{\rm AB}= 21.63\pm 0.01$) and B ($R_{\rm AB}=  
21.28\pm 0.01$), but also reveals the presence of three additional objects:
C ($R_{\rm AB}= 24.71\pm 0.07$) and E ($R_{\rm AB}= 25.44\pm 0.15$)
within $1r_0$, as well as D ($R_{\rm AB}= 24.57\pm 0.08$) slightly outside
$1r_0$.

In the FORS image, object A has a PSF that is compatible with a point source,
while  C and D appear fuzzy and could be galaxies. Object E is very faint,
close to the detection limit. It is difficult to decide whether it is a galaxy. 
Object B, which is just outside $2r_0$,  is a  galaxy
($5\farcs6\,\times\,4\farcs5$ in the FORS1 image) with a relatively large
$(g^\prime-r^\prime)_{\rm AB}$ color of 2.1 mag. 

For objects C and D, the probability of finding a galaxy with the corresponding
$R_C$-band magnitude inside a circle of radius $1r_0$ and $2r_0$ on the sky is
$p=0.23$ and 0.61, respectively. For object E, the probability of finding a galaxy 
inside a circle of radius $1r_0$ is 0.37.
  Given that C and D could be an interacting
pair, which partly extends into the 90\% c.l. error circle,  we consider
both as GRB host galaxy candidates. 
 If object E is a galaxy, it is the only one well within $1r_0$,
thus we also consider E as a host galaxy candidate. 

\paragraph{\bf \object{GRB 081012}}

The field is at high Galactic latitude ($b=-71^\circ$) and is not  crowded
with stars. The Galactic reddening is very small, at $E(B-V)=0.02$ mag, among the
lowest in our sample. The 90\% c.l. XRT error circle has  $r_0=1\farcs8$.

\begin{figure}[t!]
\includegraphics[width=0.48\textwidth]{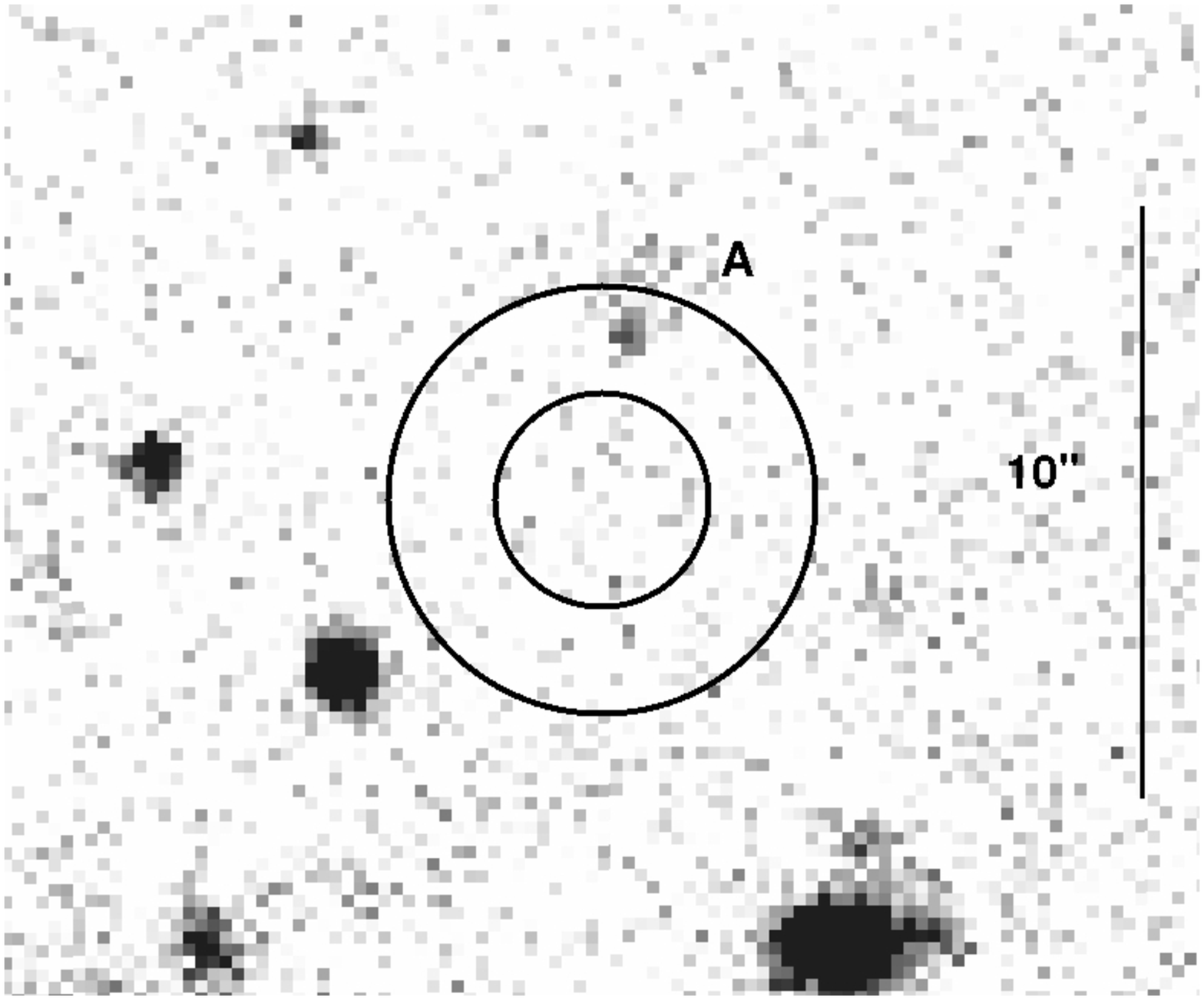}
\caption{VIMOS $R_C$-band image of the field of GRB 081012. 
It shows the 90\% c.l. XRT error circle ($r_0=1\farcs8$), as well as a circle of 
radius $2r_0$.}
\label{fig:081012}
\end{figure}

We observed the field with VIMOS and ISAAC about one year  after the burst. Our
deep VIMOS $R_C$-band image shows no source within the 90\% c.l. error circle
down to $R_{\rm AB}$=26.7. One object (A, Fig.~\ref{fig:081012}) is detected
between $1r_0$ and $2r_0$. It has a cometary shape ($1\farcs8\,\times\,
1\farcs5$) and a magnitude of $R_{\rm AB}=25.16\pm0.17$. It is possible that
this is  an irregular galaxy or a galaxy with a Galactic foreground star
superposed on its southern part. The object is not visible in our ISAAC image
down to $K_{\rm AB}=23.9$. This yields an upper limit of $(R-K)_{\rm AB}<1.2$
mag, but given the potential foreground star, this color should be considered
with caution. The field was also observed by GROND while searching for the
afterglow at a mean time of 19.3 h after the burst. Neither object A nor any
transient source were detected in any band (\citealt{Filgas2008};
Tables~\ref{tab:PhotomGROND}, \ref{tab:darkULs}).

 Given the absence of any other source within the 90\% c.l. error circle,
object A ($p=0.3$) is the only host galaxy candidate, even though  it is a
weak candidate: The angular offset of A from the boundary of the 90\%
c.l. error circle is 1\farcs0. For a redshift of $z=1$ or 0.5, this would
correspond to  a projected distance of 8.0 kpc and 6.0 kpc,
respectively. This is a  relatively large value \citep{Bloom2002a}. If
object A is not the host, then the GRB host galaxy is fainter than $R_{\rm AB}
= 26.7$ and $K_{\rm AB} = 23.9$. 

\paragraph{\bf \object{GRB 081105}}

The field is at a moderately high Galactic latitude ($b=-58^\circ$) that is not very
crowded by stars. The Galactic reddening is very small, at $E(B-V)=0.03$
mag. The 90\% c.l. XRT error circle has $r_0=4\farcs8$.

\begin{figure}[t]
\includegraphics[width=0.48\textwidth]{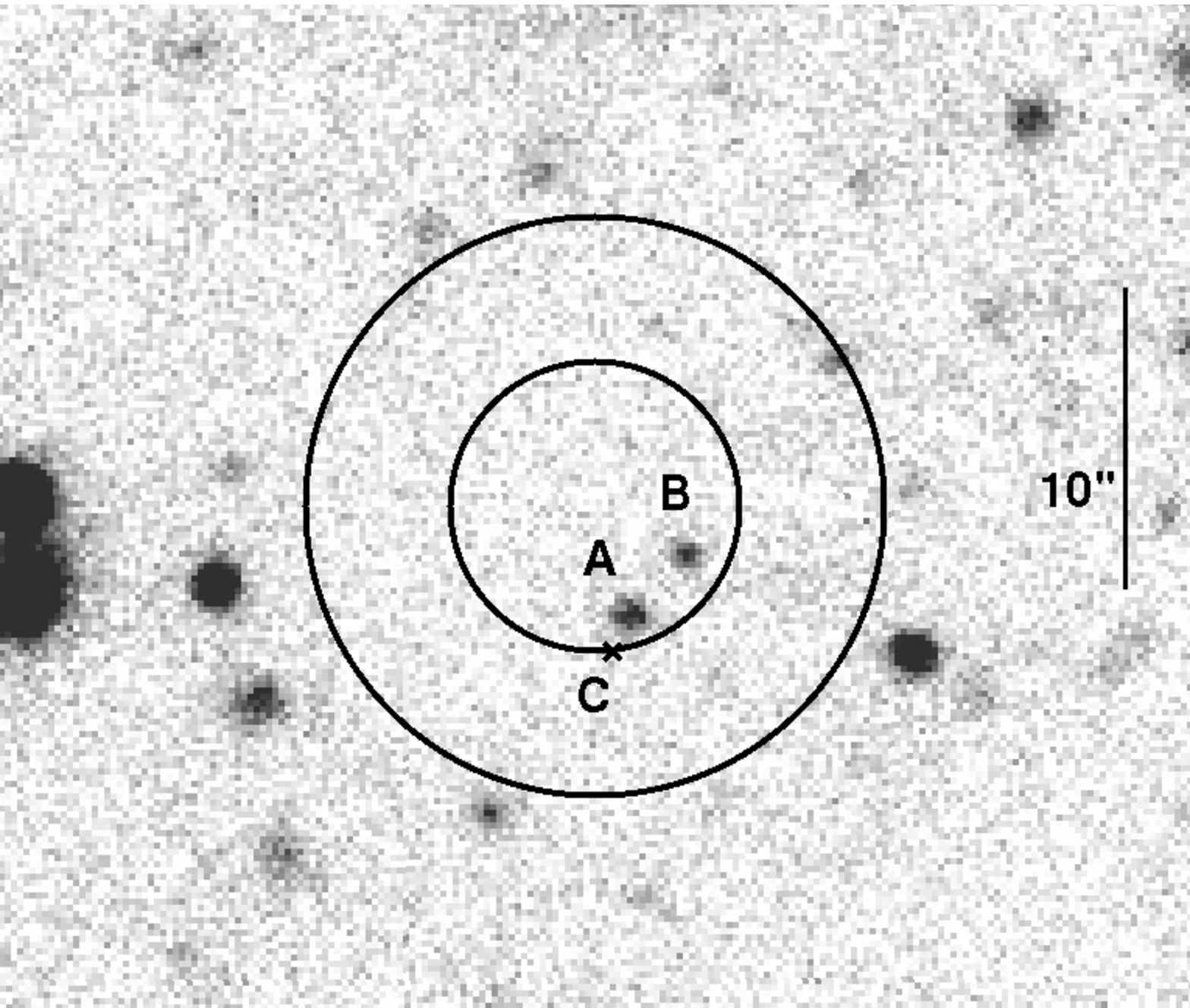}
\includegraphics[width=8.9cm]{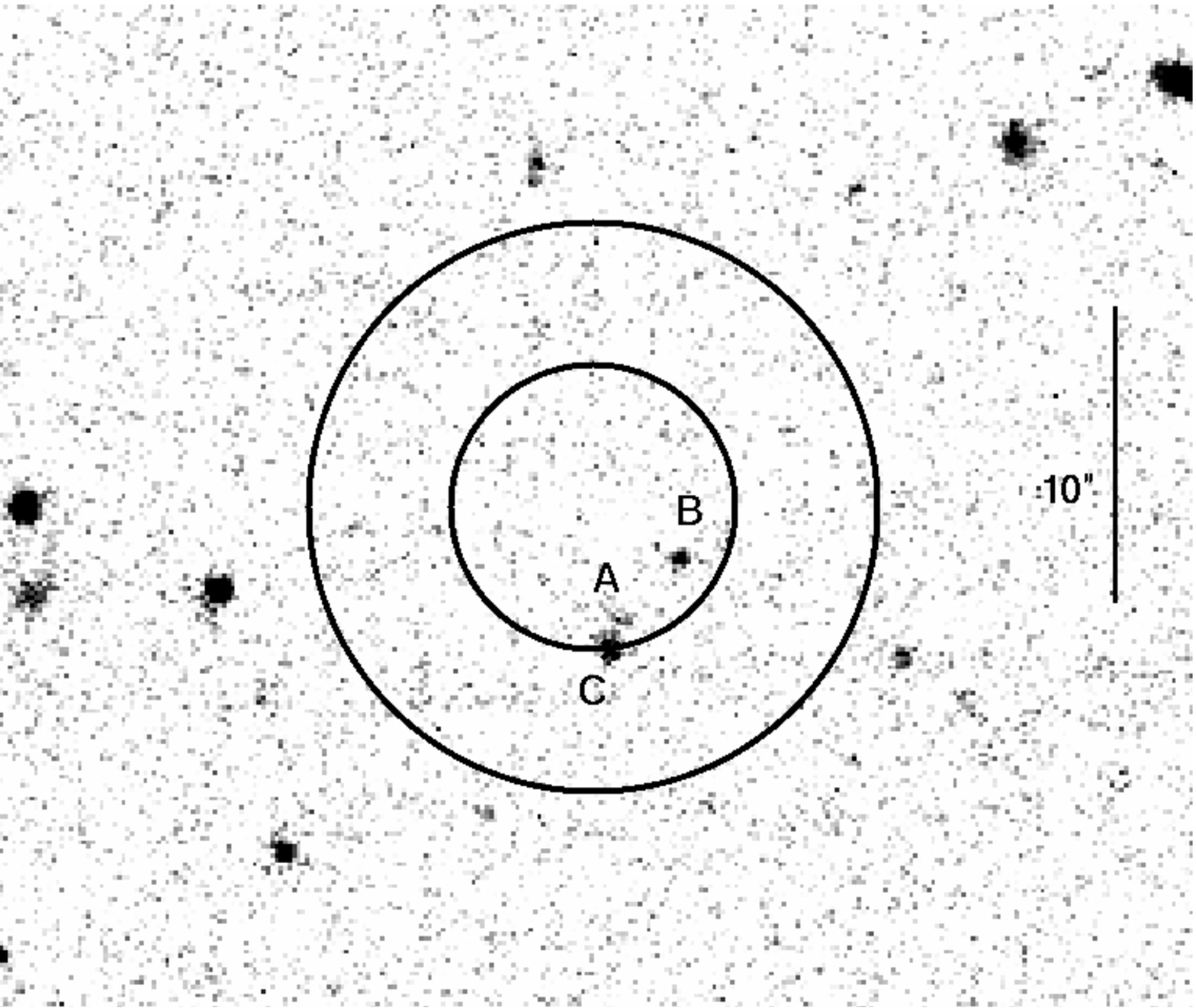}
\caption{VIMOS $R_C$-band image (top) and 
ISAAC $K_s$-band image (bottom) of the field of GRB 081105. Also shown is 
the 90\% c.l. XRT error circle ($r_0=4\farcs8$), as well as a circle
of radius $2r_0$.  Object C is an ERO.}
\label{081105K}
\end{figure}

We observed the field with VIMOS and ISAAC about one year after the burst.  In
spite of the  relatively large size of the XRT error circle, in the deep VIMOS
$R_C$-band image we detect only two objects  A and B, with AB magnitudes
$23.73\pm0.08$ and $24.34\pm 0.13$, respectively (Fig.~\ref{081105K}).
Both objects are also visible in the deep ISAAC $K_s$-band image, with AB
magnitudes  $22.78 \pm 0.18$ and $22.13 \pm 0.14$, respectively. In the
ISAAC image, object A splits into two separate objects, with the second one
(C; $K_{\rm AB}=21.74\pm0.13$) 1\farcs0 south of A.
This object C is an ERO ($(R-K)_{\rm AB}>3.5$ mag). In the $K_s$-band
image, objects A and C appear slightly extended, i.e., these might be
(interacting) galaxies. In the case of B, we cannot determine whether it is a star or a
galaxy.  

The field was also observed by GROND while searching for the afterglow,
starting about 13 h after the burst. No transient source was detected in any
band; only deep upper limits could be obtained (\citealt{Clemens2008};
Table~\ref{tab:darkULs}). None of the three objects (A,B,C) were detected
(Table~\ref{tab:PhotomGROND}).

The $(R-K)_{\rm AB}$ colors of objects A and B (about 0.9 mag and 2.1 mag,
respectively) match those of the sample of GRBHGs at a redshift of around $z=1$
(SBG09). The probability of finding a galaxy with the $R_C$-band magnitudes of
objects A and B  inside a field of radius $1r_0$ is $p=0.19$ and 0.29,
respectively.  The probability of finding an ERO like object C 
inside the same field is $p_{ERO}=0.08$.
Therefore, we consider C as the most likely host galaxy candidate. We note that
objects A and C could be a pair of galaxies.

\paragraph{\bf \object{GRB 081204}}

The field containing this object lies at moderate Galactic latitude ($b=-53^\circ$) and is not very
crowded with stars. The Galactic reddening is very small, $E(B-V)=0.03$
mag. The 90\% c.l. XRT error circle is the largest in our sample 
($r_0=5\farcs3$).

\begin{figure}[htbp]
\includegraphics[width=0.48\textwidth]{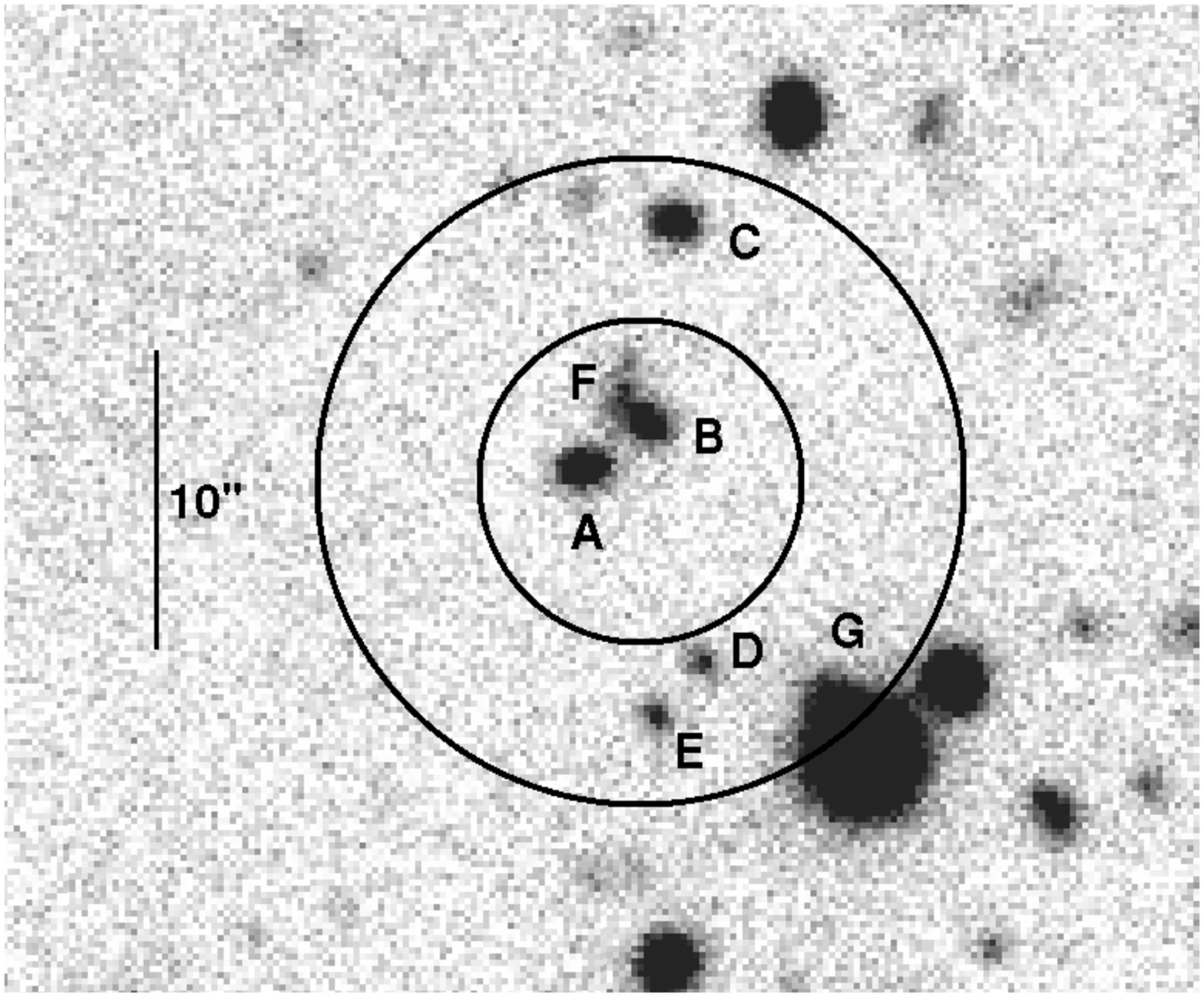}
\includegraphics[width=8.9cm]{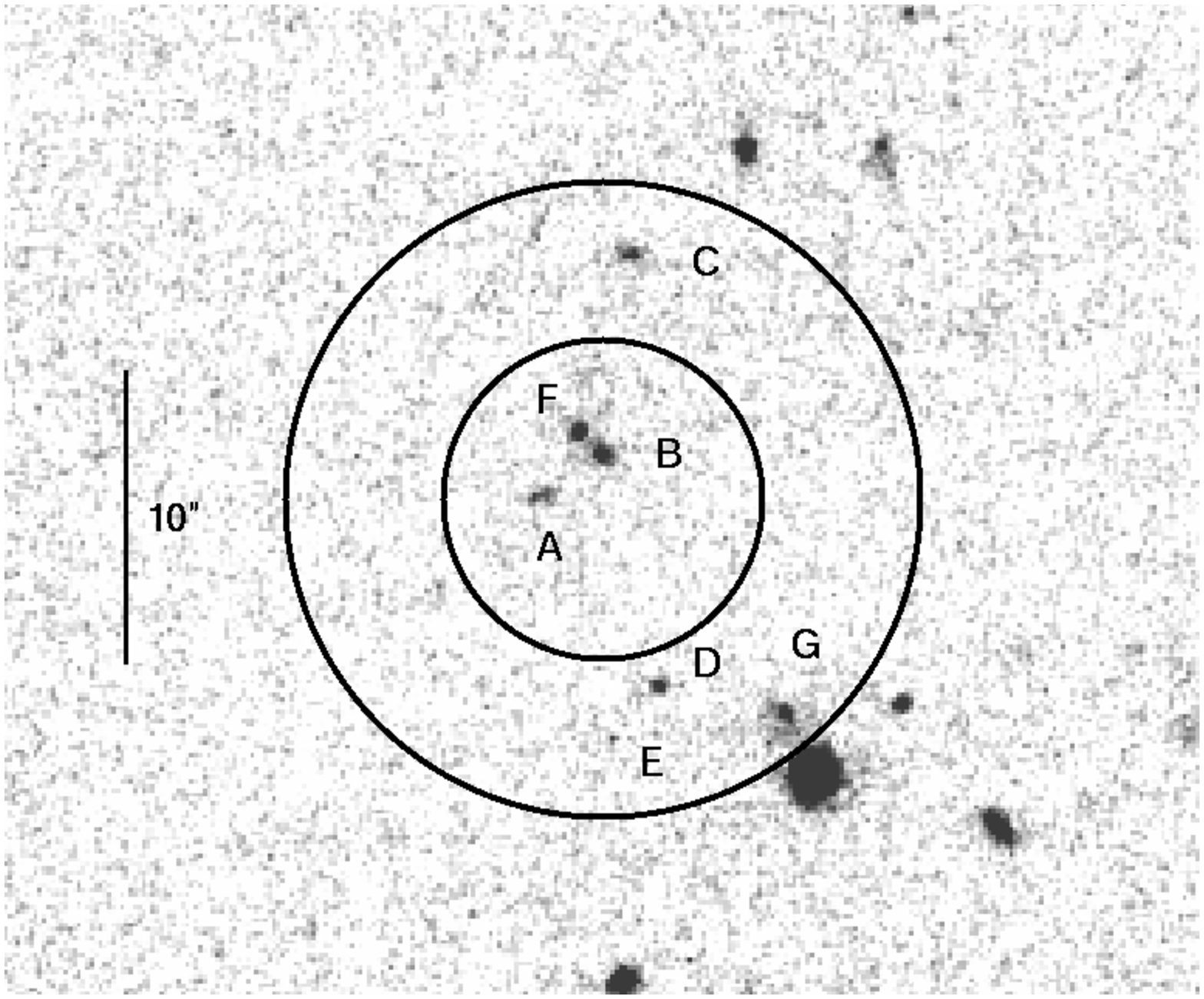}
\caption{VIMOS $R_C$-band image (top) and 
ISAAC $K_s$-band image (bottom) of the field of GRB 081204. Also shown is
the 90\% c.l. XRT error circle ($r_0=5\farcs3$), as well as a circle
of radius $2r_0$.  Object F is probably an ERO.}
\label{081204K}
\end{figure}

We observed the field with VIMOS and ISAAC about one year after the burst.
Further $J$-band imaging was performed with SOFI at the NTT nearly two years
after the event.  The field is rich in objects.  Within the 90\%
c.l. error circle lie at least three galaxies (A, B, F;
Fig.~\ref{081204K}), which are all within 5\farcs0 of each other and could
represent an interacting group. Objects A and B have similar 
magnitudes ($R_{\rm AB} \sim$ 23.2 and 23.5, respectively) and sizes. 
Object F ($R_{\rm AB}\sim24.6$) lies
very close, northeast of object B. In the VIMOS image, it is much fainter
than A and B, but in the ISAAC image it is distinctive because of its bright, point-like
core ($K_{\rm AB}=21.5$). Within the (not so small)
photometric errors, it can be classified as an extremely red object.  In
addition, in the VIMOS image, about 0\farcs7 north of F, lies another
faint, fuzzy object that is too faint for further analysis.  More
objects (C--E, G) are seen between $1r_0$ and $2r_0$. The brightest one is a
galaxy (C) of similar magnitude and size to objects A and B. Objects D and E
($R_{\rm AB}=24.2$ and 24.3, respectively) have a rather blue color of
$(R-K)_{\rm AB}<2.0$  mag and $<0.0$ mag, respectively.  Object D is not
elongated, E is only visible in the VIMOS image,  while G is blended with a
bright star. Given their faintness, it is difficult to establish their
nature. In the ISAAC image, at least G seems to be surrounded by a faint
halo, possibly indicating that this is a galaxy.

Deep follow-up observations of the field were also performed with GROND, while
(unsuccessfully) searching for the afterglow about 10 h after the burst
(Table~\ref{tab:darkULs}; \citealt{Updike2008}). Objects  A and B are detected
in $g^\prime r^\prime i^\prime z^\prime J$, while C was only
seen in $r^\prime i^\prime J$.  Galaxy A is blue, its SED is essentially flat
between $R_C$ and $K_s$ ($(R-K)_{\rm AB}= 0.8 \pm 0.2$ mag), while B is redder
($(R-K)_{\rm AB}=1.8\pm0.1$ mag). Unfortunately, photometric redshift
estimates are  not very accurate for these galaxies. 

\begin{figure}[htbp]
\includegraphics[width=8.9cm]{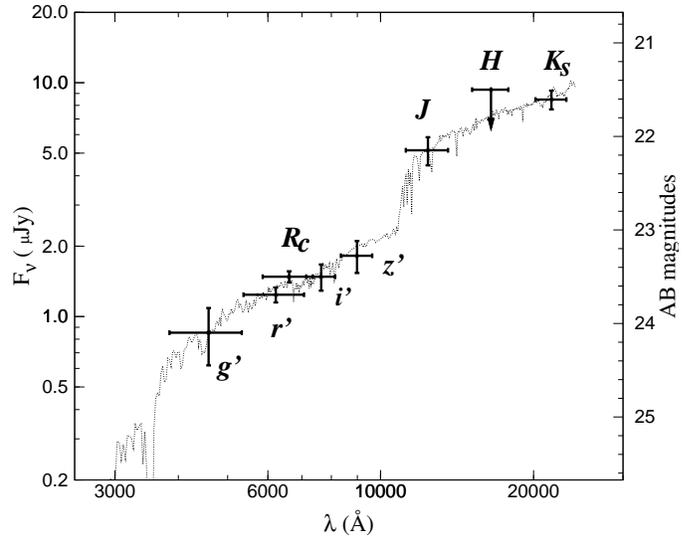}
\caption{
\emph{Hyperz} best-fit solution of the broad-band SED of object B in the XRT
error circle of GRB 081204 (Tables \ref{tab:PhotomVLT}
and \ref{tab:PhotomGROND}).  From left to right: GROND $g^\prime$, GROND $r^\prime$, VIMOS $R_C$, 
GROND $i^\prime$, GROND $z^\prime$, SOFI $J$, GROND $H$, and ISAAC $K_s$.
The fit suggests that it is a spiral galaxy at a redshift of 
$z=1.8\pm0.3$ with a moderate intrinsic SMC extinction of $A_V^{\rm host}=0.3$ 
mag ($\chi^2$/d.o.f $= 0.5$).}
\label{fig:sed081204_B}
\end{figure}

The SED of object B shows a jump between the GROND-$z'$ band and the 
SOFI-$J$ band ($J_{\rm AB}=22.2$). If this is the $4000$\AA \ Balmer
break, then the redshift is $1.8\pm0.3$ (Fig.~\ref{fig:sed081204_B}). 
Such a feature is also seen in the SED of galaxy C.
We find that \emph{Hyperz} indeed 
finds solutions within the redshift interval $1<z<2$ with different sets of
extinction laws, galaxy templates, and host extinction values. 

Given the connection between long GRBs and young stellar populations, it is
interesting to discover an interacting group of galaxies within the 90\% c.l. error
circle.   The magnitude-probability criterion gives for objects  A and B
$p=0.16$ and $p=0.19$, respectively, which implies that one galaxy 
is not more likely than another.  The probability of finding a galaxy with the red color of F within
an area of radius $1r_0$ is much smaller, however, at $p_{\rm ERO}=0.09$. 
Therefore, we consider F, which is possibly interacting with B, as the most 
likely birthplace of GRB 081204.

\section{Discussion \label{disc}}

\subsection{Magnitude-probability candidates}

 In the following, we call magnitude-probability candidates those galaxies
that satisfy $p\leq0.1$  (see eq. \ref{p}, Sect.~\ref{candidates}). These
are: \object{GRB 050717}, objects A and B; \object{GRB 060211A}, objects A and B; \object{GRB
060805A}, objects A and B; \object{GRB 060923B}, objects A and C; \object{GRB 070429A}, objects B
and C; \object{GRB 070517}, object A;  \object{GRB 080207}, object B; and 
GRB 080602, object A.  We note that in
the case of GRB 070429A two galaxies have $p\leq0.1$, and our VLT data reveal
that they could constitute a tightly bound pair. We also note that the optical
afterglow of GRB 070517  has been identified in the present study
(Sect.~\ref{ind070517}) and in this case $p$ gives the corresponding
probability of finding the galaxy labeled A (Fig.~\ref{fig:070517}) at the given
angular distance from the afterglow position. 

\begin{table*}[htbp!]
\vspace{1em}
\enlargethispage{15.6cm}
\caption{Summary of the properties of the objects found in the XRT error circles.}
\renewcommand{\tabcolsep}{3.5pt}
\begin{center}
\begin{tabular*}{1.0\textwidth}{@{\extracolsep{\fill}}rlrcc crcccc}
\toprule
\#& GRB       &Object  & $R$-band size       & $K_s$-band size  & Comment & $(R-K)_{\rm AB}$ & $(R-K)_{\rm Vega}$ & XRTpos & $p$ & $p_{ERO}$\\
\midrule   
1 & 050717    & $\bullet$A      & $2.1 \times 3.9$    & not visible      &  G &  $<$  2.08 & $<$  3.70    & 2 &  0.05$\pm$0.01  &  \\      
  &           & $\bullet$B      & $1.2 \times 1.3$    & not visible      &    &  $<$  2.93 & $<$  4.55    & 1 &  0.03$\pm$0.01 &  \\[1.9mm]      
2 & 050922B   & no candidates\hspace*{-1cm} &               &                  &    &             &               &   &        & \\[1.9mm]  
3 & 060211A   &$\bullet$A       &  $1.1 \times 1.2$   & not visible      &  G  &  $<$   2.53 & $<$  4.16            & 1 &  0.03$\pm$0.01  &  \\
  &           &B                &  $3.8 \times 2.2$   &  3 sources       &  G  &        1.21 &      2.84 $\pm$ 0.21 & 3 &  0.07$\pm$0.01  &  \\   
  &           &$\bullet$C       &  only in $J$        &  --              &  G  &  --         &  --                  & 1 &  --             &  \\
  &           &$\bullet$D       &  only in $J$        &  --              &  -- &  --         &  --                  & 2 &  --             &  \\[1.9mm] 
4 & 060805A   &$\bullet$A       &  $1.5 \times 1.1$   & not visible      &  G  &  $<$   4.3  &  $<$  5.9            & 1 &  0.09$\pm$0.04  &  \\
  &           &$\bullet$B       &  $2.7 \times 1.3$   & not visible      &  G  &  $<$   2.5  &  $<$  4.1            & 2 &  0.09$\pm$0.01  &  \\[1.9mm] 
5 & 060919    &$\bullet$A       &  $1.5 \times 1.4$   & not visible      &     &  $<$   2.60 &  $<$  4.23           & 1 &  0.15$\pm$0.03  &  \\[1.9mm] 
6 & 060923B   &$\bullet$A       &  $2.1 \times 2.2$   & $2.2 \times 1.8$ &  G  &        1.05 &      2.68 $\pm$ 0.14 & 2 &  0.06$\pm$0.01  & \\  
  &           &B                &  $2.0 \times 2.0$   & $1.9 \times 1.9$ &  S  &        2.43 &      4.06 $\pm$ 0.04 & 2 &  --             & \\ 
  &           &$\bullet$C       &  $1.1 \times 1.0$   & $0.8 \times 0.8$ &     &        1.33 &      2.96 $\pm$ 0.16 & 1 &  0.04$\pm$0.01  & \\ 
  &           &$\bullet$D       &  $1.3 \times 1.3$   & $0.8 \times 0.8$ &     &        3.82 &      5.45 $\pm$ 0.13 & 2 &  0.37$\pm$0.03  & 0.04$\pm$0.01\\  
  &           &$\bullet$E       &  blended with A     & blended with A   &     &             &                      & 1 &  --             & \\[1.9mm] 
7 & 061102    &$\bullet$A       &  $2.5 \times 1.5$   & not visible      &  G  &  $<$   1.22 & $<$  2.85            & 2 &  0.34$\pm$0.01  & \\
  &           &$\bullet$B       &  $1.8 \times 1.9$   & not visible      &  G  &  $<$   1.08 & $<$  2.71            & 2 &  0.31$\pm$0.01  & \\[1.9mm] 
8 & {\bf 070429A}   &A          &  $3.8 \times 1.4$   & $1.9 \times 1.1$ &  G  &        2.10 &      3.73 $\pm$ 0.32 & 3 &  0.54$\pm$0.05  & \\
  &           &$\bullet$B       &  $2.8 \times 1.4$   & blended with C   &  G  &        1.41 &      3.04 $\pm$ 0.23 & 1 &  0.04$\pm$0.01  & \\
  &           &$\bullet$C       &  $1.7 \times 1.5$   & $1.5 \times 1.0$ &  G  &        2.09 &      3.72 $\pm$ 0.16 & 1 &  0.05$\pm$0.01  & \\
  &           &$\bullet$D       &  not visible        & $1.6 \times 1.0$ &  G  &     $>$ 3.2 &   $>$ 4.8            & 2 &  --             & 0.07$\pm$0.01 \\[1.9mm] 
9 & 070517   &$\bullet$A       &  $1.6 \times 1.3$   & not visible      &  G  &    $<$ 1.69 & $<$  3.32            & 1 &  0.07$\pm$0.03  & \\[1.9mm] 
10& {\bf 080207}    &A          &  $2.4 \times 1.3$   & not visible      &  G &        2.08 &      3.71 $\pm$ 0.43  & 2 &  0.20$\pm$0.02 & \\
  &           &$\bullet$B       &  $2.1 \times 1.1$   & $1.6 \times 0.9$ &  G &        4.66 &      6.29 $\pm$ 0.40  & 1 &  0.14$\pm$0.04 & 0.01$\pm$0.01\\[1.9mm] 
11& {\bf 080218B} &$\bullet$A   &  $2.5 \times 1.1$   & $1.4 \times 0.7$ &  G &        4.15 &      5.78 $\pm$ 0.16 & 1 &  0.12$\pm$0.01  & $0.01$ \\
  &           &B                &  $1.6 \times 1.5$   & $0.8 \times 0.8$ &    &        1.54 &      3.17 $\pm$ 0.24 & 3 &  0.28$\pm$0.01  & \\[1.9mm] 
12& 080602    &$\bullet$A       &  $2.6 \times 2.0$   & $1.2 \times 0.8$ &  G &        0.34 &      1.97 $\pm$ 0.05 & 1 &  0.01$\pm$0.01  & \\   
  &           &B                &  $2.0 \times 1.8$   & not visible      &  G &  $<$   0.44 & $<$  2.07            & 2 &  0.13$\pm$0.01  & \\
  &           &$\bullet$C       &  not visible        & $1.2 \times 0.8$ &  G &  $>$   4.35 & $>$  5.98            & 2 &  --             &  0.04$\pm$0.01 \\[1.9mm] 
13& 080727A   & no candidates\hspace*{-1cm} &               &                  &    &             &                         &   &        & \\[1.9mm]  
14& 080915A   &A                &  $3.5 \times 3.5$   & $1.2 \times 1.2$ &  S &        1.11 &      2.74 $\pm$ 0.02 & 1 &  --    & \\
  &           &B                &  $5.6 \times 4.5$   & $3.4 \times 1.8$ &  G &        1.99 &      3.62 $\pm$ 0.01 & 3 &  0.16$\pm$0.01  & \\
  &           &$\bullet$C       &  $1.2 \times 1.2$   & not visible      &  G &  $<$ 1.21 & $<$  2.83    & 1 &  0.23$\pm$0.02  & \\
  &           &$\bullet$D       &  $1.2 \times 1.2$   & not visible      &  G &  $<$ 1.07 & $<$  2.69    & 2 &  0.61$\pm$0.02  & \\
  &           &$\bullet$E                &  $1.5 \times 1.5$   & not visible      &    &  $<$   1.94 &  $<$ 3.57   & 1 &  0.37$\pm$0.04  & \\[1.9mm] 
15& 081012    &$\bullet$A       &  $1.8 \times 1.5$   & not visible      &  G &  $<$   1.21 & $<$  2.84            & 2 &  0.31$\pm$0.03  &  \\[1.9mm] 
16& 081105    &A                &  $1.2 \times 1.2$   & $1.0 \times 1.0$ &  G &        0.89 &  2.51 $\pm$ 0.20 & 1 &  0.19$\pm$0.01  & \\
  &           &B                &  $1.1 \times 1.1$   & $0.8 \times 0.8$ &    &        2.15 &      3.78 $\pm$ 0.19 & 1 &  0.29$\pm$0.02  & \\ 
  &           &$\bullet$C       &  not visible        & $0.8 \times 0.5$ &  G &  $>$   3.50 & $>$  5.13            & 1 &  --             &  0.08$\pm$0.01  \\[1.9mm]   
17& 081204    &A                &  $2.1 \times 1.4$   & $0.9 \times 0.5$ &  G &         0.78  &  2.40$\pm$ 0.16 & 1 &  0.16$\pm$0.01  & \\
  &           &B                &  $2.7 \times 1.4$   & $1.1 \times 0.8$ &  G &         1.81  &  3.43$\pm$ 0.09 & 1 &  0.19$\pm$0.01  & \\
  &           &C                &  $2.2 \times 1.7$   & $0.9 \times 0.6$ &  G &         1.04  &  2.66$\pm$ 0.13 & 2 &  0.49$\pm$0.02  &  \\
  &           &D                &  $1.2 \times 1.2$   & $0.5 \times 0.5$ &    &         1.97  &  3.59$\pm$0.18  & 2 &  0.78$\pm$0.03  & \\
  &           &E                &  $1.2 \times 1.2$   & not visible      &    &    $<-$0.04 & $<$  1.59            & 2 &  0.81$\pm$0.03  & \\
  &           &$\bullet$F       &  $\sim1$            & $0.6 \times 0.6$ &  G &        3.1  &       4.7 $\pm$ 0.50 & 1 &  0.43$\pm$0.12  &  0.09$\pm$0.01  \\ 
  &           &G                &  blended with a star& $1.5 \times 1.5$ &    &        --   &      --              & 2 &  --    & \\ 
\bottomrule 
\end{tabular*}
\label{tab:size}
\end{center} 
\tablefoot{
(1) A bullet ($\bullet$) in column \#2 indicates the most likely GRB host
candidate. If more than one candidate is marked then we cannot determine
which is the best.  For details about the selection, see Sect.~\ref{notesind}.
GRBs that are  truly dark according to J04 and V09 are highlighted in
boldface. (2)  Angular sizes are given in units of arcsec. (3)
Magnitude errors in $(R-K)_{\rm AB}$ are identical to the corresponding errors
for $(R-K)_{\rm Vega}$. Colors are corrected  for Galactic extinction. (4)
The  last two columns gives the chance probability $p$ of finding a galaxy of
the  corresponding extinction-corrected (Vega) $R$-band magnitude on the sky
in a region with the size of the corresponding X-ray error circle with a
radius $r=$ XRTpos $\,\times\, r_0$ (Eq.~\ref{p}). Thereby, the first column
refers to number counts of galaxies of all kinds. If the object is for sure a
star, then no value is given. The second column refers to number counts of
EROs only (see Sect.~\ref{EROsII}). (5) Comment ``G'' stands for galaxy, ``S''
for star; if no letter is given then we could not determine if this is a star
or a galaxy. (6) In case of GRB 070517, the probability $p$ is based on the
distance between the afterglow and galaxy A ($1\farcs6$;
Sect.~\ref{ind070517}).  GRB 070429A/object D, and GRB 081204/object F are
EROs within the large $1\sigma$ photometric error 
(see Table~\ref{tab:PhotomVLT}).}
\enlargethispage{0cm}
\end{table*}


The other cases in our sample have $p>0.1$, either because the detected
galaxies are too faint, the XRT error circles are too big, or a mixture of
both. In addition, if more than one galaxy is found inside an XRT error circle, this
criterion does not tend to select one candidate over the other, as
the differences in the corresponding $p$-values are insufficiently large
(e.g., GRB 050717 and GRB 061102; Table~\ref{tab:size}).  This situation
changes, however, if we consider number counts of extremely red objects.

\subsection{Extremely red objects as candidates \label{red}}

Long bursts trace the birth places of the most massive stars (e.g.,
\citealt{Fruchter2006a}), which leads to the expectation that a certain
percentage of hosts of long bursts are dust-enshrouded, starburst
galaxies. Among them, the most extreme cases are classified as EROs.  
To date, only a small number of GRB hosts  have been found that enter this category: 
GRB 020127 (\citealt{Berger2007a}), GRB 030115 (\citealt{Levan2006a,
Dullighan2004a}), GRB 080207 (\citealt{Hunt2011a,Svensson2011}), as well as
GRB 080325  (\citealt{Hashimoto2010}). 

In our sample, {seven} objects fall (within their $1\sigma$ magnitude error)
into this category (Table~\ref{tab:size}). These are: \object{GRB 060923B},  object D
with $(R-K)_{\rm AB} = 3.82 \pm 0.13$ mag; \object{GRB 070429A}, object D with
$(R-K)_{\rm AB}>3.5$ mag (within the $1\sigma$ error in $K_s$); \object{GRB 080207},
object B with $(R-K)_{\rm AB}=4.66 \pm 0.40$ mag; \object{GRB 080218B}, object A with
$(R-K)_{\rm AB}=4.15 \pm 0.16$ mag; \object{GRB 080602}, object C with $(R-K)_{\rm
  AB}>4.3$ mag; \object{GRB 081105}, object C with $(R-K)_{\rm AB}>3.5$ mag; and \object{GRB
081204}, object F with $(R-K)_{\rm AB}=3.1 \pm 0.5$ mag.  Three of them lie
within the corresponding 90\% c.l. XRT error circle (GRB 080207, 080218B, and
081204), one object lying at 1$r_0$ (GRB 081105), the other three objects lying
within less than 1.5$r_0$. All objects have very small $p$-values based on
number counts of EROs on the sky (\citealt{Gonzales2009a}). 

\subsection{Lyman-dropout candidates \label{Lymans}}

For two of the 17 bursts investigated here (GRBs 050922B and 080727A), we could
not find any galaxy inside 2$r_0$, where $r_0$ is the corresponding 90\%
c.l. XRT error circle radius. We therefore consider the optical afterglows of
these events as Lyman dropout candidates, though we cannot rule out very
faint hosts at $z\lesssim5$. 

An additional, but weaker Lyman dropout candidate, is the  optical afterglow
of GRB 081012. Here we find only one galaxy between 1$r_0$ and $2r_0$.  As we
have noted in Sect.~\ref{notesind}, if we were to consider this as the host, then  the
offset of the afterglow from the center of this galaxy would be quite
large. Therefore, an alternative interpretation would be that the host is not
detected in our deep VLT images, i.e. it could be a Lyman drop-out.

Finally, among the seven ERO galaxies discovered in our sample, four  have
been detected in $R$, i.e. they are not Lyman drop-outs. On the other hand,
these three ERO galaxies with no $R$-band detection could lie at higher
redshifts (GRB 070429A/object D, GRB 080602/object C, and 
GRB 081105/object C).  

We conclude that in our sample we have at best six Lyman-dropout candidates ($\sim$
30\%), while in all other cases this interpretation is not required
\footnote{We did not consider the two cases where 
a galaxy is found to be no closer than $1r_0$ from the center of
the corresponding error circle (GRBs 060923B and 061102), but the galaxy's
outer parts extend to within the 90\% c.l.  XRT error circle. In other words,
here the afterglow could have been placed well inside $1r_0$.}.
This result is in qualitative agreement with other studies.
\citet{Perley2009} concluded that in their uniform sample of 29 \swift\ bursts
observed with the robotic Palomar 60 inch telescope, which contains 14 dark
events, at most two bursts could be dark owing to a redshift $z>4.5$. Similarily,
\citet{Greiner2011} found that the fraction of high-$z$ events  among 39 dark
long-duration GRBs observed with GROND in $g'r'i'z'JHK_s$ is on the order of
25\%; extinction by dust in combination with a modest redshift is the main cause
of the optical dimness of these events.  \citet{Melandri2012a}
confirmed this picture using a uniform  sample of 58 bright \swift\ GRBs,
among which about one-third were classified as dark. They provided strong arguments
that high redshift cannot be the main reason for optically dark events.

\subsection{Interacting pairs of galaxies as candidates}

Since long bursts are related to star formation, their host galaxies could be
interacting, morphologically disturbed galaxies, where a starburst was
triggered by galaxy-galaxy-interaction (e.g.,
\citealt{Fruchter1999b,Chen2012a}). In our images, we find four such potential
cases where at least one partner lies inside the 90\% c.l. XRT error
circle. These are: GRB 070429A, objects B and C ($R_C$ band;
Fig.~\ref{070429AK}); GRB 080602, objects A and C ($K_s$ band;
Fig.~\ref{080602}); GRB 080915A, objects C and D ($R_C$ and $K_s$ band;
Fig.~\ref{080915K}); as well as GRB 081204, objects B and F
($R_C$ and $K_s$ band; Fig.~\ref{081204K}). In addition, object B
in the field of GRB 061102 looks morphologically disturbed
(Fig.~\ref{fig:061102A}) but no other galaxy very close to it is seen in our
images. 

There are no statistics at hand that  could provide chance probability values
for finding an interacting pair of galaxies in a randomly chosen area on the
sky. Nevertheless, we conclude that the search for an interacting pair could
be an effective means of finding GRB host galaxy candidates (see also
\citealt{Wainwright2007,Chen2012a}).

\subsection{Normal candidates}

The fields of five bursts  (\object{GRB 050717}, \object{GRB 060211A}, \object{GRB 060805A}, 
\object{GRB 060919}, and \object{GRB 081012})
are of particular interest in our sample. In these case, galaxy candidates are seen in the
corresponding XRT error circle, at least within $2r_0$, but no EROs are found,
nor is there evidence of an interacting pair.  We must conclude that if  one
of these is the host, then it is a normal galaxy, i.e., typical of the host
galaxies of non-extinguished GRBs. This agrees with
\cite{Perley2009}, who found that in their sample most of the hosts galaxies of dark GRBs 
do not differ phenomenologically from the hosts of
bursts with optically detected afterglows.

A special case is \object{GRB 070517}, for which we were able to identify the optical afterglow
based on the early observations by \cite{Fox6420}. Here only one object is
found within $1r_0$, but this object is offset from the position of the
optical afterglow by 1\farcs6.  If this is the host then a 
redshift $z\lesssim0.5$ is required in
order to avoid a projected offset in kpc which is rather large compared to
the observed mean of the GRB offset distribution (\citealt{Bloom2002a}). This
redshift then provides further evidence of a subluminous galaxy (see
Sect.~\ref{redshiftsEstimate}), although  no strong constraints on its
properties can be obtained. 

\subsection{Redshift estimates \label{redshiftsEstimate}}

No precise redshifts are known for the galaxies we have found in the XRT error
circles; only rough estimates can be obtained.
A first approach is the estimation of the photometric redshift by
SED fitting using \emph{Hyperz}. We decided to apply this method only to the objects 
inside a XRT error circle that are considered to be the GRB host candidates. Since this approach
requires a well-sampled optical/NIR SED  (Tables~\ref{tab:PhotomGROND} and \ref{tab:size}),
it was possible to use it in only two cases.
For GRB
080602/object A we obtain $z=1.40^{+0.30}_{-0.15}$
(Fig.~\ref{080602sed}). One could then speculate that this is also the
redshift of the ERO just 1\farcs3 north of A (object C;
Fig.~\ref{080602}). For GRB 081204/object B, we find  $z=1.8\pm0.3$
(Fig.~\ref{fig:sed081204_B}). One can then again speculate that this is
also the redshift of the ERO about 1$''$ northeast of it (object F;
Fig.~\ref{081204K}).

A second approach for estimating $z$ is more statistical.  In
Table~\ref{tab:Color}, we provide estimated redshifts by assuming  that
the galaxies have absolute magnitudes of $M_R=-22, -20$, and $-18$,
respectively. The first value is about 1 mag below the most luminous galaxies
found in the Las Campanas redshift survey ($M_R = -23;$ \citealt{Lin1996}).
The middle value  is approximately the characteristic $M^\star$ of the
corresponding Schechter $r$-band luminosity function. The third value roughly
corresponds to the absolute magnitude of the Large Magellanic Cloud. By
adopting a power-law spectrum for the SED of the form $F_\nu \propto
\nu^{-\beta}$, we then calculated the corresponding redshift for two different
spectral slopes ($\beta = 0.0$ and $1.0$).\footnote{Thereby,  the absolute
$R$-band magnitude is given by  $M_R = m_R - \mu - k$, where $\mu$ is the
distance modulus and $k$ is the cosmological $k$-correction, $k= -2.5\,
(1-\beta) \log (1+z$). If the deduced redshift is higher than 5, the Lyman
dropout in the $R$-band could have affected the apparent magnitudes and no
values for $z$ are given.}  We find that most galaxies would lie at redshifts
$z<2$ if their luminosity were lower than that of the Milky Way.

Finally, we used the  $E_{\rm peak}$-$E_{\rm iso}$ correlation for long GRBs
\citep{Amati2006a} by analyzing the \swift/BAT or \textit{Konus-WIND} spectrum.
Unfortunately, for the majority of the bursts the required spectral
information for this analysis is not available. Only in five cases did we
obtain results, though they are not tight constraints
(Table~\ref{tab:epeiso}). In the case of GRB 080207 and GRB 080602, the
$E_{\rm peak}$-$E_{\rm iso}$ relation constrains the redshift  to be $z > 0.9$
($2\sigma$) and $z > 1$ ($2\sigma$), respectively, in agreement with the
 redshifts found for \object{GRB 080207} ($z\sim2.2$; 
\citealt{Hunt2011a} and Sect.~\ref{ind080207}) and \object{GRB 080602} ($z\sim1.4$;
Sect.~\ref{ind080602}).

\subsection{Host galaxy candidates of truly dark bursts \label{sec:darkhost}}

\cite{Fynbo2009a} shows that at least 39\% of optically dim GRB afterglows are
dark according to the J04 criterion. In our sample, \object{GRB 070429A},  \object{GRB 080207}, and
\object{GRB 080218B} are truly optically dark (Sect.~\ref{sec:darksample}) and also belong
to our small subsample of bursts with extremely red host galaxy candidates
(Sect.~\ref{red}). This supports the idea that global dust
extinction in their host galaxies  was responsible for dimming the afterglow
in the optical bands. This holds especially for \object{GRB 080207},
for which we can be sure that the host galaxy is object B, thanks to the
precise localization of its X-ray afterglow by {\chandra}.

In the case of \object{GRB 080218B}, the host galaxy candidate is visible in the
$R_C$-band, constraining its redshift to be $\lesssim 5$. For this event,
\citet{Greiner2011} found that different pairs of ($z, A_V$) solutions can
explain the non-detection of the optical/NIR afterglow by GROND. For example,
for a redshift of 3.5 a host extinction of $A_{\rm V}^{\rm host}=1.5$ mag is
required. A lower redshift would increase the deduced amount of host
extinction even more; however, studies of {\it optically detected} afterglows
(e.g., \citealt{Kann2010a}) show that these extinction values would be very
high compared to the average. Nevertheless, host galaxies with low extinction
may be linked to the optical detection of GRB afterglows;  more statistics are
needed to establish how common high values of extinction really are.

Another ERO,  the host candidate D of \object{GRB 070429A} that was not detected
in $R_C$, lies just on the border of the 90\% c.l. XRT error circle.  It is a
less compelling host candidate because it is
also very faint in $K_s$, with a correspondingly large photometric error of
0.4 mag. In this case, there are also two more galaxies within the 90\%
c.l. error circle. Their ($R-K$), however, colors are notparticularly red. If
the very red galaxy were to be confirmed as an ERO, then  our study 
would indicate that there is a strong link between optically dark GRBs and ERO galaxies.

\subsection{EROs as an important subpopulation of GRB host galaxies 
\label{EROsII}}

The seven EROs that we have identified have magnitudes between $K_{\rm AB}$ =
 21.5 and 23.0, i.e. $K_{\rm Vega}$ =  19.6 to 21.1
(Table~\ref{tab:PhotomVLT}). For these $K$-band magnitudes, the number density
of EROs on the sky is on the order of 1 per 1000 arcsec$^2$
(\citealt{Gonzales2009a,Hempel2011a,Kim2011a}).\footnote{More precisely, we
used figure 10 in \cite{Gonzales2009a} to calculate chance probability values
for EROs (Table~\ref{tab:size}).} 
Our findings then imply that there is an overdensity of EROs in the XRT
error circles that we have studied here.  Four of the seven EROs that we have
found lie inside their corresponding 90\% c.l. XRT error circle.  In the
remaining three cases (GRBs 060923B, 070429A, and 080602), the ERO
lies just close to the border of the 90\% c.l. error circle.

Since long GRBs are thought to trace the formation of massive stars
(passively evolving ellipticals cannot be their hosts), 
the results obtained with our study suggest that
bursts with optically
non-detected afterglows (but with rapid and deep follow-up observations) 
trace a subpopulation of massive galaxies undergoing violent star
formation. This holds for dark bursts in particular: all three bursts
investigated here that belong to this class have an ERO within or close to
their 90\% c.l. XRT error circle (\object{GRB 070429A}, \object{GRB 080207}, and \object{GRB 080218B}).  If we
consider as dark all GRBs that follow the J04 
{\it or} V09 criterion (but keeping in
mind that this now includes events where the X-ray data are not so easily
interpreted, see Sect.~\ref{sec:darksample}), then we have eight of these GRBs in
our sample (Table~\ref{tab:box}); five of them have an ERO within or close to
their 90\% c.l. error circle (in addition GRBs 080602 and 081204). It
should be stressed that, in principle, all GRBs studied here  except GRB
070517 (for which we identified the afterglow; Sect.~\ref{ind070517}) could
be truly dark bursts according to the criterion from J04 and V09; we just do
not have sufficiently deep optical limits to be certain.

Several previous studies have targeted GRBHs (e.g.,
\citealt{LeFloch2003,Christensen2004a,Fruchter2006a,Ovaldsen2007a,Svensson2010a,Levesque2010a}
and SBG09).  They have focused on the low-redshift regime (up to $z\sim1.5$) and
showed that most hosts are subluminous ($L < L^*$), blue, of low metallicity
and with a moderate star formation rate ($\sim
1-10\,M_\odot$\,yr$^{-1}$). However, our results suggest that an
infrared-bright subpopulation of very dusty GRBHs exists, which stands
out from the main GRB host galaxy population.

Redshift measurements for the EROs that we have identified here are missing in most
cases. However, for the ERO related to GRB 080207, a  photometric
redshift was derived by \citet{Hunt2011a}. The observed broad-band
SED is indicative of a very luminous ($M_{K}\sim24.4$), infrared-bright galaxy, very
different from  the sample of GRBHs compiled by SBG09. This host galaxy is
similar in color, luminosity, and redshift to the hosts of the dark bursts
 GRB 020127 (\citealt{Berger2007a}), GRB 030115 (\citealt{Levan2006a}), and
GRB 080325 \citep{Hashimoto2010}.  There is possibly a bias in
the GRB host samples studied so far, which are dominated by host galaxies of
optically detected afterglows. This conclusion  is supported by 
recent work on dark
bursts observed with GROND \citep{Kruhler2011A&A534}, where it is shown that
highly extinguished afterglows trace a subpopulation of luminous, massive,
metal-rich, and chemically evolved GRBHs that were not previously associated
with GRBs.

\section{Summary \label{Sum}}

Motivated by the non-detection of the optical afterglows of a substantial
fraction of \swift\ bursts with well-observed X-ray afterglows, we have
selected 17 of these events with small \swift/XRT error circles (defined by their
individual 90\% c.l. radius $r_0$) and searched for the potential host
galaxies of these bursts using deep multi-color imaging.  Our primary
telescope was the VLT equipped with FORS1, FORS2, and VIMOS for $R_C$-band
imaging and ISAAC and HAWK-I for $K_s$-band imaging. These data were supplemented by
observations with the seven-channel imager GROND mounted at the 2.2-m MPG/ESO
telescope on La Silla and the infrared imager NEWFIRM mounted at the  4-m
Mayall telescope  on Kitt Peak. The limiting magnitudes we achieved are deep,
at usually $R_{\rm AB}=26.5$ and $K_{\rm AB}=23.5$ as well as $g^\prime r^\prime
i^\prime z^\prime JHK=25.5$, 25, 24.5, 24, 22.5, 21.5, and 21 for GROND. The
latter data include late-time imaging as well as data gained in rapid response
mode, where we did not find evidence of a fading afterglow.

We have discovered up to six events, about one-third of our sample, where the corresponding
GRB host galaxy could be Lyman dropped out in the $R_C$ band. In two cases, we
do not see any object within 
an area of twice the radius of each associated 90\% c.l. \swift/XRT error circle
 down to deep flux limits (GRBs 050922B and
080727A); in one event, there is only one galaxy within $1r_0$ and $2r_0$ (GRB
081012); and three bursts have a very red galaxy detected only in the
$K_s$-band (\object{GRB 070429A}/object D, \object{GRB 080602}/object C, \object{GRB 081105}/object C).
These three bursts  belong to a subsample of seven bursts for which we found
that an ERO, which we recall are defined as having $(R-K)_{\rm AB}>3.5$ mag, was the
confirmed or likely host galaxy. In particular, all three bursts in our sample
that are classified as securely dark according to their observed X-ray flux
(following J04 and V09) belong to this group. Even though these are small number
statistics, our findings imply that a non-negligible fraction of optically dim
bursts may be located in globally dust-enshrouded galaxies. 

While the $(R-K)$ color of galaxies has emerged as a powerful criterion for
identifying host galaxy candidates, we also considered  chance-probability
constraints based on published number counts of (all types of) galaxies on the
sky.  In nine bursts, the chance probability $p$ of finding a galaxy of the given
$R_C$-band magnitude in the corresponding 90\% c.l. XRT error circle is
$\leq$10\% (within $1\sigma$), which makes them good host galaxy
candidates. In the remaining cases (about 1/2 of our sample), galaxies were
identified but they are not special in any way, in terms of either their 
($R-K$) colors, their magnitudes, or their $p$-values. However, for four
bursts,  we have discovered possibly interacting galaxies in the XRT error
circle, which is potentially a sign of triggered star-formation.

The connection between star-forming activity and dark bursts is even more
intriguing for the seven EROs in our sample.  This is the most outstanding
result of our study.  It points to the  existence of a subpopulation of GRBHs,
characterized by violent star formation, that is missed by host galaxy surveys
of bursts with detected optical afterglows.  The putative host of \object{GRB 080207}
is the most remarkable example ($(R-K)_{\rm AB}=4.66 \pm 0.40$ mag;
\citealt{Hunt2011a,Svensson2011}).  The possibility that a non-negligible
fraction of optically dim bursts are highly dust-enshrouded and possibly
submm-bright galaxies makes these bursts interesting cosmological tools 
for achieving a deeper insight into the optically obscured star-formation history of
the Universe (\citealt{Berger2003a,Tanvir2004a}).
 
\begin{acknowledgements}

A.R. and S.K. acknowledge the support of DFG Kl 766/11-3, 13-2, and
16-1. A.R. acknowledges the support of the BLANCEFLOR Boncompagni-Ludovisi,
n\'ee Bildt foundation and by the Jenaer Graduierten\-akademie.  
A.R., S.K., and A.N.G. acknowledge the support of  the
Deutscher Akademischer Austausch-Dienst (DAAD; project D/08/15024).
P.F. acknowledge the support of the MICINN Proyecto Internacional
ref. AIB2010DE-00287.  S.S. acknowledges the support of a Grant of Excellence from
the Icelandic Research Fund.  L.A.A. and E.G. acknowledge the DAAD RISE
program. T.K. acknowledges the support of the DFG cluster of excellence  'Origin
and Structure of the Universe'. A.N.G. and D.A.K. acknowledge the support of DFG
grant Kl 766/16-1. E.P. and N.M. acknowledge the support of the AIT Vigoni program
2008-2009. L.K.H. and E.P. gratefully acknowledge a financial
contribution  from the agreement ASI-INAF I/009/10/0.  F.O.E. acknowledges
funding of his Ph.D. through the DAAD.   J. Go. and A.J.C.T.  are funded by
the Spanish research programmes AYA-2007-63677, AYA-2008-03467/ESP and
AYA-2009-14000-C03-01.  M.N. and P.S. acknowledge the support of DFG grant SA
2001/2-1.  E. Pian acknowledges the support of the grant ASI I/088/06/0.
A.d.P. acknowledges the support of the DNRF. Part of the funding for GROND (both
hardware as well as personnel) was generously granted from the Leibniz-Prize
to Prof. G. Hasinger (DFG grant HA 1850/28-1).  A.R., A.N.G., D.A.K. and
A.C.U. are grateful for travel funding support through MPE.  This research has
made use of the NASA/IPAC Extragalactic Database (NED) which is operated by
the Jet Propulsion Laboratory, California Institute of Technology, under
contract with the National Aeronautics and Space Administration. A.R. and
S.K. thank Johan Fynbo for a careful reading of the manuscript. This work made
use of data supplied by the UK Swift Science Data Centre at the University of
Leicester.  We thank the ESO staff for performing the VLT service
observations. We thank the referee  for a very careful reading of the
manuscript and very valuable comments, which helped to improve the text
substantially. 

\end{acknowledgements}


\bibliographystyle{aa}
\bibliography{paperbib}

\begin{thebibliography}{209}
\expandafter\ifx\csname natexlab\endcsname\relax\def\natexlab#1{#1}\fi

\bibitem[{{Amati}(2006)}]{Amati2006a}
{Amati}, L. 2006, \mnras, 372, 233

\bibitem[{{Andreev} {et~al.}(2008){Andreev}, {Kurenya}, \&
  {Pozanenko}}]{Andreev08_GCN7333}
{Andreev}, M., {Kurenya}, A., \& {Pozanenko}, A. 2008, \gcn, 7333

\bibitem[{{Appenzeller} {et~al.}(1998){Appenzeller}, {Fricke}, {F{\"u}rtig},
  {G{\"a}ssler}, {H{\"a}fner}, {Harke}, {Hess}, {Hummel}, {J{\"u}rgens},
  {Kudritzki}, {Mantel}, {Meisl}, {Muschielok}, {Nicklas}, {Rupprecht},
  {Seifert}, {Stahl}, {Szeifert}, \& {Tarantik}}]{Appenzeller1998a}
{Appenzeller}, I., {Fricke}, K., {F{\"u}rtig}, W., {et~al.} 1998, The
  Messenger, 94, 1

\bibitem[{{Autry} {et~al.}(2003){Autry}, {Probst}, {Starr}, {Abdel-Gawad},
  {Blakley}, {Daly}, {Dominguez}, {Hileman}, {Liang}, {Pearson}, {Shaw}, \&
  {Tody}}]{Autry2003a}
{Autry}, R.~G., {Probst}, R.~G., {Starr}, B.~M., {et~al.} 2003, in Society of
  Photo-Optical Instrumentation Engineers (SPIE) Conference Series, Vol. 4841,
  Society of Photo-Optical Instrumentation Engineers (SPIE) Conference Series,
  ed. M.~{Iye} \& A.~F.~M. {Moorwood}, 525--539

\bibitem[{{Barbier} {et~al.}(2006{\natexlab{a}}){Barbier}, {Barthelmy},
  {Cummings}, {Fenimore}, {Gehrels}, {Hullinger}, {Koss}, {Krimm}, {Markwardt},
  {Palmer}, {Parsons}, {Sakamoto}, {Sato}, {Stamatikos}, \&
  {Tueller}}]{Barbier2006GCN5403}
{Barbier}, L., {Barthelmy}, S., {Cummings}, J., {et~al.} 2006{\natexlab{a}},
  \gcn, 5403

\bibitem[{{Barbier} {et~al.}(2006{\natexlab{b}}){Barbier}, {Barthelmy},
  {Cummings}, {Fenimore}, {Gehrels}, {Hullinger}, {Krimm}, {Markwardt},
  {Palmer}, {Parsons}, {Sakamoto}, {Sato}, {Stamatikos}, \&
  {Tueller}}]{Barbier2006GCN5595}
{Barbier}, L., {Barthelmy}, S.~D., {Cummings}, J., {et~al.} 2006{\natexlab{b}},
  \gcn, 5595

\bibitem[{{Barthelmy} {et~al.}(2005){Barthelmy}, {Barbier}, {Cummings},
  {Fenimore}, {Gehrels}, {Hullinger}, {Krimm}, {Markwardt}, {Palmer},
  {Parsons}, {Sato}, {Suzuki}, {Takahashi}, {Tashiro}, \&
  {Tueller}}]{Barthelmy2005a}
{Barthelmy}, S.~D., {Barbier}, L.~M., {Cummings}, J.~R., {et~al.} 2005, Space
  Sci. Rev., 120, 143

\bibitem[{{Barthelmy} {et~al.}(2007){Barthelmy}, {Markwardt}, {Page}, {Palmer},
  {Sato}, {Stamatikos}, \& {Starling}}]{Barthelmy2007GCN6355}
{Barthelmy}, S.~D., {Markwardt}, C.~B., {Page}, K.~L., {et~al.} 2007, \gcn,
  6355

\bibitem[{{Beardmore} {et~al.}(2008{\natexlab{a}}){Beardmore}, {Barthelmy},
  {Cummings}, {Gehrels}, {Holland}, {Hoversten}, {Kennea}, {Markwardt},
  {Marshall}, {McLean}, {Page}, {Palmer}, \& {Ukwatta}}]{Beardmore2008GCN7781}
{Beardmore}, A.~P., {Barthelmy}, S.~D., {Cummings}, J.~R., {et~al.}
  2008{\natexlab{a}}, \gcn, 7781

\bibitem[{{Beardmore} {et~al.}(2008{\natexlab{b}}){Beardmore}, {Burrows}, \&
  {Cummings}}]{Beardmore2008}
{Beardmore}, A.~P., {Burrows}, D.~N., \& {Cummings}, J.~R. 2008{\natexlab{b}},
  \gcn, 8522

\bibitem[{{Beardmore} \& {Cummings}(2008)}]{Beardmore2008GCN8487}
{Beardmore}, A.~P. \& {Cummings}, J. 2008, \gcn, 8487

\bibitem[{{Beardmore} {et~al.}(2008{\natexlab{c}}){Beardmore}, {Evans}, {Goad},
  \& {Osborne}}]{Beardmore08_GCN7782}
{Beardmore}, A.~P., {Evans}, P.~A., {Goad}, M.~R., \& {Osborne}, J.~P.
  2008{\natexlab{c}}, \gcn, 7782

\bibitem[{{Beardmore} {et~al.}(2008{\natexlab{d}}){Beardmore}, {Page}, \&
  {Evans}}]{Beardmore08_GCN7785}
{Beardmore}, A.~P., {Page}, K.~L., \& {Evans}, P.~A. 2008{\natexlab{d}}, \gcn,
  7785

\bibitem[{{Beardmore} {et~al.}(2008{\natexlab{e}}){Beardmore}, {Page}, {Evans},
  {Hoversten}, {Barthelmy}, {Burrows}, {Roming}, {Gehrels}, {Beckmann},
  {Mereghetti}, {Shaw}, \& {Tuerler}}]{Beardmore2008GCNR145}
{Beardmore}, A.~P., {Page}, K.~L., {Evans}, P.~A., {et~al.} 2008{\natexlab{e}},
  GCN Report, 145

\bibitem[{{Berger} {et~al.}(2003){Berger}, {Cowie}, {Kulkarni}, {Frail},
  {Aussel}, \& {Barger}}]{Berger2003a}
{Berger}, E., {Cowie}, L.~L., {Kulkarni}, S.~R., {et~al.} 2003, \apj, 588, 99

\bibitem[{{Berger} {et~al.}(2007){Berger}, {Fox}, {Kulkarni}, {Frail}, \&
  {Djorgovski}}]{Berger2007a}
{Berger}, E., {Fox}, D.~B., {Kulkarni}, S.~R., {Frail}, D.~A., \& {Djorgovski},
  S.~G. 2007, \apj, 660, 504

\bibitem[{{Berger} \& {Lopez-Morales}(2005)}]{Berger2005GCN3639}
{Berger}, E. \& {Lopez-Morales}, M. 2005, \gcn, 3639

\bibitem[{{Berger} {et~al.}(2005){Berger}, {Lopez-Morales}, \&
  {Osip}}]{Berger2005GCN3643}
{Berger}, E., {Lopez-Morales}, M., \& {Osip}, D. 2005, \gcn, 3643

\bibitem[{{Berger} \& {Rest}(2008)}]{Berger08_GCN8624}
{Berger}, E. \& {Rest}, A. 2008, \gcn, 8624

\bibitem[{{Bissaldi}(2008)}]{Bissaldi08_GCN8370}
{Bissaldi}, E. 2008, \gcn, 8370

\bibitem[{{Bloom} {et~al.}(2002){Bloom}, {Kulkarni}, \&
  {Djorgovski}}]{Bloom2002a}
{Bloom}, J.~S., {Kulkarni}, S.~R., \& {Djorgovski}, S.~G. 2002, \aj, 123, 1111

\bibitem[{{Blustin} {et~al.}(2005){Blustin}, {Hurkett}, {Smale}, \&
  {Cominsky}}]{Blustin2005GCN3638}
{Blustin}, A., {Hurkett}, C., {Smale}, A., \& {Cominsky}, L. 2005, \gcn, 3638

\bibitem[{{Bolzonella} {et~al.}(2000){Bolzonella}, {Miralles}, \&
  {Pell{\'o}}}]{Bolzonella2000}
{Bolzonella}, M., {Miralles}, J.-M., \& {Pell{\'o}}, R. 2000, \aap, 363, 476

\bibitem[{{Breeveld} \& {Guidorzi}(2006)}]{Breeveld2006GCN5580}
{Breeveld}, A. \& {Guidorzi}, C. 2006, \gcn, 5580

\bibitem[{{Breeveld} \& {Oates}(2008)}]{Breeveld2008GCN8232}
{Breeveld}, A. \& {Oates}, S.~R. 2008, \gcn, 8232

\bibitem[{{Burrows} {et~al.}(2005){Burrows}, {Hill}, {Nousek}, {Kennea},
  {Wells}, {Osborne}, {Abbey}, {Beardmore}, {Mukerjee}, {Short}, {Chincarini},
  {Campana}, {Citterio}, {Moretti}, {Pagani}, {Tagliaferri}, {Giommi},
  {Capalbi}, {Tamburelli}, {Angelini}, {Cusumano}, {Br{\"a}uninger}, {Burkert},
  \& {Hartner}}]{Burrows2005a}
{Burrows}, D.~N., {Hill}, J.~E., {Nousek}, J.~A., {et~al.} 2005, Space Sci.
  Rev., 120, 165

\bibitem[{{Butler}(2007)}]{Butler2007}
{Butler}, N.~R. 2007, \aj, 133, 1027

\bibitem[{{Cannizzo} {et~al.}(2007){Cannizzo}, {Barbier}, {Barthelmy},
  {Cummings}, {Fenimore}, {Gehrels}, {Krimm}, {Markwardt}, {Palmer}, {Parsons},
  {Sakamoto}, {Sato}, {Stamatikos}, \& {Tueller}}]{Cannizzo2007GCN6362}
{Cannizzo}, J., {Barbier}, L., {Barthelmy}, S.~D., {et~al.} 2007, \gcn, 6362

\bibitem[{{Castro-Tirado} {et~al.}(2007){Castro-Tirado}, {Bremer}, {McBreen},
  {Gorosabel}, {Guziy}, {Fakthullin}, {Sokolov}, {Gonz{\'a}lez Delgado},
  {Bihain}, {Pandey}, {Jel{\'{\i}}nek}, {de Ugarte Postigo}, {Misra}, {Sagar},
  {Bama}, {Kamble}, {Anupama}, {Licandro}, {P{\'e}rez-Ram{\'{\i}}rez},
  {Bhattacharya}, {Aceituno}, \& {Neri}}]{Castro-Tirado2007a}
{Castro-Tirado}, A.~J., {Bremer}, M., {McBreen}, S., {et~al.} 2007, \aap, 475,
  101

\bibitem[{{Castro-Tirado} {et~al.}(2006){Castro-Tirado}, {Cunniffe}, {de Ugarte
  Postigo}, {Jel{\'{\i}}nek}, {Vitek}, {Kub{\'a}nek}, {Gorosabel}, {Castillo
  Carri{\'o}n}, {Mateo Sanguino}, {Riva}, {Conconi}, {di Caprio}, {Zerbi},
  {Amado}, {C{\'a}rdenas}, {Claret}, {Guziy}, {Mart{\'{\i}}n-Ruiz},
  {S{\'a}nchez}, {Garc{\'{\i}}a Teodoro}, {Castro Cer{\'o}n}, {D{\'{\i}}az
  Verdejo}, {Hudec}, {L{\'o}pez Soler}, {Bern{\'a} Galiano}, {Casares},
  {Fabregat}, {P{\'a}ta}, {S{\'a}nchez Fern{\'a}ndez}, {Sabau-Graziati},
  {Trigo-Rodr{\'{\i}}guez}, \& {Vitali}}]{Castro-Tirado2006a}
{Castro-Tirado}, A.~J., {Cunniffe}, R., {de Ugarte Postigo}, A., {et~al.} 2006,
  in Presented at the Society of Photo-Optical Instrumentation Engineers (SPIE)
  Conference, Vol. 6267, Society of Photo-Optical Instrumentation Engineers
  (SPIE) Conference Series

\bibitem[{{Cenko}(2006)}]{Cenko2006GCN5401}
{Cenko}, S.~B. 2006, \gcn, 5401

\bibitem[{{Cenko} {et~al.}(2009){Cenko}, {Kelemen}, {Harrison}, {Fox},
  {Kulkarni}, {Kasliwal}, {Ofek}, {Rau}, {Gal-Yam}, {Frail}, \&
  {Moon}}]{Cenko2009a}
{Cenko}, S.~B., {Kelemen}, J., {Harrison}, F.~A., {et~al.} 2009, \apj, 693,
  1484

\bibitem[{{Chen}(2012)}]{Chen2012a}
{Chen}, H.-W. 2012, \mnras, 419, 3039

\bibitem[{{Christensen} {et~al.}(2004){Christensen}, {Hjorth}, \&
  {Gorosabel}}]{Christensen2004a}
{Christensen}, L., {Hjorth}, J., \& {Gorosabel}, J. 2004, \aap, 425, 913

\bibitem[{{Clemens} {et~al.}(2008){Clemens}, {Filgas}, {Greiner}, {Kruehler},
  {Yolda\c{s} }, {K{\"u}pc{\"u} Yolda\c{s} }, \& {Szokoly}}]{Clemens2008}
{Clemens}, C., {Filgas}, R., {Greiner}, J., {et~al.} 2008, \gcn, 8492

\bibitem[{{Cobb}(2008{\natexlab{a}})}]{Cobb2008GCN7318}
{Cobb}, B.~E. 2008{\natexlab{a}}, \gcn, 7318

\bibitem[{{Cobb}(2008{\natexlab{b}})}]{Cobb2008GCN8248}
{Cobb}, B.~E. 2008{\natexlab{b}}, \gcn, 8248

\bibitem[{{Conselice} {et~al.}(2008){Conselice}, {Bundy}, {U}, {Eisenhardt},
  {Lotz}, \& {Newman}}]{Conselice2008MNRAS383}
{Conselice}, C.~J., {Bundy}, K., {U}, V., {et~al.} 2008, \mnras, 383, 1366

\bibitem[{{Covino} {et~al.}(2008{\natexlab{a}}){Covino}, {D'Avanzo},
  {Antonelli}, {Fugazza}, {Calzoletti}, {Campana}, {Chincarini}, {Conciatore},
  {Cutini}, {D'Elia}, {D'Alessio}, {Fiore}, {Goldoni}, {Guetta}, {Guidorzi},
  {Israel}, {Maiorano}, {Masetti}, {Melandri}, {Meurs}, {Nicastro}, {Palazzi},
  {Pian}, {Piranomonte}, {Stella}, {Stratta}, {Tagliaferri}, {Tosti}, {Testa},
  {Vergani}, \& {Vitali}}]{Covino2008GCN7322}
{Covino}, S., {D'Avanzo}, P., {Antonelli}, L.~A., {et~al.} 2008{\natexlab{a}},
  \gcn, 7322

\bibitem[{{Covino} {et~al.}(2008{\natexlab{b}}){Covino}, {D'Avanzo},
  {Antonelli}, {Malesani}, {Fugazza}, {Calzoletti}, {Campana}, {Chincarini},
  {Conciatore}, {Cutini}, {D'Elia}, {D'Alessio}, {Fiore}, {Goldoni}, {Guetta},
  {Guidorzi}, {Israel}, {Maiorano}, {Masetti}, {Melandri}, {Meurs}, {Nicastro},
  {Palazzi}, {Pian}, {Piranomonte}, {Stella}, {Stratta}, {Tagliaferri},
  {Tosti}, {Testa}, {Vergani}, \& {Vitali}}]{Covino2008GCN8233}
{Covino}, S., {D'Avanzo}, P., {Antonelli}, L.~A., {et~al.} 2008{\natexlab{b}},
  \gcn, 8233

\bibitem[{{Cucchiara} \& {Fox}(2008)}]{Cucchiara08_GCN7276}
{Cucchiara}, A. \& {Fox}, D.~B. 2008, \gcn, 7276

\bibitem[{{Cucchiara} {et~al.}(2011){Cucchiara}, {Levan}, {Fox}, {Tanvir},
  {Ukwatta}, {Berger}, {Kr{\"u}hler}, {K{\"u}pc{\"u} Yolda{\c s}}, {Wu},
  {Toma}, {Greiner}, {Olivares}, {Rowlinson}, {Amati}, {Sakamoto}, {Roth},
  {Stephens}, {Fritz}, {Fynbo}, {Hjorth}, {Malesani}, {Jakobsson}, {Wiersema},
  {O'Brien}, {Soderberg}, {Foley}, {Fruchter}, {Rhoads}, {Rutledge}, {Schmidt},
  {Dopita}, {Podsiadlowski}, {Willingale}, {Wolf}, {Kulkarni}, \&
  {D'Avanzo}}]{Cucchiara2011}
{Cucchiara}, A., {Levan}, A.~J., {Fox}, D.~B., {et~al.} 2011, \apj, 736, 7

\bibitem[{{Cucchiara} \& {Racusin}(2008)}]{Cucchiara2008GCN7268}
{Cucchiara}, A. \& {Racusin}, J. 2008, \gcn, 7268

\bibitem[{{Cummings} {et~al.}(2005{\natexlab{a}}){Cummings}, {Barbier},
  {Barthelmy}, {Fenimore}, {Gehrels}, {Gendreau}, {Hullinger}, {Krimm},
  {Markwardt}, {Meszaros}, {Mitani}, {Palmer}, {Parsons}, {Sakamoto}, {Sato},
  {Suzuki}, \& {Tueller}}]{Cummings2005GCN3637}
{Cummings}, J., {Barbier}, L., {Barthelmy}, S., {et~al.} 2005{\natexlab{a}},
  \gcn, 3637

\bibitem[{{Cummings} {et~al.}(2005{\natexlab{b}}){Cummings}, {Barbier},
  {Barthelmy}, {Fenimore}, {Gehrels}, {Hullinger}, {Krimm}, {Markwardt},
  {Palmer}, {Parsons}, {Sakamoto}, {Sato}, \& {Tueller}}]{Cummings05_GCN4033}
{Cummings}, J., {Barbier}, L., {Barthelmy}, S., {et~al.} 2005{\natexlab{b}},
  \gcn, 4033

\bibitem[{{Cummings} {et~al.}(2008){Cummings}, {Barthelmy}, {Gehrels}, {Krimm},
  {Palmer}, {Golenetskii}, {Aptekar}, {Mazets}, {Pal'shin}, {Frederiks},
  {Hurley}, {Cline}, {Yamaoka}, {Ohno}, {Fukazawa}, {Takahashi}, {Tashiro},
  {Terada}, {Murakami}, {Makishima}, {Kienlin}, {Lichti}, {Rau}, {Marisaldi},
  {Fuschino}, {Galli}, {Labanti}, {Monte}, {Lazzarotto}, {Pacciani}, \&
  {Soffitta}}]{Cummings08_GCN8484}
{Cummings}, J., {Barthelmy}, S., {Gehrels}, N., {et~al.} 2008, \gcn, 8484

\bibitem[{{Curran} {et~al.}(2008){Curran}, {Schady}, \&
  {Cummings}}]{Curran08_GCN8488}
{Curran}, P.~A., {Schady}, P., \& {Cummings}, J. 2008, \gcn, 8488

\bibitem[{{D'Avanzo} {et~al.}(2008){D'Avanzo}, {Antonelli}, {Covino},
  {Fugazza}, {Calzoletti}, {Campana}, {Chincarini}, {Conciatore}, {Cutini},
  {D'Elia}, {Dalessio}, {Fiore}, {Goldoni}, {Guetta}, {Guidorzi}, {Israel},
  {Masetti}, {Melandri}, {Meurs}, {Nicastro}, {Palazzi}, {Pian}, {Piranomonte},
  {Stella}, {Stratta}, {Tagliaferri}, {Tosti}, {Testa}, {Vergani}, \&
  {Vitali}}]{D'Avanzo08_GCN7269}
{D'Avanzo}, P., {Antonelli}, L.~A., {Covino}, S., {et~al.} 2008, \gcn, 7269

\bibitem[{{De Pasquale} {et~al.}(2003){De Pasquale}, {Piro}, {Perna}, {Costa},
  {Feroci}, {Gandolfi}, {in't Zand}, {Nicastro}, {Frontera}, {Antonelli},
  {Fiore}, \& {Stratta}}]{DePasquale2003a}
{De Pasquale}, M., {Piro}, L., {Perna}, R., {et~al.} 2003, ApJ, 592, 1018

\bibitem[{{de Ugarte Postigo} {et~al.}(2007){de Ugarte Postigo},
  {Castro-Tirado}, {Jelinek}, {Kubanek}, {Cunniffe}, {Vitek}, {Gorosabel},
  {Skillen}, \& {Sabau-Graziati}}]{Postigo2007GCN6361}
{de Ugarte Postigo}, A., {Castro-Tirado}, A.~J., {Jelinek}, M., {et~al.} 2007,
  \gcn, 6361

\bibitem[{{de Ugarte Postigo} \& {Malesani}(2008)}]{Postigo08_GCN8366}
{de Ugarte Postigo}, A. \& {Malesani}, D. 2008, \gcn, 8366

\bibitem[{{Doherty} {et~al.}(2005){Doherty}, {Bunker}, {Ellis}, \&
  {McCarthy}}]{Doherty2005a}
{Doherty}, M., {Bunker}, A.~J., {Ellis}, R.~S., \& {McCarthy}, P.~J. 2005,
  \mnras, 361, 525

\bibitem[{{Dullighan} {et~al.}(2004){Dullighan}, {Ricker}, {Butler}, \&
  {Vanderspek}}]{Dullighan2004a}
{Dullighan}, A., {Ricker}, G., {Butler}, N., \& {Vanderspek}, R. 2004, in
  American Institute of Physics Conference Series, Vol. 727, Gamma-Ray Bursts:
  30 Years of Discovery, ed. {E.~Fenimore \& M.~Galassi}, 467--470

\bibitem[{{Elston} {et~al.}(1988){Elston}, {Rieke}, \& {Rieke}}]{Elston1988}
{Elston}, R., {Rieke}, G.~H., \& {Rieke}, M.~J. 1988, \apjl, 331, L77

\bibitem[{{Evans} {et~al.}(2010){Evans}, {Primini}, {Glotfelty}, {Anderson},
  {Bonaventura}, {Chen}, {Davis}, {Doe}, {Evans}, {Fabbiano}, {Galle}, {Gibbs},
  {Grier}, {Hain}, {Hall}, {Harbo}, {(Helen He}, {Houck}, {Karovska},
  {Kashyap}, {Lauer}, {McCollough}, {McDowell}, {Miller}, {Mitschang},
  {Morgan}, {Mossman}, {Nichols}, {Nowak}, {Plummer}, {Refsdal}, {Rots},
  {Siemiginowska}, {Sundheim}, {Tibbetts}, {Van Stone}, {Winkelman}, \&
  {Zografou}}]{Evans2010}
{Evans}, I.~N., {Primini}, F.~A., {Glotfelty}, K.~J., {et~al.} 2010, \apjs,
  189, 37

\bibitem[{{Evans}(2011{\natexlab{a}})}]{Evans12250}
{Evans}, P.~A. 2011{\natexlab{a}}, GCN, 12250

\bibitem[{{Evans}(2011{\natexlab{b}})}]{Evans12273}
{Evans}, P.~A. 2011{\natexlab{b}}, GCN, 12273

\bibitem[{{Evans} {et~al.}(2009){Evans}, {Beardmore}, {Page}, {Osborne},
  {O'Brien}, {Willingale}, {Starling}, {Burrows}, {Godet}, {Vetere}, {Racusin},
  {Goad}, {Wiersema}, {Angelini}, {Capalbi}, {Chincarini}, {Gehrels}, {Kennea},
  {Margutti}, {Morris}, {Mountford}, {Pagani}, {Perri}, {Romano}, \&
  {Tanvir}}]{Evans2009a}
{Evans}, P.~A., {Beardmore}, A.~P., {Page}, K.~L., {et~al.} 2009, \mnras, 397,
  1177

\bibitem[{{Evans} {et~al.}(2007){Evans}, {Beardmore}, {Page}, {Tyler},
  {Osborne}, {Goad}, {O'Brien}, {Vetere}, {Racusin}, {Morris}, {Burrows},
  {Capalbi}, {Perri}, {Gehrels}, \& {Romano}}]{Evans2007a}
{Evans}, P.~A., {Beardmore}, A.~P., {Page}, K.~L., {et~al.} 2007, \aap, 469,
  379

\bibitem[{{Evans} {et~al.}(2008){Evans}, {Goad}, {Osborne}, \&
  {Beardmore}}]{Evans08_GCN8391}
{Evans}, P.~A., {Goad}, M.~R., {Osborne}, J.~P., \& {Beardmore}, A.~P. 2008,
  \gcn, 8391

\bibitem[{{Evans} \& {Oates}(2008)}]{Evans2008GCN8231}
{Evans}, P.~A. \& {Oates}, S.~R. 2008, \gcn, 8231

\bibitem[{{Filgas} {et~al.}(2008){Filgas}, {Kruehler}, {Greiner}, {Clemens},
  {Yolda\c{s} }, {Rossi}, {K{\"u}pc{\"u} Yolda\c{s} }, \&
  {Szokoly}}]{Filgas2008}
{Filgas}, R., {Kruehler}, T., {Greiner}, J., {et~al.} 2008, \gcn, 8373

\bibitem[{{Fontanot} \& {Monaco}(2010)}]{Fontanot2010a}
{Fontanot}, F. \& {Monaco}, P. 2010, \mnras, 405, 705

\bibitem[{{Fox} {et~al.}(2007){Fox}, {Price}, \& {Berger}}]{Fox6420}
{Fox}, D.~B., {Price}, P.~A., \& {Berger}, E. 2007, \gcn, 6420

\bibitem[{{French} {et~al.}(2008){French}, {Jelinek}, {Kubanek}, \& {de Ugarte
  Postigo}}]{French2008GCN7316}
{French}, J., {Jelinek}, M., {Kubanek}, P., \& {de Ugarte Postigo}, A. 2008,
  \gcn, 7316

\bibitem[{{Fruchter} {et~al.}(2006){Fruchter}, {Levan}, {Strolger},
  {Vreeswijk}, {Thorsett}, {Bersier}, {Burud}, {Castro Cer{\'o}n},
  {Castro-Tirado}, {Conselice}, {Dahlen}, {Ferguson}, {Fynbo}, {Garnavich},
  {Gibbons}, {Gorosabel}, {Gull}, {Hjorth}, {Holland}, {Kouveliotou}, {Levay},
  {Livio}, {Metzger}, {Nugent}, {Petro}, {Pian}, {Rhoads}, {Riess}, {Sahu},
  {Smette}, {Tanvir}, {Wijers}, \& {Woosley}}]{Fruchter2006a}
{Fruchter}, A.~S., {Levan}, A.~J., {Strolger}, L., {et~al.} 2006, \nat, 441,
  463

\bibitem[{{Fruchter} {et~al.}(1999){Fruchter}, {Thorsett}, {Metzger}, {Sahu},
  {Petro}, {Livio}, {Ferguson}, {Pian}, {Hogg}, {Galama}, {Gull},
  {Kouveliotou}, {Macchetto}, {van Paradijs}, {Pedersen}, \&
  {Smette}}]{Fruchter1999b}
{Fruchter}, A.~S., {Thorsett}, S.~E., {Metzger}, M.~R., {et~al.} 1999, \apjl,
  519, L13

\bibitem[{{Fugazza} {et~al.}(2008){Fugazza}, {D'Elia}, {D'Avanzo}, {Covino}, \&
  {Tagliaferri}}]{Fugazza08_GCN7293}
{Fugazza}, D., {D'Elia}, V., {D'Avanzo}, P., {Covino}, S., \& {Tagliaferri}, G.
  2008, \gcn, 7293

\bibitem[{{Fynbo} {et~al.}(2009){Fynbo}, {Jakobsson}, {Prochaska}, {Malesani},
  {Ledoux}, {de Ugarte Postigo}, {Nardini}, {Vreeswijk}, {Wiersema}, {Hjorth},
  {Sollerman}, {Chen}, {Th{\"o}ne}, {Bj{\"o}rnsson}, {Bloom}, {Castro-Tirado},
  {Christensen}, {De Cia}, {Fruchter}, {Gorosabel}, {Graham}, {Jaunsen},
  {Jensen}, {Kann}, {Kouveliotou}, {Levan}, {Maund}, {Masetti},
  {Milvang-Jensen}, {Palazzi}, {Perley}, {Pian}, {Rol}, {Schady}, {Starling},
  {Tanvir}, {Watson}, {Xu}, {Augusteijn}, {Grundahl}, {Telting}, \&
  {Quirion}}]{Fynbo2009a}
{Fynbo}, J.~P.~U., {Jakobsson}, P., {Prochaska}, J.~X., {et~al.} 2009, \apjs,
  185, 526

\bibitem[{{Fynbo} {et~al.}(2001){Fynbo}, {Jensen}, {Gorosabel}, {Hjorth},
  {Pedersen}, {M{\o}ller}, {Abbott}, {Castro-Tirado}, {Delgado}, {Greiner},
  {Henden}, {Magazz{\`u}}, {Masetti}, {Merlino}, {Masegosa}, {{\O}stensen},
  {Palazzi}, {Pian}, {Schwarz}, {Cline}, {Guidorzi}, {Goldsten}, {Hurley},
  {Mazets}, {McClanahan}, {Montanari}, {Starr}, \& {Trombka}}]{Fynbo2001a}
{Fynbo}, J.~U., {Jensen}, B.~L., {Gorosabel}, J., {et~al.} 2001, A\&A, 369, 373

\bibitem[{{Gehrels} {et~al.}(2004){Gehrels}, {Chincarini}, {Giommi}, {Mason},
  {Nousek}, {Wells}, {White}, {Barthelmy}, {Burrows}, {Cominsky}, {Hurley},
  {Marshall}, {M{\'e}sz{\'a}ros}, {Roming}, {Angelini}, {Barbier}, {Belloni},
  {Campana}, {Caraveo}, {Chester}, {Citterio}, {Cline}, {Cropper}, {Cummings},
  {Dean}, {Feigelson}, {Fenimore}, {Frail}, {Fruchter}, {Garmire}, {Gendreau},
  {Ghisellini}, {Greiner}, {Hill}, {Hunsberger}, {Krimm}, {Kulkarni}, {Kumar},
  {Lebrun}, {Lloyd-Ronning}, {Markwardt}, {Mattson}, {Mushotzky}, {Norris},
  {Osborne}, {Paczynski}, {Palmer}, {Park}, {Parsons}, {Paul}, {Rees},
  {Reynolds}, {Rhoads}, {Sasseen}, {Schaefer}, {Short}, {Smale}, {Smith},
  {Stella}, {Tagliaferri}, {Takahashi}, {Tashiro}, {Townsley}, {Tueller},
  {Turner}, {Vietri}, {Voges}, {Ward}, {Willingale}, {Zerbi}, \&
  {Zhang}}]{Gehrels2004}
{Gehrels}, N., {Chincarini}, G., {Giommi}, P., {et~al.} 2004, \apj, 611, 1005

\bibitem[{{Gilmore}(2007)}]{Gilmore07_GCN6412}
{Gilmore}, A.~C. 2007, \gcn, 6412

\bibitem[{{Godet} {et~al.}(2005){Godet}, {Page}, {Osborne}, {Burrows},
  {Gehrels}, {Hurley}, \& {Chester}}]{Godet05_GCN4031}
{Godet}, O., {Page}, K.~L., {Osborne}, J.~P., {et~al.} 2005, \gcn, 4031

\bibitem[{{Golenetskii} {et~al.}(2005){Golenetskii}, {Aptekar}, {Mazets},
  {Pal'Shin}, {Frederiks}, \& {Cline}}]{Golenetskii2005GCN3640}
{Golenetskii}, S., {Aptekar}, R., {Mazets}, E., {et~al.} 2005, \gcn, 3640

\bibitem[{{Golenetskii} {et~al.}(2008){Golenetskii}, {Aptekar}, {Mazets},
  {Pal'shin}, {Frederiks}, \& {Cline}}]{Golenetskii08_GCN7784}
{Golenetskii}, S., {Aptekar}, R., {Mazets}, E., {et~al.} 2008, \gcn, 7784

\bibitem[{{Gomboc} {et~al.}(2006){Gomboc}, {Guidorzi}, {Steele}, {Kobayashi},
  {Mundell}, {Monfardini}, {Melandri}, {Mottram}, {Smith}, {Bersier}, {Carter},
  {Bode}, {O'Brien}, {Rol}, \& {Bannister}}]{Gomboc2006GCN4738}
{Gomboc}, A., {Guidorzi}, C., {Steele}, I.~A., {et~al.} 2006, \gcn, 4738

\bibitem[{{Gonzalez-Perez} {et~al.}(2009){Gonzalez-Perez}, {Baugh}, {Lacey}, \&
  {Almeida}}]{Gonzales2009a}
{Gonzalez-Perez}, V., {Baugh}, C.~M., {Lacey}, C.~G., \& {Almeida}, C. 2009,
  \mnras, 398, 497

\bibitem[{{Gonzalez-Perez} {et~al.}(2011){Gonzalez-Perez}, {Baugh}, {Lacey}, \&
  {Kim}}]{Gonzalez-Perez2011a}
{Gonzalez-Perez}, V., {Baugh}, C.~M., {Lacey}, C.~G., \& {Kim}, J.-W. 2011,
  \mnras, 417, 517

\bibitem[{{Gorosabel} {et~al.}(2003){Gorosabel}, {Christensen}, {Hjorth},
  {Fynbo}, {Pedersen}, {Jensen}, {Andersen}, {Lund}, {Jaunsen}, {Castro
  Cer{\'o}n}, {Castro-Tirado}, {Fruchter}, {Greiner}, {Pian}, {Vreeswijk},
  {Burud}, {Frontera}, {Kaper}, {Klose}, {Kouveliotou}, {Masetti}, {Palazzi},
  {Rhoads}, {Rol}, {Salamanca}, {Tanvir}, {Wijers}, \& {van den
  Heuvel}}]{Gorosabel2003a}
{Gorosabel}, J., {Christensen}, L., {Hjorth}, J., {et~al.} 2003, \aap, 400, 127

\bibitem[{{G\"otz} {et~al.}(2008){G\"otz}, {Mereghetti}, {Paizis}, {Galis},
  {Beckmann}, {Beck}, \& {Borkowski}}]{Gotz08_GCN8614}
{G\"otz}, D., {Mereghetti}, S., {Paizis}, A., {et~al.} 2008, \gcn, 8614

\bibitem[{{Greiner} {et~al.}(2008){Greiner}, {Bornemann}, {Clemens}, {Deuter},
  {Hasinger}, {Honsberg}, {Huber}, {Huber}, {Krauss}, {Kr{\"u}hler},
  {K{\"u}pc{\"u} Yolda{\c s}}, {Mayer-Hasselwander}, {Mican}, {Primak},
  {Schrey}, {Steiner}, {Szokoly}, {Th{\"o}ne}, {Yolda{\c s}}, {Klose}, {Laux},
  \& {Winkler}}]{Greiner2008a}
{Greiner}, J., {Bornemann}, W., {Clemens}, C., {et~al.} 2008, \pasp, 120, 405

\bibitem[{{Greiner} {et~al.}(2009){Greiner}, {Kr{\"u}hler}, {Fynbo}, {Rossi},
  {Schwarz}, {Klose}, {Savaglio}, {Tanvir}, {McBreen}, {Totani}, {Zhang}, {Wu},
  {Watson}, {Barthelmy}, {Beardmore}, {Ferrero}, {Gehrels}, {Kann}, {Kawai},
  {Yolda{\c s}}, {M{\'e}sz{\'a}ros}, {Milvang-Jensen}, {Oates}, {Pierini},
  {Schady}, {Toma}, {Vreeswijk}, {Yolda{\c s}}, {Zhang}, {Afonso}, {Aoki},
  {Burrows}, {Clemens}, {Filgas}, {Haiman}, {Hartmann}, {Hasinger}, {Hjorth},
  {Jehin}, {Levan}, {Liang}, {Malesani}, {Pyo}, {Schulze}, {Szokoly}, {Terada},
  \& {Wiersema}}]{Greiner2009}
{Greiner}, J., {Kr{\"u}hler}, T., {Fynbo}, J.~P.~U., {et~al.} 2009, \apj, 693,
  1610

\bibitem[{{Greiner} {et~al.}(2011){Greiner}, {Kr{\"u}hler}, {Klose}, {Afonso},
  {Clemens}, {Filgas}, {Hartmann}, {K{\"u}pc{\"u} Yolda{\c s}}, {Nardini},
  {Olivares E.}, {Rau}, {Rossi}, {Schady}, \& {Updike}}]{Greiner2011}
{Greiner}, J., {Kr{\"u}hler}, T., {Klose}, S., {et~al.} 2011, \aap, 526, A30

\bibitem[{{Guidorzi} {et~al.}(2006{\natexlab{a}}){Guidorzi}, {Barthelmy},
  {Evans}, {Gehrels}, {Gronwall}, {Kennea}, {Krimm}, {Mangano}, {Moretti},
  {Page}, {Palmer}, {Romano}, \& {Vergani}}]{Guidorzi2006GCN5575}
{Guidorzi}, C., {Barthelmy}, S.~D., {Evans}, P.~A., {et~al.}
  2006{\natexlab{a}}, \gcn, 5575

\bibitem[{{Guidorzi} {et~al.}(2006{\natexlab{b}}){Guidorzi}, {Romano},
  {Moretti}, \& {Vergani}}]{Guidorzi2006GCN5577}
{Guidorzi}, C., {Romano}, P., {Moretti}, A., \& {Vergani}, S.
  2006{\natexlab{b}}, \gcn, 5577

\bibitem[{{Guziy} {et~al.}(2005){Guziy}, {Jelinek}, {Gorosabel},
  {Castro-Tirado}, {Postigo}, {Flores}, \& {Vijanen}}]{GuziyGCN4025}
{Guziy}, S., {Jelinek}, M., {Gorosabel}, J., {et~al.} 2005, \gcn, 4025

\bibitem[{{Haislip} {et~al.}(2006){Haislip}, {Nysewander}, {Reichart}, {Levan},
  {Tanvir}, {Cenko}, {Fox}, {Price}, {Castro-Tirado}, {Gorosabel}, {Evans},
  {Figueredo}, {MacLeod}, {Kirschbrown}, {Jelinek}, {Guziy}, {Postigo},
  {Cypriano}, {Lacluyze}, {Graham}, {Priddey}, {Chapman}, {Rhoads}, {Fruchter},
  {Lamb}, {Kouveliotou}, {Wijers}, {Bayliss}, {Schmidt}, {Soderberg},
  {Kulkarni}, {Harrison}, {Moon}, {Gal-Yam}, {Kasliwal}, {Hudec}, {Vitek},
  {Kubanek}, {Crain}, {Foster}, {Clemens}, {Bartelme}, {Canterna}, {Hartmann},
  {Henden}, {Klose}, {Park}, {Williams}, {Rol}, {O'Brien}, {Bersier}, {Prada},
  {Pizarro}, {Maturana}, {Ugarte}, {Alvarez}, {Fernandez}, {Jarvis}, {Moles},
  {Alfaro}, {Ivarsen}, {Kumar}, {Mack}, {Zdarowicz}, {Gehrels}, {Barthelmy}, \&
  {Burrows}}]{Haislip2006}
{Haislip}, J.~B., {Nysewander}, M.~C., {Reichart}, D.~E., {et~al.} 2006, \nat,
  440, 181

\bibitem[{{Hashimoto} {et~al.}(2010){Hashimoto}, {Ohta}, {Aoki}, {Tanaka},
  {Yabe}, {Kawai}, {Aoki}, {Furusawa}, {Hattori}, {Iye}, {Kawabata},
  {Kobayashi}, {Komiyama}, {Kosugi}, {Minowa}, {Mizumoto}, {Niino}, {Nomoto},
  {Noumaru}, {Ogasawara}, {Pyo}, {Sakamoto}, {Sekiguchi}, {Shirasaki},
  {Suzuki}, {Tajitsu}, {Takata}, {Tamagawa}, {Terada}, {Totani}, {Watanabe},
  {Yamada}, \& {Yoshida}}]{Hashimoto2010}
{Hashimoto}, T., {Ohta}, K., {Aoki}, K., {et~al.} 2010, \apj, 719, 378

\bibitem[{{Hempel} {et~al.}(2011){Hempel}, {Crist{\'o}bal-Hornillos}, {Prieto},
  {Trujillo}, {Balcells}, {L{\'o}pez-Sanjuan}, {Abreu}, {Eliche-Moral}, \&
  {Dom{\'{\i}}nguez Palmero}}]{Hempel2011a}
{Hempel}, A., {Crist{\'o}bal-Hornillos}, D., {Prieto}, M., {et~al.} 2011,
  \mnras, 414, 2246

\bibitem[{{Hogg} {et~al.}(1997){Hogg}, {Pahre}, {McCarthy}, {Cohen},
  {Blandford}, {Smail}, \& {Soifer}}]{Hogg1997}
{Hogg}, D.~W., {Pahre}, M.~A., {McCarthy}, J.~K., {et~al.} 1997, \mnras, 288,
  404

\bibitem[{{Holland} \& {Cucchiara}(2006)}]{Holland2006GCN5603}
{Holland}, S. \& {Cucchiara}, A. 2006, \gcn, 5603

\bibitem[{{Holland}(2006)}]{Holland2006GCN5784}
{Holland}, S.~T. 2006, \gcn, 5784

\bibitem[{{Holland} {et~al.}(2006){Holland}, {Barthelmy}, {Chester}, {Gehrels},
  {Gronwall}, {Kennea}, {Marshall}, {McLean}, {Page}, {Palmer}, {Parsons},
  {Stamatikos}, \& {Starling}}]{Holland2006GCN5776}
{Holland}, S.~T., {Barthelmy}, S.~D., {Chester}, M.~M., {et~al.} 2006, \gcn,
  5776

\bibitem[{{Holland} {et~al.}(2010){Holland}, {Sbarufatti}, {Shen}, {Schady},
  {Cummings}, {Fonseca}, {Fynbo}, {Jakobsson}, {Leitet}, {Linn{\'e}}, {Roming},
  {Still}, \& {Zhang}}]{Holland2010}
{Holland}, S.~T., {Sbarufatti}, B., {Shen}, R., {et~al.} 2010, \apj, 717, 223

\bibitem[{{Hunt} {et~al.}(2011){Hunt}, {Palazzi}, {Rossi}, {Savaglio},
  {Cresci}, {Klose}, {Micha{\l}owski}, \& {Pian}}]{Hunt2011a}
{Hunt}, L., {Palazzi}, E., {Rossi}, A., {et~al.} 2011, \apjl, 736, L36

\bibitem[{{Hurkett} {et~al.}(2006){Hurkett}, {Beardmore}, {Godet}, {Kennea},
  {Krimm}, {Marshall}, {Osborne}, {Palmer}, \& {Parsons}}]{Hurkett2006GCN4736}
{Hurkett}, C., {Beardmore}, A., {Godet}, O., {et~al.} 2006, \gcn, 4736

\bibitem[{{Hurkett} {et~al.}(2005{\natexlab{a}}){Hurkett}, {Page}, {Burrows},
  {Chester}, {Angelini}, \& {Gehrels}}]{Hurkett2005GCN3636}
{Hurkett}, C., {Page}, K., {Burrows}, D., {et~al.} 2005{\natexlab{a}}, \gcn,
  3636

\bibitem[{{Hurkett} {et~al.}(2005{\natexlab{b}}){Hurkett}, {Page}, {Kennea},
  {Burrows}, {Blustin}, {Barbier}, {Markwardt}, {Parsons}, \&
  {Gehrels}}]{Hurkett2005GCN3633}
{Hurkett}, C., {Page}, K., {Kennea}, J., {et~al.} 2005{\natexlab{b}}, \gcn,
  3633

\bibitem[{{Immler} {et~al.}(2008){Immler}, {Baumgartner}, {Gehrels},
  {Hunsberger}, {Landsman}, {Markwardt}, {Page}, {Palmer}, {Parsons}, {Perez},
  {Sakamoto}, \& {Ukwatta}}]{Immler08_GCN8021}
{Immler}, S., {Baumgartner}, W.~H., {Gehrels}, N., {et~al.} 2008, \gcn, 8021

\bibitem[{{Jakobsson} {et~al.}(2004){Jakobsson}, {Hjorth}, {Fynbo}, {Watson},
  {Pedersen}, {Bj{\"o}rnsson}, \& {Gorosabel}}]{Jakobsson2004a}
{Jakobsson}, P., {Hjorth}, J., {Fynbo}, J.~P.~U., {et~al.} 2004, ApJ, 617, L21
  (J04)

\bibitem[{{Kann} {et~al.}(2011){Kann}, {Klose}, {Zhang}, {Covino}, {Butler},
  {Malesani}, {Nakar}, {Wilson}, {Antonelli}, {Chincarini}, {Cobb}, {D'Avanzo},
  {D'Elia}, {Della Valle}, {Ferrero}, {Fugazza}, {Gorosabel}, {Israel},
  {Mannucci}, {Piranomonte}, {Schulze}, {Stella}, {Tagliaferri}, \&
  {Wiersema}}]{Kann2011a}
{Kann}, D.~A., {Klose}, S., {Zhang}, B., {et~al.} 2011, \apj, 734, 96

\bibitem[{{Kann} {et~al.}(2010){Kann}, {Klose}, {Zhang}, {Malesani}, {Nakar},
  {Pozanenko}, {Wilson}, {Butler}, {Jakobsson}, {Schulze}, {Andreev},
  {Antonelli}, {Bikmaev}, {Biryukov}, {B{\"o}ttcher}, {Burenin}, {Castro
  Cer{\'o}n}, {Castro-Tirado}, {Chincarini}, {Cobb}, {Covino}, {D'Avanzo},
  {D'Elia}, {Della Valle}, {de Ugarte Postigo}, {Efimov}, {Ferrero}, {Fugazza},
  {Fynbo}, {G{\aa}lfalk}, {Grundahl}, {Gorosabel}, {Gupta}, {Guziy}, {Hafizov},
  {Hjorth}, {Holhjem}, {Ibrahimov}, {Im}, {Israel}, {Je{\'l}inek}, {Jensen},
  {Karimov}, {Khamitov}, {Kizilo{\v g}lu}, {Klunko}, {Kub{\'a}nek}, {Kutyrev},
  {Laursen}, {Levan}, {Mannucci}, {Martin}, {Mescheryakov}, {Mirabal},
  {Norris}, {Ovaldsen}, {Paraficz}, {Pavlenko}, {Piranomonte}, {Rossi},
  {Rumyantsev}, {Salinas}, {Sergeev}, {Sharapov}, {Sollerman}, {Stecklum},
  {Stella}, {Tagliaferri}, {Tanvir}, {Telting}, {Testa}, {Updike}, {Volnova},
  {Watson}, {Wiersema}, \& {Xu}}]{Kann2010a}
{Kann}, D.~A., {Klose}, S., {Zhang}, B., {et~al.} 2010, \apj, 720, 1513

\bibitem[{{Kawai} {et~al.}(2006){Kawai}, {Kosugi}, {Aoki}, {Yamada}, {Totani},
  {Ohta}, {Iye}, {Hattori}, {Aoki}, {Furusawa}, {Hurley}, {Kawabata},
  {Kobayashi}, {Komiyama}, {Mizumoto}, {Nomoto}, {Noumaru}, {Ogasawara},
  {Sato}, {Sekiguchi}, {Shirasaki}, {Suzuki}, {Takata}, {Tamagawa}, {Terada},
  {Watanabe}, {Yatsu}, \& {Yoshida}}]{Kawai2006}
{Kawai}, N., {Kosugi}, G., {Aoki}, K., {et~al.} 2006, \nat, 440, 184

\bibitem[{{Kennea} {et~al.}(2005){Kennea}, {Burrows}, {Hurkett}, {Page}, \&
  {Gehrels}}]{Kennea2005GCN3634}
{Kennea}, J.~A., {Burrows}, D.~N., {Hurkett}, C., {Page}, K., \& {Gehrels}, N.
  2005, \gcn, 3634

\bibitem[{{Kennea} \& {Stroh}(2008)}]{Kennea08_GCN8364}
{Kennea}, J.~A. \& {Stroh}, M. 2008, \gcn, 8364

\bibitem[{{Khamitov} {et~al.}(2008){Khamitov}, {Kose}, {Yakut}, {Eker},
  {Kiziloglu}, {Gogus}, {Sakhibullin}, {Pavlinsky}, \&
  {Sunyaev}}]{Khamitov08_GCN7270}
{Khamitov}, I., {Kose}, O., {Yakut}, K., {et~al.} 2008, \gcn, 7270

\bibitem[{{Kim} {et~al.}(2011){Kim}, {Edge}, {Wake}, \& {Stott}}]{Kim2011a}
{Kim}, J.-W., {Edge}, A.~C., {Wake}, D.~A., \& {Stott}, J.~P. 2011, \mnras,
  410, 241

\bibitem[{{Kissler-Patig} {et~al.}(2008){Kissler-Patig}, {Pirard}, {Casali},
  {Moorwood}, {Ageorges}, {Alves de Oliveira}, {Baksai}, {Bedin}, {Bendek},
  {Biereichel}, {Delabre}, {Dorn}, {Esteves}, {Finger}, {Gojak}, {Huster},
  {Jung}, {Kiekebush}, {Klein}, {Koch}, {Lizon}, {Mehrgan}, {Petr-Gotzens},
  {Pritchard}, {Selman}, \& {Stegmeier}}]{Kissler-Patig2008a}
{Kissler-Patig}, M., {Pirard}, J.-F., {Casali}, M., {et~al.} 2008, \aap, 491,
  941

\bibitem[{{Klose} {et~al.}(1996){Klose}, {Eisl\"offel}, \&
  {Richter}}]{Klose1996}
{Klose}, S., {Eisl\"offel}, J., \& {Richter}, S. 1996, \apjl, 470, L93

\bibitem[{{Klose} {et~al.}(2004){Klose}, {Greiner}, {Rau}, {Henden},
  {Hartmann}, {Zeh}, {Ries}, {Masetti}, {Malesani}, {Guenther}, {Gorosabel},
  {Stecklum}, {Antonelli}, {Brinkworth}, {Castro Cer{\'o}n}, {Castro-Tirado},
  {Covino}, {Fruchter}, {Fynbo}, {Ghisellini}, {Hjorth}, {Hudec},
  {Jel{\'{\i}}nek}, {Kaper}, {Kouveliotou}, {Lindsay}, {Maiorano}, {Mannucci},
  {Nysewander}, {Palazzi}, {Pedersen}, {Pian}, {Reichart}, {Rhoads}, {Rol},
  {Smail}, {Tanvir}, {de Ugarte Postigo}, {Vreeswijk}, {Wijers}, \& {van den
  Heuvel}}]{Klose2004}
{Klose}, S., {Greiner}, J., {Rau}, A., {et~al.} 2004, \aj, 128, 1942

\bibitem[{{Klotz} {et~al.}(2006){Klotz}, {Boer}, \&
  {Atteia}}]{Klotz2006GCN5576}
{Klotz}, A., {Boer}, M., \& {Atteia}, J.~L. 2006, \gcn, 5576

\bibitem[{{Klotz} {et~al.}(2008){Klotz}, {Boer}, \&
  {Atteia}}]{Klotz2008GCN7267}
{Klotz}, A., {Boer}, M., \& {Atteia}, J.~L. 2008, \gcn, 7267

\bibitem[{{Kong} {et~al.}(2009){Kong}, {Fang}, {Arimoto}, \&
  {Wang}}]{Kong2009ApJ702}
{Kong}, X., {Fang}, G., {Arimoto}, N., \& {Wang}, M. 2009, \apj, 702, 1458

\bibitem[{{Kornienko} {et~al.}(2005){Kornienko}, {Rumyantsev}, \&
  {Pozanenko}}]{Kornienko05_GCN4047}
{Kornienko}, G., {Rumyantsev}, V., \& {Pozanenko}, A. 2005, \gcn, 4047

\bibitem[{{Kouveliotou} {et~al.}(1993){Kouveliotou}, {Meegan}, {Fishman},
  {Bhat}, {Briggs}, {Koshut}, {Paciesas}, \& {Pendleton}}]{Kouveliotou1993a}
{Kouveliotou}, C., {Meegan}, C.~A., {Fishman}, G.~J., {et~al.} 1993, \apjl,
  413, L101

\bibitem[{{Krimm} {et~al.}(2006{\natexlab{a}}){Krimm}, {Barbier}, {Barthelmy},
  {Cummings}, {Fenimore}, {Gehrels}, {Hullinger}, {Hurkett}, {Markwardt},
  {Palmer}, {Parsons}, {Sakamoto}, {Sato}, \& {Tueller}}]{Krimm2006GCN4757}
{Krimm}, H., {Barbier}, L., {Barthelmy}, S., {et~al.} 2006{\natexlab{a}}, \gcn,
  4757

\bibitem[{{Krimm} {et~al.}(2006{\natexlab{b}}){Krimm}, {Hurkett}, {Pal'shin},
  {Norris}, {Zhang}, {Barthelmy}, {Burrows}, {Gehrels}, {Golenetskii},
  {Osborne}, {Parsons}, {Perri}, \& {Willingale}}]{Krimm2006a}
{Krimm}, H.~A., {Hurkett}, C., {Pal'shin}, V., {et~al.} 2006{\natexlab{b}},
  ApJ, 648, 1117

\bibitem[{{Kr{\"u}hler} {et~al.}(2011){Kr{\"u}hler}, {Greiner}, {Schady},
  {Savaglio}, {Afonso}, {Clemens}, {Elliott}, {Filgas}, {Gruber}, {Kann},
  {Klose}, {K{\"u}pc{\"u}-Yolda{\c s}}, {McBreen}, {Olivares}, {Pierini},
  {Rau}, {Rossi}, {Nardini}, {Nicuesa Guelbenzu}, {Sudilovsky}, \&
  {Updike}}]{Kruhler2011A&A534}
{Kr{\"u}hler}, T., {Greiner}, J., {Schady}, P., {et~al.} 2011, \aap, 534, A108

\bibitem[{{Kr{\"u}hler} {et~al.}(2008){Kr{\"u}hler}, {K{\"u}pc{\"u} Yolda{\c
  s}}, {Greiner}, {Clemens}, {McBreen}, {Primak}, {Savaglio}, {Yolda{\c s}},
  {Szokoly}, \& {Klose}}]{Kruhler2008a}
{Kr{\"u}hler}, T., {K{\"u}pc{\"u} Yolda{\c s}}, A., {Greiner}, J., {et~al.}
  2008, \apj, 685, 376

\bibitem[{{Kr{\"u}hler} {et~al.}(2012){Kr{\"u}hler}, {Malesani},
  {Milvang-Jensen}, {Fynbo}, {Hjorth}, {Jakobsson}, {Levan}, {Sparre},
  {Tanvir}, \& {Watson}}]{Kruhler2012a}
{Kr{\"u}hler}, T., {Malesani}, D., {Milvang-Jensen}, B., {et~al.} 2012, ArXiv
  e-prints, (arXiv:1205.4036)

\bibitem[{{Kuin} \& {Stroh}(2008)}]{Kuin08_GCN8365}
{Kuin}, N.~P.~M. \& {Stroh}, M.~C. 2008, \gcn, 8365

\bibitem[{{K{\"u}pc{\"u} Yolda{\c s}} {et~al.}(2010){K{\"u}pc{\"u} Yolda{\c
  s}}, {Greiner}, {Klose}, {Kr{\"u}hler}, \& {Savaglio}}]{KupcuYoldas2010AA515}
{K{\"u}pc{\"u} Yolda{\c s}}, A., {Greiner}, J., {Klose}, S., {Kr{\"u}hler}, T.,
  \& {Savaglio}, S. 2010, \aap, 515, L2

\bibitem[{{K{\"u}pc{\"u} Yolda\c{s}} {et~al.}(2008){K{\"u}pc{\"u} Yolda\c{s}},
  {Yolda\c{s}}, {Greiner}, {Kruehler}, {Klose}, \&
  {Szokoly}}]{Yoldas2008GCN7279}
{K{\"u}pc{\"u} Yolda\c{s}}, A., {Yolda\c{s}}, A., {Greiner}, J., {et~al.} 2008,
  \gcn, 7279

\bibitem[{{Landsman} \& {Immler}(2008)}]{Landsman08_GCN8027}
{Landsman}, W.~B. \& {Immler}, S. 2008, \gcn, 8027

\bibitem[{{Le F{\`e}vre} {et~al.}(2003){Le F{\`e}vre}, {Saisse}, {Mancini},
  {Brau-Nogue}, {Caputi}, {Castinel}, {D'Odorico}, {Garilli}, {Kissler-Patig},
  {Lucuix}, {Mancini}, {Pauget}, {Sciarretta}, {Scodeggio}, {Tresse}, \&
  {Vettolani}}]{LeFevre2003a}
{Le F{\`e}vre}, O., {Saisse}, M., {Mancini}, D., {et~al.} 2003, in Society of
  Photo-Optical Instrumentation Engineers (SPIE) Conference Series, Vol. 4841,
  Society of Photo-Optical Instrumentation Engineers (SPIE) Conference Series,
  ed. M.~{Iye} \& A.~F.~M. {Moorwood}, 1670--1681

\bibitem[{{Le Floc'h} {et~al.}(2003){Le Floc'h}, {Duc}, {Mirabel}, {Sanders},
  {Bosch}, {Diaz}, {Donzelli}, {Rodrigues}, {Courvoisier}, {Greiner},
  {Mereghetti}, {Melnick}, {Maza}, \& {Minniti}}]{LeFloch2003}
{Le Floc'h}, E., {Duc}, P., {Mirabel}, I.~F., {et~al.} 2003, \aap, 400, 499

\bibitem[{{Levan} {et~al.}(2006){Levan}, {Fruchter}, {Rhoads}, {Mobasher},
  {Tanvir}, {Gorosabel}, {Rol}, {Kouveliotou}, {Dell'Antonio}, {Merrill},
  {Bergeron}, {Castro Cer{\'o}n}, {Masetti}, {Vreeswijk}, {Antonelli},
  {Bersier}, {Castro-Tirado}, {Fynbo}, {Garnavich}, {Holland}, {Hjorth},
  {Nugent}, {Pian}, {Smette}, {Thomsen}, {Thorsett}, \& {Wijers}}]{Levan2006a}
{Levan}, A., {Fruchter}, A., {Rhoads}, J., {et~al.} 2006, \apj, 647, 471

\bibitem[{{Levan} \& {Wiersema}(2008)}]{Levan08_GCN8048}
{Levan}, A.~J. \& {Wiersema}, K. 2008, \gcn, 8048

\bibitem[{{Levesque} {et~al.}(2010){Levesque}, {Berger}, {Kewley}, \&
  {Bagley}}]{Levesque2010a}
{Levesque}, E.~M., {Berger}, E., {Kewley}, L.~J., \& {Bagley}, M.~M. 2010, \aj,
  139, 694

\bibitem[{{Li}(2006)}]{Li2006GCN5400}
{Li}, W. 2006, \gcn, 5400

\bibitem[{{Lin} {et~al.}(1996){Lin}, {Kirshner}, {Shectman}, {Landy}, {Oemler},
  {Tucker}, \& {Schechter}}]{Lin1996}
{Lin}, H., {Kirshner}, R.~P., {Shectman}, S.~A., {et~al.} 1996, \apj, 464, 60

\bibitem[{{Lipunov} {et~al.}(2006){Lipunov}, {Kornilov}, {Kuvshinov},
  {Tyurina}, {Belinski}, {Gorbovskoy}, {Krylov}, {Borisov}, {Sankovich},
  {Antipov}, {Vladimirov}, {Sinitsin}, \& {Gritsyk}}]{Lipunov2006GCN4741}
{Lipunov}, V., {Kornilov}, V., {Kuvshinov}, D., {et~al.} 2006, \gcn, 4741

\bibitem[{{Luckas} {et~al.}(2005){Luckas}, {Trondal}, \&
  {Schwartz}}]{Luckas2005GCN3642}
{Luckas}, P., {Trondal}, O., \& {Schwartz}, M. 2005, \gcn, 3642

\bibitem[{{MacLeod} {et~al.}(2005){MacLeod}, {Kirschbrown}, {Haislip},
  {Nysewander}, {Crain}, \& {Reichart}}]{MacLeod2005GCN3652}
{MacLeod}, C., {Kirschbrown}, J., {Haislip}, J., {et~al.} 2005, \gcn, 3652

\bibitem[{{Malesani} {et~al.}(2008{\natexlab{a}}){Malesani}, {Jakobsson},
  {Levan}, {Rol}, \& {Fynbo}}]{Malesani08_GCN8039}
{Malesani}, D., {Jakobsson}, P., {Levan}, A.~J., {Rol}, E., \& {Fynbo},
  J.~P.~U. 2008{\natexlab{a}}, \gcn, 8039

\bibitem[{{Malesani} {et~al.}(2008{\natexlab{b}}){Malesani}, {Quirion},
  {Fynbo}, \& {Jakobsson}}]{Malesani2008GCN7783}
{Malesani}, D., {Quirion}, P., {Fynbo}, J.~P.~U., \& {Jakobsson}, P.
  2008{\natexlab{b}}, \gcn, 7783

\bibitem[{{Malesani} {et~al.}(2008{\natexlab{c}}){Malesani}, {Quirion},
  {Fynbo}, \& {Jakobsson}}]{Malesani08_GCN7783}
{Malesani}, D., {Quirion}, P., {Fynbo}, J.~P.~U., \& {Jakobsson}, P.
  2008{\natexlab{c}}, \gcn, 7783

\bibitem[{{Mangano} {et~al.}(2008{\natexlab{a}}){Mangano}, {Sbarufatti}, {La
  Parola}, \& {Baumgartner}}]{Mangano2008GCN8620}
{Mangano}, V., {Sbarufatti}, B., {La Parola}, V., \& {Baumgartner}, W.~H.
  2008{\natexlab{a}}, \gcn, 8620

\bibitem[{{Mangano} {et~al.}(2008{\natexlab{b}}){Mangano}, {Sbarufatti}, \&
  {Parola}}]{Mangano08_GCN8616}
{Mangano}, V., {Sbarufatti}, B., \& {Parola}, V.~L. 2008{\natexlab{b}}, \gcn,
  8616

\bibitem[{{Mar\`in} {et~al.}(2008){Mar\`in}, {Sabater}, {Castro-Tirado},
  {Gorosabel}, {Jel\`inek}, \& {Postigo}}]{Marin08_GCN7291}
{Mar\`in}, V.~M., {Sabater}, J., {Castro-Tirado}, A.~J., {et~al.} 2008, \gcn,
  7291

\bibitem[{{McLean} {et~al.}(2008){McLean}, {Barthelmy}, {Baumgartner},
  {Cummings}, {Fenimore}, {Gehrels}, {Krimm}, {Markwardt}, {Palmer}, {Parsons},
  {Sakamoto}, {Sato}, {Stamatikos}, {Tueller}, \& {Ukwatta}}]{McLean2008}
{McLean}, K., {Barthelmy}, S.~D., {Baumgartner}, W., {et~al.} 2008, \gcn, 8029

\bibitem[{{Melandri} {et~al.}(2006){Melandri}, {Gomboc}, {Smith}, \&
  {Tanvir}}]{Melandri2006GCN5579}
{Melandri}, A., {Gomboc}, A., {Smith}, R.~J., \& {Tanvir}, N. 2006, \gcn, 5579

\bibitem[{{Melandri} {et~al.}(2012){Melandri}, {Sbarufatti}, {D'Avanzo},
  {Salvaterra}, {Campana}, {Covino}, {Vergani}, {Nava}, {Ghisellini},
  {Ghirlanda}, {Fugazza}, {Mangano}, {Capalbi}, \&
  {Tagliaferri}}]{Melandri2012a}
{Melandri}, A., {Sbarufatti}, B., {D'Avanzo}, P., {et~al.} 2012, \mnras, 421,
  1265

\bibitem[{{Moin} {et~al.}(2008){Moin}, {Tingay}, {Phillips}, {Taylor},
  {Wieringa}, \& {Martin}}]{Moin2008GCN8466}
{Moin}, A., {Tingay}, S., {Phillips}, C., {et~al.} 2008, \gcn, 8466

\bibitem[{{Moorwood} {et~al.}(1998{\natexlab{a}}){Moorwood}, {Cuby},
  {Biereichel}, {Brynnel}, {Delabre}, {Devillard}, {van Dijsseldonk}, {Finger},
  {Gemperlein}, {Gilmozzi}, {Herlin}, {Huster}, {Knudstrup}, {Lidman}, {Lizon},
  {Mehrgan}, {Meyer}, {Nicolini}, {Petr}, {Spyromilio}, \&
  {Stegmeier}}]{Moorwood1998a}
{Moorwood}, A., {Cuby}, J.-G., {Biereichel}, P., {et~al.} 1998{\natexlab{a}},
  The Messenger, 94, 7

\bibitem[{{Moorwood} {et~al.}(1998{\natexlab{b}}){Moorwood}, {Cuby}, \&
  {Lidman}}]{Moorwood1998b}
{Moorwood}, A., {Cuby}, J.-G., \& {Lidman}, C. 1998{\natexlab{b}}, The
  Messenger, 91, 9

\bibitem[{{Muehlegger} {et~al.}(2006){Muehlegger}, {Duscha}, {Stefanescu},
  {Kanbach}, {Primak}, {Schrey}, \& {Steinle}}]{Muehlegger2006GCN5405}
{Muehlegger}, M., {Duscha}, S., {Stefanescu}, A., {et~al.} 2006, \gcn, 5405

\bibitem[{{Norris} {et~al.}(2005){Norris}, {Barbier}, {Barthelmy}, {Boyd},
  {Burrows}, {Cummings}, {Gehrels}, {Holland}, {Kennea}, {Krimm}, {Marshall},
  {Godet}, {Palmer}, \& {Sakamoto}}]{Norris05_GCN4008}
{Norris}, J., {Barbier}, L., {Barthelmy}, S., {et~al.} 2005, \gcn, 4008

\bibitem[{{Norris} {et~al.}(2006){Norris}, {Kutyrev}, {Ganguly}, {Canterna}, \&
  {Pierce}}]{Norris2006GCN4760}
{Norris}, J., {Kutyrev}, A., {Ganguly}, R., {Canterna}, R., \& {Pierce}, M.
  2006, \gcn, 4760

\bibitem[{{Oates} {et~al.}(2008{\natexlab{a}}){Oates}, {Beardmore}, {Cummings},
  {Evans}, {Gronwall}, {Holland}, {Kennea}, {Marshall}, {Page}, {Palmer},
  {Parsons}, {Starling}, \& {Ukwatta}}]{Oates2008GCN8227}
{Oates}, S.~R., {Beardmore}, A.~P., {Cummings}, J.~R., {et~al.}
  2008{\natexlab{a}}, \gcn, 8227

\bibitem[{{Oates} {et~al.}(2008{\natexlab{b}}){Oates}, {Ukwatta}, {Evans}, \&
  {Breeveld}}]{Oates2008GCNrep168}
{Oates}, S.~R., {Ukwatta}, T.~N., {Evans}, P., \& {Breeveld}, A.
  2008{\natexlab{b}}, GCN Report, 168, 1

\bibitem[{{Ovaldsen} {et~al.}(2007){Ovaldsen}, {Jaunsen}, {Fynbo}, {Hjorth},
  {Th{\"o}ne}, {F{\'e}ron}, {Xu}, {Selj}, \& {Teuber}}]{Ovaldsen2007a}
{Ovaldsen}, J.-E., {Jaunsen}, A.~O., {Fynbo}, J.~P.~U., {et~al.} 2007, ApJ,
  662, 294

\bibitem[{{Pandey} {et~al.}(2006){Pandey}, {Page}, {Ziaeepour}, \&
  {Oates}}]{Pandey2006GCN5402}
{Pandey}, S.~B., {Page}, M.~J., {Ziaeepour}, H.~Z., \& {Oates}, S.~R. 2006,
  \gcn, 5402

\bibitem[{{Pasquale} {et~al.}(2005){Pasquale}, {Norris}, {Kennedy}, {Mason}, \&
  {Gehrels}}]{Pasquale05_GCN4028}
{Pasquale}, M.~D., {Norris}, J., {Kennedy}, T., {Mason}, K., \& {Gehrels}, N.
  2005, \gcn, 4028

\bibitem[{{P{\'e}rez-Ram{\'{\i}}rez} {et~al.}(2010){P{\'e}rez-Ram{\'{\i}}rez},
  {de Ugarte Postigo}, {Gorosabel}, {Aloy}, {J{\'o}hannesson}, {Guerrero},
  {Osborne}, {Page}, {Warwick}, {Horv{\'a}th}, {Veres}, {Jel{\'{\i}}nek},
  {Kub{\'a}nek}, {Guziy}, {Bremer}, {Winters}, {Riva}, \&
  {Castro-Tirado}}]{PerezRamirez2010a}
{P{\'e}rez-Ram{\'{\i}}rez}, D., {de Ugarte Postigo}, A., {Gorosabel}, J.,
  {et~al.} 2010, \aap, 510, A105+

\bibitem[{{Perley} {et~al.}(2009){Perley}, {Cenko}, {Bloom}, {Chen}, {Butler},
  {Kocevski}, {Prochaska}, {Brodwin}, {Glazebrook}, {Kasliwal}, {Kulkarni},
  {Lopez}, {Ofek}, {Pettini}, {Soderberg}, \& {Starr}}]{Perley2009}
{Perley}, D.~A., {Cenko}, S.~B., {Bloom}, J.~S., {et~al.} 2009, \aj, 138, 1690

\bibitem[{{Piro} {et~al.}(2002){Piro}, {Frail}, {Gorosabel}, {Garmire},
  {Soffitta}, {Amati}, {Andersen}, {Antonelli}, {Berger}, {Frontera}, {Fynbo},
  {Gandolfi}, {Garcia}, {Hjorth}, {in 't Zand}, {Jensen}, {Masetti},
  {M{\o}ller}, {Pedersen}, {Pian}, \& {Wieringa}}]{Piro2002a}
{Piro}, L., {Frail}, D.~A., {Gorosabel}, J., {et~al.} 2002, \apj, 577, 680

\bibitem[{{Price}(2007)}]{Price2007GCN6371}
{Price}, P.~A. 2007, \gcn, 6371

\bibitem[{{Racusin} {et~al.}(2008){Racusin}, {Barthelmy}, {Baumgartner},
  {Brown}, {Burrows}, {Cummings}, {Evans}, {Hunsberger}, {Kennea}, {Markwardt},
  {Marshall}, {O'Brien}, {Palmer}, {Sakamoto}, {Starling}, {Stroh}, \&
  {Ukwatta}}]{Racusin2008GCN7264}
{Racusin}, J.~L., {Barthelmy}, S.~D., {Baumgartner}, W.~H., {et~al.} 2008,
  \gcn, 7264

\bibitem[{{Rieke} \& {Lebofsky}(1985)}]{Rieke1985}
{Rieke}, G.~H. \& {Lebofsky}, M.~J. 1985, \apj, 288, 618

\bibitem[{{Rol} \& {Page}(2006)}]{Rol2006GCN5406}
{Rol}, E. \& {Page}, K.~L. 2006, \gcn, 5406

\bibitem[{{Rol} {et~al.}(2007){Rol}, {van der Horst}, {Wiersema}, {Patel},
  {Levan}, {Nysewander}, {Kouveliotou}, {Wijers}, {Tanvir}, {Reichart},
  {Fruchter}, {Graham}, {Ovaldsen}, {Jaunsen}, {Jonker}, {van Ham}, {Hjorth},
  {Starling}, {O'Brien}, {Fynbo}, {Burrows}, \& {Strom}}]{Rol2007}
{Rol}, E., {van der Horst}, A., {Wiersema}, K., {et~al.} 2007, \apj, 669, 1098

\bibitem[{{Rol} {et~al.}(2005){Rol}, {Wijers}, {Kouveliotou}, {Kaper}, \&
  {Kaneko}}]{Rol2005a}
{Rol}, E., {Wijers}, R.~A.~M.~J., {Kouveliotou}, C., {Kaper}, L., \& {Kaneko},
  Y. 2005, ApJ, 624, 868

\bibitem[{{Roming} {et~al.}(2005){Roming}, {Kennedy}, {Mason}, {Nousek}, {Ahr},
  {Bingham}, {Broos}, {Carter}, {Hancock}, {Huckle}, {Hunsberger}, {Kawakami},
  {Killough}, {Koch}, {McLelland}, {Smith}, {Smith}, {Soto}, {Boyd},
  {Breeveld}, {Holland}, {Ivanushkina}, {Pryzby}, {Still}, \&
  {Stock}}]{Roming2005a}
{Roming}, P.~W.~A., {Kennedy}, T.~E., {Mason}, K.~O., {et~al.} 2005, \ssr, 120,
  95

\bibitem[{{Rossi} {et~al.}(2008{\natexlab{a}}){Rossi}, {de Ugarte Postigo},
  {Ferrero}, {Kann}, {Klose}, {Schulze}, {Greiner}, {Schady}, {Filgas},
  {Gonsalves}, {K{\"u}pc{\"u} Yolda{\c s}}, {Kr{\"u}hler}, {Szokoly}, {Yolda{\c
  s}}, {Afonso}, {Clemens}, {Bloom}, {Perley}, {Fynbo}, {Castro-Tirado},
  {Gorosabel}, {Kub{\'a}nek}, {Updike}, {Hartmann}, {Giuliani}, {Holland},
  {Hanlon}, {Bremer}, {French}, {Melady}, \&
  {Garc{\'{\i}}a-Hern{\'a}ndez}}]{Rossi2008AA491}
{Rossi}, A., {de Ugarte Postigo}, A., {Ferrero}, P., {et~al.}
  2008{\natexlab{a}}, \aap, 491, L29

\bibitem[{{Rossi} {et~al.}(2008{\natexlab{b}}){Rossi}, {Greiner},
  {K{\"u}pc{\"u} Yolda\c{s} }, \& {Yolda\c{s} }}]{Rossi2008GCN7319}
{Rossi}, A., {Greiner}, J., {K{\"u}pc{\"u} Yolda\c{s} }, A., \& {Yolda\c{s} },
  A. 2008{\natexlab{b}}, \gcn, 7319

\bibitem[{{Rossi} {et~al.}(2008{\natexlab{c}}){Rossi}, {Kruehler}, {Greiner},
  {Yolda\c{s} }, {Clemens}, {Filgas}, {Yolda\c{s} }, \&
  {Szokoly}}]{Rossi2008GCN8268}
{Rossi}, A., {Kruehler}, T., {Greiner}, J., {et~al.} 2008{\natexlab{c}}, \gcn,
  8268

\bibitem[{{Rossi} {et~al.}(2011){Rossi}, {Schulze}, {Klose}, {Kann}, {Rau},
  {Krimm}, {J{\'o}hannesson}, {Panaitescu}, {Yuan}, {Ferrero}, {Kr{\"u}hler},
  {Greiner}, {Schady}, {Pandey}, {Amati}, {Afonso}, {Akerlof}, {Arnold},
  {Clemens}, {Filgas}, {Hartmann}, {K{\"u}pc{\"u} Yolda{\c s}}, {McBreen},
  {McKay}, {Nicuesa Guelbenzu}, {Olivares}, {Paciesas}, {Rykoff}, {Szokoly},
  {Updike}, \& {Yolda{\c s}}}]{Rossi2011a}
{Rossi}, A., {Schulze}, S., {Klose}, S., {et~al.} 2011, \aap, 529, A142

\bibitem[{{Rujopakarn} {et~al.}(2006){Rujopakarn}, {Rykoff}, {Schaefer},
  {Yuan}, \& {Yost}}]{Rujopakarn2006GCN4737}
{Rujopakarn}, W., {Rykoff}, E.~S., {Schaefer}, B.~E., {Yuan}, F., \& {Yost},
  S.~A. 2006, \gcn, 4737

\bibitem[{{Rujopakarn} {et~al.}(2008){Rujopakarn}, {Swan}, \&
  {Guver}}]{Rujopakarn2008GCN8228}
{Rujopakarn}, W., {Swan}, H., \& {Guver}, T. 2008, \gcn, 8228

\bibitem[{{Rykoff} {et~al.}(2007){Rykoff}, {Schaefer}, \&
  {Swan}}]{Rykoff2007GCN6356}
{Rykoff}, E.~S., {Schaefer}, B.~E., \& {Swan}, H. 2007, \gcn, 6356

\bibitem[{{Salvaterra} {et~al.}(2009){Salvaterra}, {Della Valle}, {Campana},
  {Chincarini}, {Covino}, {D'Avanzo}, {Fern{\'a}ndez-Soto}, {Guidorzi},
  {Mannucci}, {Margutti}, {Th{\"o}ne}, {Antonelli}, {Barthelmy}, {de Pasquale},
  {D'Elia}, {Fiore}, {Fugazza}, {Hunt}, {Maiorano}, {Marinoni}, {Marshall},
  {Molinari}, {Nousek}, {Pian}, {Racusin}, {Stella}, {Amati}, {Andreuzzi},
  {Cusumano}, {Fenimore}, {Ferrero}, {Giommi}, {Guetta}, {Holland}, {Hurley},
  {Israel}, {Mao}, {Markwardt}, {Masetti}, {Pagani}, {Palazzi}, {Palmer},
  {Piranomonte}, {Tagliaferri}, \& {Testa}}]{Salvaterra2009}
{Salvaterra}, R., {Della Valle}, M., {Campana}, S., {et~al.} 2009, \nat, 461,
  1258

\bibitem[{{Sari} {et~al.}(1998){Sari}, {Piran}, \& {Narayan}}]{Sari1998a}
{Sari}, R., {Piran}, T., \& {Narayan}, R. 1998, ApJ, 497, L17

\bibitem[{{Sato} {et~al.}(2006{\natexlab{a}}){Sato}, {Barbier}, {Barthelmy},
  {Cummings}, {Fenimore}, {Gehrels}, {Hullinger}, {Hurkett}, {Krimm},
  {Markwardt}, {Palmer}, {Parsons}, {Sakamoto}, \& {Tueller}}]{Sato2006GCN4751}
{Sato}, G., {Barbier}, L., {Barthelmy}, S., {et~al.} 2006{\natexlab{a}}, \gcn,
  4751

\bibitem[{{Sato} {et~al.}(2006{\natexlab{b}}){Sato}, {Barbier}, {Barthelmy},
  {Cummings}, {Fenimore}, {Gehrels}, {Guidorzi}, {Hullinger}, {Krimm},
  {Markwardt}, {Palmer}, {Parsons}, {Sakamoto}, {Stamatikos}, \&
  {Tueller}}]{Sato2006GCN5578}
{Sato}, G., {Barbier}, L., {Barthelmy}, S.~D., {et~al.} 2006{\natexlab{b}},
  \gcn, 5578

\bibitem[{{Savaglio} {et~al.}(2009){Savaglio}, {Glazebrook}, \& {Le
  Borgne}}]{Savaglio2009a}
{Savaglio}, S., {Glazebrook}, K., \& {Le Borgne}, D. 2009, \apj, 691, 182
  (SBG09)

\bibitem[{{Schady} {et~al.}(2008{\natexlab{a}}){Schady}, {Barthelmy},
  {Baumgartner}, {Beardmore}, {Burrows}, {Cummings}, {Evans}, {Holland},
  {Kennea}, {Krimm}, {Marshall}, {Pagani}, {Page}, {Palmer}, {Sakamoto},
  {Sato}, {Stamatikos}, {Starling}, {vanden Berk}, \&
  {Ward}}]{Schady2008GCN7314}
{Schady}, P., {Barthelmy}, S.~D., {Baumgartner}, W.~H., {et~al.}
  2008{\natexlab{a}}, \gcn, 7314

\bibitem[{{Schady} \& {Cannizzo}(2007)}]{Schady2007GCN6364}
{Schady}, P. \& {Cannizzo}, J. 2007, \gcn, 6364

\bibitem[{{Schady} {et~al.}(2008{\natexlab{b}}){Schady}, {Evans}, \&
  {Krimm}}]{Schady2008GCNR117}
{Schady}, P., {Evans}, P.~A., \& {Krimm}, H. 2008{\natexlab{b}}, GCN Report,
  117

\bibitem[{{Schaefer} {et~al.}(1998){Schaefer}, {Cline}, {Hurley}, \&
  {Laros}}]{Schaefer1998}
{Schaefer}, B.~E., {Cline}, T.~L., {Hurley}, K.~C., \& {Laros}, J.~G. 1998,
  \apjs, 118, 353

\bibitem[{{Schaefer} {et~al.}(2005){Schaefer}, {Quimby}, {Yost}, \&
  {Rujopakarn}}]{Schaefer05_GCN4010}
{Schaefer}, B.~E., {Quimby}, R., {Yost}, S.~A., \& {Rujopakarn}, W. 2005, \gcn,
  4010

\bibitem[{{Schlegel} {et~al.}(1998){Schlegel}, {Finkbeiner}, \&
  {Davis}}]{Schlegel1998}
{Schlegel}, D.~J., {Finkbeiner}, D.~P., \& {Davis}, M. 1998, \apj, 500, 525

\bibitem[{{Sharapov} {et~al.}(2006){Sharapov}, {Ibrahimov}, {Pozanenko}, \&
  {Rumyantsev}}]{Sharapov2006GCN4927}
{Sharapov}, D., {Ibrahimov}, M., {Pozanenko}, A., \& {Rumyantsev}, V. 2006,
  \gcn, 4927

\bibitem[{{Sonoda} {et~al.}(2005){Sonoda}, {Maeno}, {Tokunaga}, \&
  {Yamauchi}}]{Sonoda05_GCN4009}
{Sonoda}, E., {Maeno}, S., {Tokunaga}, Y., \& {Yamauchi}, M. 2005, \gcn, 4009

\bibitem[{{Spergel} {et~al.}(2003){Spergel}, {Verde}, {Peiris}, {Komatsu},
  {Nolta}, {Bennett}, {Halpern}, {Hinshaw}, {Jarosik}, {Kogut}, {Limon},
  {Meyer}, {Page}, {Tucker}, {Weiland}, {Wollack}, \& {Wright}}]{Spergel2003}
{Spergel}, D.~N., {Verde}, L., {Peiris}, H.~V., {et~al.} 2003, \apjs, 148, 175

\bibitem[{{Stamatikos} {et~al.}(2006){Stamatikos}, {Barthelmy}, {Burrows},
  {Capalbi}, {Conciatore}, {Gehrels}, {Guidorzi}, {Holland}, {Kennea},
  {Markwardt}, {Palmer}, {Perri}, {Sakamoto}, {vanden Berk}, \&
  {Ziaeepour}}]{Stamatikos2006GCN5590}
{Stamatikos}, M., {Barthelmy}, S.~D., {Burrows}, D.~N., {et~al.} 2006, \gcn,
  5590

\bibitem[{{Stamatikos} {et~al.}(2008){Stamatikos}, {Barthelmy}, {Cummings},
  {Fenimore}, {Gehrels}, {Krimm}, {Markwardt}, {Palmer}, {Racusin}, {Sakamoto},
  {Sato}, {Tueller}, \& {Ukwatta}}]{Stamatikos08_GCN7277}
{Stamatikos}, M., {Barthelmy}, S.~D., {Cummings}, J., {et~al.} 2008, \gcn, 7277

\bibitem[{{Starling} {et~al.}(2006){Starling}, {Page}, \&
  {Holland}}]{Starling2006GCN5783}
{Starling}, R.~L.~C., {Page}, K.~L., \& {Holland}, S.~T. 2006, \gcn, 5783

\bibitem[{{Svensson} {et~al.}(2010){Svensson}, {Levan}, {Tanvir}, {Fruchter},
  \& {Strolger}}]{Svensson2010a}
{Svensson}, K.~M., {Levan}, A.~J., {Tanvir}, N.~R., {Fruchter}, A.~S., \&
  {Strolger}, L. 2010, \mnras, 405, 57

\bibitem[{{Svensson} {et~al.}(2011){Svensson}, {Tanvir}, {Perley},
  {Michalowski}, {Page}, {Bloom}, {Cenko}, {Hjorth}, {Jakobsson}, {Watson}, \&
  {Wheatley}}]{Svensson2011}
{Svensson}, K.~M., {Tanvir}, N.~R., {Perley}, D.~A., {et~al.} 2011, \mnras, in
  press (arXiv:1109.3167)

\bibitem[{{Tagliaferri} {et~al.}(2005){Tagliaferri}, {Antonelli}, {Chincarini},
  {Fern{\'a}ndez-Soto}, {Malesani}, {Della Valle}, {D'Avanzo}, {Grazian},
  {Testa}, {Campana}, {Covino}, {Fiore}, {Stella}, {Castro-Tirado},
  {Gorosabel}, {Burrows}, {Capalbi}, {Cusumano}, {Conciatore}, {D'Elia},
  {Filliatre}, {Fugazza}, {Gehrels}, {Goldoni}, {Guetta}, {Guziy}, {Held},
  {Hurley}, {Israel}, {Jel{\'{\i}}nek}, {Lazzati}, {L{\'o}pez-Echarri},
  {Melandri}, {Mirabel}, {Moles}, {Moretti}, {Mason}, {Nousek}, {Osborne},
  {Pellizza}, {Perna}, {Piranomonte}, {Piro}, {de Ugarte Postigo}, \&
  {Romano}}]{Tagliaferri2005}
{Tagliaferri}, G., {Antonelli}, L.~A., {Chincarini}, G., {et~al.} 2005, \aap,
  443, L1

\bibitem[{{Tanvir} {et~al.}(2004){Tanvir}, {Barnard}, {Blain}, {Fruchter},
  {Kouveliotou}, {Natarajan}, {Ramirez-Ruiz}, {Rol}, {Smith}, {Tilanus}, \&
  {Wijers}}]{Tanvir2004a}
{Tanvir}, N.~R., {Barnard}, V.~E., {Blain}, A.~W., {et~al.} 2004, \mnras, 352,
  1073

\bibitem[{{Tanvir} {et~al.}(2009){Tanvir}, {Fox}, {Levan}, {Berger},
  {Wiersema}, {Fynbo}, {Cucchiara}, {Kr{\"u}hler}, {Gehrels}, {Bloom},
  {Greiner}, {Evans}, {Rol}, {Olivares}, {Hjorth}, {Jakobsson}, {Farihi},
  {Willingale}, {Starling}, {Cenko}, {Perley}, {Maund}, {Duke}, {Wijers},
  {Adamson}, {Allan}, {Bremer}, {Burrows}, {Castro-Tirado}, {Cavanagh}, {de
  Ugarte Postigo}, {Dopita}, {Fatkhullin}, {Fruchter}, {Foley}, {Gorosabel},
  {Kennea}, {Kerr}, {Klose}, {Krimm}, {Komarova}, {Kulkarni}, {Moskvitin},
  {Mundell}, {Naylor}, {Page}, {Penprase}, {Perri}, {Podsiadlowski}, {Roth},
  {Rutledge}, {Sakamoto}, {Schady}, {Schmidt}, {Soderberg}, {Sollerman},
  {Stephens}, {Stratta}, {Ukwatta}, {Watson}, {Westra}, {Wold}, \&
  {Wolf}}]{Tanvir2009}
{Tanvir}, N.~R., {Fox}, D.~B., {Levan}, A.~J., {et~al.} 2009, \nat, 461, 1254

\bibitem[{{Tody}(1993)}]{Tody1993}
{Tody}, D. 1993, in Astronomical Society of the Pacific Conference Series,
  Vol.~52, Astronomical Data Analysis Software and Systems II, ed.
  {R.~J.~Hanisch, R.~J.~V.~Brissenden, \& J.~Barnes}, 173

\bibitem[{{Torii}(2005)}]{Torii05_GCN4024}
{Torii}, K. 2005, \gcn, 4024

\bibitem[{{Tueller} {et~al.}(2006){Tueller}, {Barbier}, {Barthelmy},
  {Cummings}, {Fenimore}, {Gehrels}, {Holland}, {Hullinger}, {Krimm},
  {Markwardt}, {Palmer}, {Parsons}, {Sakamoto}, {Sato}, \&
  {Stamatikos}}]{Tueller2006GCN5777}
{Tueller}, J., {Barbier}, L., {Barthelmy}, S.~D., {et~al.} 2006, \gcn, 5777

\bibitem[{{Ukwatta} {et~al.}(2008){Ukwatta}, {Barthelmy}, {Baumgartner},
  {Cummings}, {Fenimore}, {Gehrels}, {Krimm}, {Markwardt}, {McLean}, {Oates},
  {Palmer}, {Parsons}, {Sakamoto}, {Sato}, {Stamatikos}, \&
  {Tueller}}]{Ukwatta2008GCN8230}
{Ukwatta}, T., {Barthelmy}, S.~D., {Baumgartner}, W., {et~al.} 2008, \gcn, 8230

\bibitem[{{Updike} {et~al.}(2008{\natexlab{a}}){Updike}, {Clemens}, \&
  {Greiner}}]{Updike2008}
{Updike}, A., {Clemens}, C., \& {Greiner}, J. 2008{\natexlab{a}}, \gcn, 8627

\bibitem[{{Updike} {et~al.}(2008{\natexlab{b}}){Updike}, {Milne}, {Williams},
  \& {Hartmann}}]{Updike08_GCN7273}
{Updike}, A.~C., {Milne}, P.~A., {Williams}, G.~G., \& {Hartmann}, D.~H.
  2008{\natexlab{b}}, \gcn, 7273

\bibitem[{{Urata} {et~al.}(2006){Urata}, {Kuwahara}, {Tashiro}, {Abe}, {Onda},
  {Kodaka}, {Masuno}, {Usui}, \& {Tamagawa}}]{Urata2006GCN5204}
{Urata}, Y., {Kuwahara}, M., {Tashiro}, M., {et~al.} 2006, \gcn, 5204

\bibitem[{{van der Horst} {et~al.}(2009){van der Horst}, {Kouveliotou},
  {Gehrels}, {Rol}, {Wijers}, {Cannizzo}, {Racusin}, \&
  {Burrows}}]{VanderHorst2009a}
{van der Horst}, A.~J., {Kouveliotou}, C., {Gehrels}, N., {et~al.} 2009, \apj,
  699, 1087 (V09)

\bibitem[{{Vergani} {et~al.}(2007{\natexlab{a}}){Vergani}, {Barthelmy},
  {Beardmore}, {Burrows}, {Campana}, {Chester}, {Cusumano}, {Evans}, {Gehrels},
  {Kennea}, {Krimm}, {Landsman}, {Mangano}, {Moretti}, {O'Brien}, {Osborne},
  {Page}, {Palmer}, {Romano}, {Sato}, {Sbarufatti}, {Stamatikos}, {Starling},
  {Tagliaferri}, {Troja}, \& {Ziaeepour}}]{Vergani07_grb070517}
{Vergani}, S.~D., {Barthelmy}, S.~D., {Beardmore}, A.~P., {et~al.}
  2007{\natexlab{a}}, \gcn, 6411

\bibitem[{{Vergani} {et~al.}(2007{\natexlab{b}}){Vergani}, {Romano},
  {Guidorzi}, {Moretti}, {Krimm}, {Chester}, {Landsman}, {Barthelmy},
  {Burrows}, {Roming}, \& {Gehrels}}]{Vergani07_GCNR56.2}
{Vergani}, S.~D., {Romano}, P., {Guidorzi}, C., {et~al.} 2007{\natexlab{b}},
  GCN Report, 56.2

\bibitem[{{Vrba} {et~al.}(1995){Vrba}, {Hartmann}, \& {Jennings}}]{Vrba1995}
{Vrba}, F.~J., {Hartmann}, D.~H., \& {Jennings}, M.~C. 1995, \apj, 446, 115

\bibitem[{{Vrba} {et~al.}(1999){Vrba}, {Luginbuhl}, {Jennings}, \&
  {Hartmann}}]{Vrba1999}
{Vrba}, F.~J., {Luginbuhl}, C.~B., {Jennings}, M.~C., \& {Hartmann}, D.~H.
  1999, \apj, 511, 298

\bibitem[{{Vreeswijk} {et~al.}(2008){Vreeswijk}, {Fynbo}, {Milvang-Jensen},
  {Malesani}, {Hjorth}, {Jakobsson}, {Jaunsen}, \&
  {Tanvir}}]{Vreeswijk2008GCN7327}
{Vreeswijk}, P., {Fynbo}, J., {Milvang-Jensen}, B., {et~al.} 2008, \gcn, 7327

\bibitem[{{Wainwright} {et~al.}(2007){Wainwright}, {Berger}, \&
  {Penprase}}]{Wainwright2007}
{Wainwright}, C., {Berger}, E., \& {Penprase}, B.~E. 2007, \apj, 657, 367

\bibitem[{{Yolda{\c s}} {et~al.}(2008){Yolda{\c s}}, {Kr{\"u}hler}, {Greiner},
  {Yolda{\c s}}, {Clemens}, {Szokoly}, {Primak}, \&
  {Klose}}]{KupcuYoldas2008AIPC}
{Yolda{\c s}}, A.~K., {Kr{\"u}hler}, T., {Greiner}, J., {et~al.} 2008, in
  American Institute of Physics Conference Series, Vol. 1000, American
  Institute of Physics Conference Series, ed. {M.~Galassi, D.~Palmer, \&
  E.~Fenimore}, 227--231

\bibitem[{{Ziaeepour} {et~al.}(2006){Ziaeepour}, {Barthelmy}, {Cummings},
  {Gehrels}, {Guidorzi}, {Kennea}, {Krimm}, {Markwardt}, {Marshall}, {McLean},
  {Page}, {Page}, {Palmer}, \& {Sakamoto}}]{Ziaeepour2006GCN5398}
{Ziaeepour}, H.~Z., {Barthelmy}, S.~D., {Cummings}, J.~R., {et~al.} 2006, \gcn,
  5398

\end{thebibliography}


\Online

\begin{appendix} 

\section{Additional notes on individual targets: observations
by Swift and other facilities}\label{sec:addnotes}

\subsection{\object{GRB 050717}}

GRB 050717 triggered \swift/BAT at 10:30:52 UT \citep{Hurkett2005GCN3633}. It
was a long burst with a duration of $T_{90}(15-350$ keV) = $(86\pm 2)$ s
\citep{Cummings2005GCN3637} that was also  detected  by \textit{Konus-WIND}
\citep{Golenetskii2005GCN3640}. \swift/XRT began observing 79 s after the
trigger and found a bright, fading X-ray source, while simultaneous 
\swift Ultra-Violet/Optical Telescope (UVOT, \citealt{Roming2005a}) observations started 78 s after the trigger and  resulted only in
upper limits \citep{Hurkett2005GCN3633,Blustin2005GCN3638}.  Unfortunately,
XRT was unable to automatically  centre on the burst, leading to a delay of
2.5 h in the determination of the X-ray position (error circle radius
6\farcs0; \citealt{Kennea2005GCN3634};  see also
\citealt{Hurkett2005GCN3636}). The burst is discussed by \cite{Krimm2006a} in
detail; it was very luminous, has one of the highest-ever measured peak
energies, and a probable redshift $z>2.7$.  Deep ground-based $K$-band
follow-up observations were performed with the du Pont 100-inch telescope at
Las Campanas Observatory with a first run starting 37.7 h after the burst. No
fading NIR source was detected  \citep{Berger2005GCN3639,Berger2005GCN3643}.
Optical observations with the Tenagra 0.35-m telescope at Perth, Australia,
did not find a new source down to the limit of the DSS2 red survey
\citep{Luckas2005GCN3642}. In addition, PROMPT-5 at Cerro Tololo Inter-American
Observatory in Chile automatically observed the field starting 13 h after the
burst. No fading source was found down to $R_C=21.7$ and $I_C=21.5$
\citep{MacLeod2005GCN3652}.

\swift/UVOT obtained an upper limit of $v>19.0$ for any  afterglow at 420~s (mid-time) after the onset of the burst \citep{Blustin2005GCN3638},
corresponding to $v>18.3$ after correction for Galactic extinction. Using the
observed constraint on the spectral slope $\beta_{\rm OX}<0.40$ at the time of
the UVOT observations, this corresponds to an upper limit of $R_{\rm
  AB}>18.2$. In the same way, following \cite{Rol2005a}, at the time of the
optical observation the observed  (mean) X-ray flux together with the observed
(mean) spectral slope $\beta_{\rm X}$  predicts  a non-extinguished $R_{\rm
  AB}$-band magnitude of between $14.5\pm1.5$ and $18.2\pm1.5$, where the
brighter magnitude corresponds to $\nu_c = \nu_{\rm opt}$ and the fainter
magnitude to $\nu_c = \nu_{\rm X}$  ($\nu_c$ is the cooling frequency;
\citealt{Sari1998a}).   According to the criterion of J04, which uses
$\beta_{\rm OX}$, the burst is dark, while according to the criterion of V09,
which uses $\beta_{\rm OX}$ and $\beta_{\rm X}$,  the burst is not dark
(Table~\ref{tab:box}). 

\subsection{\object{GRB 050922B}}

\swift/BAT detected the burst at 15:02:00 UT.  It was an image trigger lasting
for 168 s \citep{Norris05_GCN4008}. \citet{Cummings05_GCN4033} measured
$T_{90}(15-150$ keV) = $(250\pm 20)$ s. Because of the image trigger history,
\citet{Norris05_GCN4008} speculated that it could be a high-redshift event
similar to GRB 050904. \swift/XRT started observing 342 s after the trigger,
and UVOT one second later \citep{Norris05_GCN4008}. A decaying X-ray afterglow was
detected \citep{Godet05_GCN4031}  but no optical counterpart
\citep{Pasquale05_GCN4028}. Several ground-based small telescopes responded to
the trigger but also failed to find any afterglow candidate, namely ROTSE IIIa (upper limit $CR=17.3$ at 3 min; \citealt{Schaefer05_GCN4010}), the 14-inch Automated Response Telescope at the University of Osaka, Japan (upper limit $CR=15.1$ at 3 min; \citealt{Torii05_GCN4024}), the 0.4-m telescope of Ussuriysk
Astrophysical Observatory, Russia (upper limit $CR=16.0$ at 15 min;
\citealt{Kornienko05_GCN4047}), and the 30-cm telescope at University of
Miyazaki, Japan (upper limit $CR=16.1$ at 21 min;
\citealt{Sonoda05_GCN4009}). 

The INT 2.5-m telescope at Observatorio del Roque de los Muchachos on La Palma
obtained  an upper limit to the afterglow of $r'>22.5$ at 49~ks  (mid-time)
after the onset of the burst \citep{GuziyGCN4025}, corresponding to $r'>22.4$
after correction for Galactic extinction. There are no X-ray data for the time
between about $t=10$ ks and 100 ks after the burst, but there are  for
observations from $t\sim100$ ks to about 1 Ms. The latter data can be used to
extrapolate the X-ray flux to $t=49$ ks. The spectral slope is then
$\beta_{\rm OX} <0.39$, corresponding to an upper limit of about $R_{\rm
  AB}>22.3$. Similarly, the observed X-ray flux together with the observed
spectral slope $\beta_{\rm X}$ at $t=49$ ks predicts a non-extinguished
$R_{\rm AB}$-band magnitude between $11.1\pm2.9$ and $14.8\pm3.0$. Using
$\beta_{\rm OX}$ and $\beta_{\rm X}$, the burst is dark according to the
criterion of J04 as well as V09 (Table~\ref{tab:box}).

\subsection{\object{GRB 060211A}}

\swift/BAT was triggered by GRB 060211A at 09:39:11 UT
\citep{Hurkett2006GCN4736}. It was a long burst with a duration of $T_{90}$
(15-350 keV)$=126\pm5$ s \citep{Sato2006GCN4751, Krimm2006GCN4757}.  The
spacecraft slewed promptly to the BAT position and \swift/XRT found a bright,
fading X-ray source, while \swift/UVOT started observing 183 s after the
trigger but did not detect any afterglow candidate \citep{Hurkett2006GCN4736}.
ROTSE IIIa, located at Siding Spring Observatory, Australia, and the Moscow
Union 'Optic' MASTER robotic system responded to GRB 060211
immediately. ROTSE's automated response took the first image  147~s after the
burst, under twilight conditions, while MASTER started 202~s after the GRB
trigger. Only upper limits could be reported
(\citealt{Rujopakarn2006GCN4737,Lipunov2006GCN4741}; see also
\citealt{Urata2006GCN5204}). In addition, the 2-m Faulkes Telescope North robotically
followed-up GRB 060211 starting 5.4 min after the trigger. No fading optical
counterpart down to $R\approx18.5$ was found \citep{Gomboc2006GCN4738}. Deep
upper limits were also reported by \cite{Norris2006GCN4760}, $J>19.1$ at 17 h
after the burst, and \cite{Sharapov2006GCN4927}, $R>22$ at 5.5 h after the
burst.

The 1.5-m telescope of Maidanak Astronomical Observatory obtained  for the
afterglow an upper limit of $R=22.0$ at $\sim20$~ks (mid-time) after the onset
of the burst \citep{Sharapov2006GCN4927}, corresponding to $R=21.6$ after
correction for Galactic extinction. This corresponds to an upper limit of
$R_{\rm AB}>21.8$. Among all available upper limits, this
observation provides the tightest constraints on the spectral properties of
the afterglow from the optical to the X-ray band.  According to these data,
however, GRB 060211A cannot be classified as a dark burst
(Table~\ref{tab:box}).

\subsection{\object{GRB 060805A}}

The burst triggered \swift/BAT on May 8, 2006 at 04:47:49 UT
\citep{Ziaeepour2006GCN5398}. It had a duration of $T_{90}(15-350$ keV) $=5.4
\pm 0.5$ s \citep{Barbier2006GCN5403}. \swift/XRT began taking data 93 seconds
after the BAT trigger. A ground analysis revealed a faint, uncatalogued X-ray
source. \swift/UVOT started observing 97 s after the trigger  but no
afterglow candidate was detected in any band \citep{Ziaeepour2006GCN5398,
  Pandey2006GCN5402}. Additional ground-based observations could only provide
upper limits. The robotic 0.76-m Katzman Automatic Imaging Telescope (KAIT) at
Lick  Observatory started observing the field 119 s after the BAT trigger but
no afterglow was found ($V>16.8, I>16.7$; \citealt{Li2006GCN5400}).  The
automated Palomar 60-inch telescope responded to GRB 060805A and started
observing 3 min after the burst trigger. No source down to $R>19$ was found in
the XRT error circle (\citealt{Cenko2006GCN5401}).  Additional upper limits
were obtained by  the 1.3-m Skinakas Observatory (University of Crete,
Heraklion, Greece) of $R>21.5$ at 14 h after the burst
(\citealt{Muehlegger2006GCN5405}) and by the 2-m Liverpool Telescope on La
Palma of $r^\prime>22.9$ and $i^\prime>22.6$ at 0.725 days and 0.748 days,
respectively, after the burst (\citealt{Rol2006GCN5406}).

The 2-m Liverpool Telescope observations  correspond to $r^\prime>22.7$, after
correction for Galactic extinction.  Using $\beta_{\rm OX} <1.00$, this
corresponds to an upper limit of $R_{\rm AB}>22.7$. Among all available upper
limits for this burst, this observation provides the tightest constraint on
$\beta_{\rm OX}$ and $\beta_{\rm X}$ (Table~\ref{tab:box}).  However, these
constraints do not qualify GRB 060805A as a dark burst, especially because the
X-ray afterglow itself was very subluminous.

\subsection{\object{GRB 060919}}

GRB 060919 triggered \swift/BAT at 07:48:38 UT  \citep{Guidorzi2006GCN5575}.
It was a long burst with a duration of $T_{90}=(15-350$ keV)$=9.1\pm 0.2$ s
\citep{Sato2006GCN5578}. \swift/XRT began taking data 87 s  after the BAT
trigger.  Ground analysis revealed a faint X-ray source with an revised error
circle of $r=4\farcs1$ \citep{Guidorzi2006GCN5575,
Guidorzi2006GCN5577}. \swift/UVOT started observing the field 73 s 
after the burst but did not detect an optical counterpart in any band down to
deep flux limits \citep{Breeveld2006GCN5580}.  The robotic TAROT telescope on
La Silla started observing 28 s after the trigger. No optical transient was
found down to $R>15.4$ in the first 60 s  of observations.  An upper
limit of $R>15.8$ could be set for any transient up to 382 s after the trigger
\citep{Klotz2006GCN5576}. The Faulkes Telescope South started observing about
2.8 h after the event. No optical transient was detected down to a
limiting magnitude of $R>19.5$ \citep{Melandri2006GCN5579}.

The UVOT upper limit at 918~s corresponds to $v>20.0$, after correction for
Galactic extinction.  Using $\beta_{\rm OX} <0.68$, this corresponds to an
upper limit of $R_{\rm AB}>19.8$. As for to the previous two bursts,  this
observation provides the tightest constraints achieved to date 
on the spectral properties of the afterglow. 
On the basis of these data, GRB 060919 is not a dark burst
(Table~\ref{tab:box}).

\subsection{\object{GRB 060923B}}

\swift/BAT triggered on GRB 060923B at 11:38:06 UT
\citep{Stamatikos2006GCN5590}. It was a single-peaked burst with a duration
of  $T_{90} (15-350$ keV)$=8.8 \pm 0.1$ s \citep{Barbier2006GCN5595}. 
\swift/XRT began observing the field 114 s after the BAT trigger 
and found an uncatalogued X-ray source with a positional accuracy of  
$2\farcs8$. \swift/UVOT started observing 122 s after the burst 
with the $white$ filter but could not detect an afterglow candidate 
\citep{Stamatikos2006GCN5590,  Holland2006GCN5603}. No further ground-based 
follow-up observations were reported in the literature. 

\swift/UVOT  obtained an afterglow upper limit of $v>18.1$ at 295~s (mid-time) after
the burst \citep{Holland2006GCN5603}, corresponding to $v>18.0$ after
correction for Galactic extinction.  Using $\beta_{\rm OX} <0.62$, this
corresponds to an upper limit of $R_{\rm AB}>17.9$. Among all available upper
limits for this burst this observation provides the tightest constraint on
$\beta_{\rm OX}$  and  $\beta_{\rm X}$ (Table~\ref{tab:box}). These
constraints do not classify GRB 060923B as a dark event.

\subsection{\object{GRB 061102}}

GRB 061102 triggered \swift/BAT at 01:00:31 UT \citep{Holland2006GCN5776}.  It
was a long burst with a duration of $T_{90}(15-350$ keV)$=17.6 \pm1$~s
\citep{Tueller2006GCN5777}. \swift/XRT began observing the field 100 s
after the BAT trigger and found an uncatalogued, fading X-ray source
\citep{Holland2006GCN5776,Starling2006GCN5783}. \swift/UVOT started observing 
 110 s after the trigger with the $white$ filter but no afterglow
candidate was seen down to a $3~\sigma$ upper limit of $white<18.5$
\citep{Holland2006GCN5776}.  Continued observations provided only upper limits
in all UVOT bands \citep{Holland2006GCN5784}.  No further ground-based
follow-up observations of this event are reported in the literature.

\swift/UVOT obtained an even deeper upper limit of $v>20.5$ at 1480~s (mid-time)
after the onset of the burst \citep{Holland2006GCN5784}, corresponding to
$v>20.4$ after correction for Galactic extinction. Using the observed
$\beta_{\rm OX} <1.10$, this corresponds to an upper limit of $R_{\rm
  AB}>20.1$. Among all available upper limits, this observation provides the
tightest constraints on the afterglow SED, which do not  classify GRB 061102
as dark (Table~\ref{tab:box}).

\subsection{\object{GRB 070429A} \label{sub:070429Aappendix}}

The burst 070429A triggered \swift/BAT at 01:35:10 UT
\citep{Barthelmy2007GCN6355}. It was a long burst with  $T_{90}(15-350$
keV)$=163 \pm 5$ s \citep{Cannizzo2007GCN6362}. \swift/XRT started observing
153 s after the trigger and found a fading, uncatalogued X-ray source, while
\swift/UVOT started observing 211 s after the trigger but did not detect
an optical counterpart in any band \citep{Schady2007GCN6364}. 
The ROTSE-IIIc telescope,
located at Mt. Gamsberg, Namibia, started observing 97 s after the burst. No
afterglow candidate was found down to $CR>17.3$ (unfiltered images) for images
taken within 3 min after the trigger and down to $CR>18.0$ within 8 min
\citep{Rykoff2007GCN6356}. Additional data  were obtained with the 0.6-m
BOOTES-IR/T60 robotic telescope \citep{Castro-Tirado2006a}, starting 3.25 hr
after the burst but no afterglow was found
(\citealt{Postigo2007GCN6361}). Deep $K$-band observations with the 4.2-m
William Herschel Telescope on La Palma, beginning 4.1 h after the burst,
detected a faint source in the XRT error circle, but no fading behavior was
found (\citealt{Postigo2007GCN6361}).  Within its astrometric errors,
this source corresponds to object C detected in our observations
(see Table~\ref{tab:PhotomVLT}).

The Gemini North telescope mounted with the GMOS camera observed the field in
$i^\prime$ and $z^\prime$ 44~ks (mid-time) after the burst. No afterglow
candidate was found  \citep{Price2007GCN6371}. Unfortunately, no magnitude
limits were reported. Therefore, we used a conservative upper limit of
$R>24.0$ based on the original Gemini data available in the Gemini
archive (\textit{http://cadcwww.dao.nrc.ca/gsa/}). This corresponds to an upper
limit of $R_{\rm AB}>23.8$. Together with the  measured X-ray flux at the same
time, this leads to $\beta_{\rm OX}<0.42$ and $\beta_{\rm X}- \beta_{\rm OX} -
0.5 >0.14$. According to J04 as well as V09, the burst is dark. The observed
X-ray flux predicts a non-extinguished $R_{\rm AB}$-band magnitude between
$17.2\pm1.5$ and $21.0\pm1.5$. Since  the \swift/XRT light curve shows a
constant decay with a constant spectral slope during the time when the optical
upper limit was obtained,  GRB 070429A is one of three
events in our sample that can be securely classified as dark.

\subsection{\object{GRB 070517}}

This burst triggered \swift/BAT at 11:20:58 UT \citep{Vergani07_grb070517}
 and $T_{90}$ was $9\pm1$ s \citep{Vergani07_GCNR56.2}. \swift/XRT clearly detected
an afterglow and could even see evidence of a break in the X-ray light curve.
\swift/UVOT could not observe because of a 4 mag bright  star in the
field of view.  Ground-based optical follow-up was only reported by
\citet{Gilmore07_GCN6412} (UL = DSS2 Infrared at 2.7 h after the burst) and
\citet{Fox6420} using Gemini-South $\sim$16 h after the burst.  The latter
authors suggested that there were 
two afterglow candidates in the XRT error circle but no
further observations of these sources were reported in the literature.
Therefore, we used their faintest detection ($i'>24.5$) as  an upper limit at
57600~s. Using the corresponding $\beta_{\rm OX} <0.56$, this translates into an
upper limit of $R_{\rm AB}>24.3$, which does not classify GRB 070517 as a dark
burst.  However, in our late-time follow-up observations with VLT/FORS1 the
brighter object  reported by \citet{Fox6420} ($r'=22.1$) is no longer visible. 
Thus, we conclude that this was the optical afterglow of GRB 070517.

\subsection{\object{GRB 080207}\label{sub:080207appendix}}

GRB 080207 triggered \swift/BAT at 21:30:21 UT \citep{Racusin2008GCN7264} and
had  a duration of $T_{90}=340\pm20$ s \citep{Stamatikos08_GCN7277}. The XRT
started observing the field 124 s after the BAT trigger and detected a
bright source in WT mode.  After $\sim5000$~s, it continued observing in PC
mode, producing a light curve with a constant decay index. 
\swift/UVOT observations did not detect the
afterglow both in early observations after 140~s in a $white$ finding chart and
later deeper observations ($>1.5$ hours, \citealt{Cucchiara2008GCN7268}).
Several limiting magnitudes based on ground-based observations were
reported: $R>14.3$ at 1607~s (0.45 h), and $R>19.0$ at 5049~s (1.45 hours)
(TAROT, Calern observatory, \citealt{Klotz2008GCN7267}); $J>16.7$,
$H>15.9$, $K>13.9$ at 7.8 h, 7.7 h, and 10.1 h  after the trigger,
respectively (60-cm REM telescope, La Silla, \citealt{D'Avanzo08_GCN7269});
$R>21.8$ at 0.759 h (RTT150 1.5-m telescope, TUBITAK National
Observatory, \citealt{Khamitov08_GCN7270}); $R>20.8$ at 13.7 h (Super-LOTIS,
 Kitt Peak observatory, \citealt{Updike08_GCN7273}); GMOS camera
on the Gemini South telescope did not detect the afterglow down to
$g^\prime r^\prime i^\prime z^\prime $ = 24.1, 24.5, 24.2, 25.0  at 9.8 h
\citep{Cucchiara08_GCN7276}; $R>23.5$ at 9.75 h (MOSCA mounted at NOT, La
Palma; \citealt{Marin08_GCN7291}); $J>23.5$, $H>22.8$, and $K>21.5$ (VLT/SINFONI,
\citealt{Fugazza08_GCN7293}); $R>20.3$ at 1.49 h and $R>21.0$ at 4.94 hr
(Zeiss-600 at Mt.Terskol observatory, \citealt{Andreev08_GCN7333}). 
Based on GROND data we did not detect the afterglow in any band down to deep flux limits
(Table~\ref{tab:darkULs}).
 
The Zeiss-600 telescope upper limit at 1.69 h corresponds to an  upper limit
of $R_{\rm AB}>20.5$. The observed X-ray flux predicts a non-extinguished
$R_{\rm AB}$-band  magnitude between $11.3\pm1.4$ and $15.0\pm1.4$.  According
to the criterion of V09 as well as J04, GRB 080207 is a dark burst
(Table~\ref{tab:box}).  It is one of three
events in our sample that can be securely classified as dark.

\subsection{\object{GRB 080218B}\label{sub:080218Bappendix}}

GRB 080218B  triggered \swift/BAT at 23:57:47 UT and had a duration of
$T_{90}=6.2\pm1.2$ s \citep{Schady2008GCNR117}. \swift \ slewed immediately to
the burst and XRT found a bright, uncatalogued X-ray source that could be
localized with an uncertainty of $r=3\farcs0$. The \swift/UVOT started observing 551
seconds after the BAT trigger using the $white$ filter. No afterglow candidate
was found down to a 3$\sigma$ limiting magnitude of 20.6
\citep{Schady2008GCN7314}.  Several limiting magnitudes based on ground-based
observations were  then reported: $CR>16$, starting 60 s after the trigger
(unfiltered, 0.4-m Watcher telescope, South Africa,
\citealt{French2008GCN7316};  $I>21$ and $J>18.7$ at 3.1 h after the burst
(1.3-m SMARTS telescope equipped with ANDICAM at CTIO,
\citealt{Cobb2008GCN7318});  $H>13.7$ at 2 min and $K>12.6$ at 8 min  after
the trigger (60-cm REM telescope on La Silla, \citealt{Covino2008GCN7322});
$B>22.1$, $V>22.7$, $R>22.9$, and $I>22.6$ at about 1 h and $J>20.6$, 
$H>20.1$, and $K_s>19.4$ at about 3 h after the trigger using VLT/FORS2 and NTT/SOFI
\citep{Vreeswijk2008GCN7327}. Finally, no transient radio source was detected
in the XRT error circle two weeks after the burst (Australia Telescope Compact
Array, ATCA; \citealt{Moin2008GCN8466}).  Most importantly, the GROND imager 
did not detect
the afterglow down to deep limits in all seven bands in spite of a rapid
response time (Table~\ref{tab:darkULs}).
 
Based on GROND data we obtained an afterglow upper limit of $r^\prime>24.7$ at 11520~s
(mid-time) after the onset of the burst \citep{Rossi2008GCN7319},
corresponding to $r^\prime>24.3$ after correction for Galactic extinction.
Using the observed spectral slope  $\beta_{\rm OX} <0.18$, this corresponds to
an upper limit of $R_{\rm AB}>24.3$. The observed X-ray flux predicts a
non-extinguished $R$-band magnitude between $15.1\pm1.5$ and $18.9\pm1.5$. The
burst is dark according to the criterion of J04 as well as V09. 
 It is one of three events in our sample that can 
be securely classified as dark.

\subsection{\object{GRB 080602}}

\swift/BAT triggered the burst at 01:30:28 UT \citep{Beardmore2008GCN7781} and
$T_{90}$  was $74\pm7$ s \citep{Beardmore2008GCNR145}. The burst was
also detected by \textit{Konus-WIND}, observations of this satellite allowing
the peak energy to be constrained to be higher than 226 keV
\citep{Golenetskii08_GCN7784}.   \swift/XRT found a bright, uncatalogued X-ray
source resulting in a 5\farcs8 error circle. Evidence of substantial X-ray
absorption in excess of the Galactic value was found. \swift/UVOT started
observing $123$~s after the trigger but no afterglow candidate was
detected. The XRT error circle was finally reduced to only 1\farcs7  and
1\farcs8, respectively \citep{Beardmore08_GCN7782,Beardmore08_GCN7785}.  The
only optical follow-up observation was reported by \cite{Malesani2008GCN7783}
about 3.4 h after the trigger using the NOT telescope on La Palma.  No
afterglow candidate was found down to $R>22.3$ \citep{Malesani08_GCN7783}.

The  \swift/UVOT imager obtained an upper limit of $v>20.3$ at 504~s (mid-time) after the onset
of the burst \citep{Beardmore2008GCNR145}, corresponding to $v>20.2$ after
correction for Galactic extinction. Using the observed spectral slope of
$\beta_{\rm OX} <0.05$, this corresponds to an upper limit of $R_{\rm
  AB}>20.2$.  Following \cite{Rol2005a}, the observed X-ray flux and spectral
slope predicts a non-extinguished $R_{\rm AB}$-band magnitude of between
$13.0\pm0.7$ and $16.7\pm0.7$.  According to the criterion of J04 as well as
V09 the burst is dark (Table~\ref{tab:box}). However, because the
X-ray light curve is  rather flat instead of decaying during the time when the
optical upper limit  was obtained, the burst 
 cannot be securely classified as dark.
Unfortunately, no X-ray data was taken contemporaneously with the deep NOT
observations.

\subsection{\object{GRB 080727A}}

\swift/BAT triggered on the burst at 05:57:39 UT  with a duration ($T_{90}$)
of $4.9\pm1.0$~s. About 109 s later, \swift/XRT  began observing the
field \citep{Immler08_GCN8021}, unveiling a light  curve with constant decay
and evolving spectral index (see the XRT repository, \citealt{Evans2007a,
  Evans2009a}).  The UVOT imager started observing at 113 seconds, no afterglow being
found \citep{Landsman08_GCN8027}. In addition, UKIRT on Mauna Kea did not detect the
afterglow down to $K>19.8$ at 0.63 h after the trigger
\citep{Levan08_GCN8048}. FORS1 on ESO/Paranal observed the field at 17.5 hr
and did not detect the afterglow down to the very  deep upper limit of $R>26$
(\citealt{Malesani08_GCN8039}).

Using the observed spectral slope $\beta_{\rm OX} <0.85$, the UKIRT upper
limit corresponds to an upper limit of $R_{\rm AB}>22.8$. Following
\cite{Rol2005a}, the observed X-ray flux and spectral slope  at the time when
the optical upper limit was obtained predicts a non-extinguished $R_{\rm
  AB}$-band  magnitude of between $17.9\pm2.2$ and $21.6\pm2.2$.  According to
the criterion of V09, the burst lies at the boundary region between dark and
non-dark events ($\Delta_{\rm min}=-0.05$; Table~\ref{tab:box}).
Unfortunately, no X-ray data  was taken contemporaneously with the deep VLT
observations.  

\subsection{\object{GRB 080915A}}

GRB 080915A triggered \swift/BAT at 00:02:49 UT \citep{Oates2008GCN8227}. It
was a long burst with a duration of $T_{90}=(15-350$ keV)$=14\pm 5$ s
\citep{Ukwatta2008GCN8230}. Unfortunately, owing to an observing constraint,
\swift\ could not slew to the burst during the first hour after the event,
therefore XRT and UVOT could start observing only 3.9~ks after the trigger.
Starting at this time, \swift/UVOT did not detect the optical afterglow
\citep{Oates2008GCNrep168}. The ROTSE-IIIc telescope located on Mt. Gamsberg, Namibia,
responded to GRB 080915A automatically and took unfiltered images  starting 52
s after the GRB trigger (cloudy conditions, full Moon). No afterglow candidate
was  found in the BAT error circle  down to about $CR>14$
\citep{Rujopakarn2008GCN8228}. The robotic 60-cm REM telescope on La Silla
started observing 2 min after the trigger. No afterglow candidates fainter
than the 2MASS limits  were seen in $J, H, K$
\citep{Covino2008GCN8233}. Beginning 4.9 ks after the trigger, \swift/XRT and
\swift/UVOT started observing. XRT found a faint, fading X-ray source with an
error circle of $r=6\farcs5$ \citep{Evans2008GCN8231}. Only upper limits
could be reported for the UVOT bands  \citep{Breeveld2008GCN8232}. Deep
ground-based observations with ANDICAM on the SMARTS 1.3-m telescope at CTIO
provided only upper limits of $I>21.9$ and $J>20.1$ (mid-time of 1.9
hr  after the burst; \citealt{Cobb2008GCN8248}).  

Deep prompt follow-up observations of the field were performed with GROND 
\citep{Rossi2008GCN8268}. They started 4.9 min after
the trigger and lasted for 130 min. No evidence of a variable source was
found when splitting these observations into two data sets
(Table~\ref{tab:darkULs}).  Second-epoch observations were performed with
GROND the following night. Again, no afterglow candidate was found. Using the
GROND upper limit of  $r^\prime_{\rm AB}>22.2$ at 6840~s  (mid-time;
\citealt{Rossi2008GCN8268}), and  $\beta_{\rm OX} <0.62$
(Table~\ref{tab:box}), this corresponds to an upper limit of $R_{\rm
  AB}>22.0$.   Following \cite{Rol2005a},  we can use  the observed X-ray flux
as well as the X-ray slope  to predict the non-extinguished $R_{\rm AB}$-band
magnitude.  However, in this case owing to the small number statistics we can
only give an upper limit of $R_{\rm AB}<$21. According to the criterion of
V09, the burst is dark  (Table~\ref{tab:box}),  but the X-ray light curve is
faint and very uncertain. Therefore, this burst 
cannot be securely classified as dark.

\subsection{\object{GRB 081012}}

\swift/BAT triggered on the burst at 13:10:23 UT. $T_{90}$ (15-350 keV) was
$29\pm4$ sec. The burst was also seen by Fermi/GBM, the peak energy was
$320\pm80$ keV \citep{Bissaldi08_GCN8370}. The XRT began observing the field
49 min after the BAT trigger, an X-ray afterglow was found
\citep{Kennea08_GCN8364}, the error circle is just 1\farcs8 in size
\citep{Evans08_GCN8391}. The UVOT started observing 3 min after the XRT; no
afterglow candidate was detected \citep{Kuin08_GCN8365}. Deep ground-based
follow-up observations were performed using ROTSE IIIa (starting
39 s after the burst), the 2.5-m NOT telescope \citep{Postigo08_GCN8366}.
 
The GROND imager obtained an upper limit on any optical afterglow of $r^\prime_{\rm
  AB}>23.6$ at $\sim70$~ks (mid-time) after the onset of the burst
(\citealt{Filgas2008}; Table~\ref{tab:darkULs}), corresponding to
$r^\prime=23.5$ after correction for Galactic extinction. Using the observed
spectral slope of $\beta_{\rm OX} <0.83$, this corresponds to an upper limit
of $R_{\rm AB}>23.4$. Among all available optical upper limits, this
observation provides the tightest constraint on the SED of the afterglow
(Table~\ref{tab:box}).  On the basis of these data, this burst is not dark.

\subsection{\object{GRB 081105}}

This burst triggered \textit{Konus-WIND}, \swift, \agile, \suzaku, and
\textit{INTEGRAL} at 13:26:12 UT. It was localized only via IPN. The burst had
a single peak, about 10 s long \citep{Cummings08_GCN8484}. \swift/XRT
and UVOT started observing the field about 16 h later. An X-ray afterglow
candidate was detected with an original uncertainty of 4\farcs8
\citep{Beardmore2008GCN8487} and later confirmed \citep{Beardmore2008}.
Observations with UVOT could only provide upper limits
\citep{Curran08_GCN8488}. 
 
The GROND imager obtained an afterglow upper limit of $r^\prime>23.0$ at $\sim46$~ks
(mid-time) after the burst (\citealt{Clemens2008}; Table~\ref{tab:darkULs}),
corresponding to $r^\prime>22.9$ after correction for Galactic extinction.
Using $\beta_{\rm OX} <0.61$, this corresponds to an upper limit of $R_{\rm
  AB}>22.8$. This observation provides the tightest constraints on the
SED.  On the basis of these data, the burst is not dark (Table~\ref{tab:box}).

\subsection{\object{GRB 081204}}

The burst was detected by the \textit{INTEGRAL} satellite at 16:44:55 UT.  It
lasted for about $T_{90}=20$ s \citep{Gotz08_GCN8614}. \swift\  reacted to the
Integral alert, and started observing the field  about 2.7 h after the burst,
and found an uncatalogued X-ray source
\citep{Mangano2008GCN8620,Mangano08_GCN8616}. \swift/UVOT started observing
3~h after the trigger in  the $white$ filter but no source was detected.
\citet{Berger08_GCN8624} suggested an $r=23.5\pm0.3$ afterglow candidate based
on observations with the Magellan/Clay telescope about 9 h after the
trigger. 

The field was also observed with GROND, which also detected the afterglow
candidate observed by \citet{Berger08_GCN8624}, together with another object,
without  finding evidence of fading in either source \citep{Updike2008}.  Both
objects are discussed in this paper as host candidates (see
Sec.~\ref{notesind}).  Stacking the highest quality GROND data, we obtained the revised
upper limits reported in Table~\ref{tab:darkULs}, centered at a mid-time of
9.6~h. The GROND upper limit of $r^\prime>24.1$ corresponds to
$r^\prime>24.0$ after correction for Galactic extinction.  Using the observed
$\beta_{\rm OX} <0.55$, this corresponds to an upper limit of $R_{\rm
  AB}>23.9$. Following \cite{Rol2005a},  we can use  the observed X-ray flux
as well as the X-ray slope  to predict the non-extinguished $R_{\rm AB}$-band
magnitude.  However, in this case owing to the small number statistics we can
only give an upper limit of $R_{\rm AB}<$23 in the worse  case of a break
between optical and X-ray bands.  The burst is dark according to V09
(Table~\ref{tab:box}), but owing to the faint XRT light curve and the poorly
determined high X-ray spectral slope ($\beta_{\rm X} = 1.93^{+1.56}_{-0.77}$)
this burst cannot be securely classified as dark.

\begin{table}[hpbt!]
\caption{Redshifts estimated via the Amati relation}
\renewcommand{\tabcolsep}{6.5pt}
\begin{center}
\begin{tabular}{clcccc}
\toprule
\# & GRB       & Data &Model  & $E_{\rm peak}$(keV) & Redshift  \\
\midrule   
2 & 050922B   & BAT  & Band  & $\sim20$    & $0.1<z<3$      \\
6 & 060923B   & BAT  & CPL   & $25\pm7$    & $z > 0.4$       \\   
10& 080207    & BAT  & CPL   & $108\pm72$  & $z > 0.9$       \\  
11& 080218B   & BAT  & CPL   & $24\pm15$   & $z > 0.3$       \\
12& 080602    & KW   & CPL   & $> 226$     & $z > 1  $       \\
\bottomrule 
\end{tabular}
\label{tab:epeiso}
\end{center} 
\tablefoot{
BAT stands for \swift/BAT, KW for \emph{Konus-Wind} and CPL for 
cut-off power-law.}
\end{table}

\begin{table}[htbp!]
\begin{center}
\caption{Redshift estimates of the galaxies found in the XRT
error circles for different model assumptions about their photometric
properties.}
\begin{tabular}{llclllllll}
\toprule
\# & GRB   & Object  &   (4) &   (5) &   (6) &   (7)  &  (8)  &  (9)\\
\midrule
1 & 050717 & A &         0.5 &   0.4 &   1.2 &   0.9  &  3.2  &  1.8\\
  &        & B &         0.7 &   0.6 &   1.7 &   1.2  &  --   &  2.4\\[1mm]
  
3 & 060211 & A &         0.7 &   0.6 &   1.7 &   1.2  &  --   &  2.4\\
  &        & B &         0.4 &   0.3 &   0.9 &   0.7  &  2.3  &  1.4\\[1mm]
  
4 &060805A & A &         0.9 &   0.7 &   2.3 &   1.4  &  --   &  3.0\\
  &        & B &         0.5 &   0.4 &   1.1 &   0.8  &  3.1  &  1.8\\[1mm]
  
5 &060919  & A &         1.5 &   1.0 &   4.2 &   2.1  &  --   &  4.5\\[1mm]

6 &060923B & A &         0.4 &   0.3 &   0.9 &   0.7  &  2.4  &  1.5\\
  &        & C &         0.7 &   0.6 &   1.7 &   1.2  &  --   &  2.4\\
  &        & D &         1.2 &   0.9 &   3.3 &   1.8  &  1.5  &  3.9\\[1mm]
  
7 &061102  & A &         0.6 &   0.5 &   1.4 &   1.0  &  4.1  &  2.1\\
  &        & B &         0.5 &   0.5 &   1.3 &   1.0  &  3.8  &  2.0\\[1mm]
  
8 &070429A & A &         0.9 &   0.7 &   2.2 &   1.4  &  --   &  2.9\\
  &        & B &         0.6 &   0.5 &   1.5 &   1.0  &  4.2  &  2.1\\
  &        & C &         0.6 &   0.5 &   1.6 &   1.1  &  4.7  &  2.3\\[1mm]
  
9 &070517 & A &         1.0 &   0.8 &   2.7 &   1.6  &  --   &  3.4\\[1mm]

10&080207  & A &         0.9 &   0.7 &   2.4 &   1.5  &  --   &  3.1\\
  &        & B &         1.7 &   1.1 &   --  &   2.4  &  --   &  -- \\[1mm]
  
11&080218B & A &         1.5 &   1.1 &   4.4 &   2.2  &  --   &  4.7\\
  &        & B &         0.7 &   0.6 &   1.8 &   1.2  &  --   &  2.5\\[1mm]
  
12&080602  & A &         0.4 &   0.3 &   0.8 &   0.7  &  2.2  &  1.4\\
  &        & B &         0.6 &   0.5 &   1.4 &   1.0  &  3.9  &  2.0\\[1mm]
  
14&080915A & B &         0.2 &   0.2 &   0.4 &   0.4  &  1.0  &  0.7\\
  &        & C &         0.8 &   0.6 &   1.9 &   1.3  &  --   &  2.6\\
  &        & D &         0.7 &   0.6 &   1.8 &   1.2  &  --   &  2.5\\
  &        & E &         1.0 &   0.8 &   2.8 &   1.6  &  --   &  3.5\\[1mm]
  
15&081012  & A &         0.9 &   0.7 &   2.4 &   1.5  &  --   &  3.1\\[1mm]

16&081105  & A &         0.5 &   0.4 &   1.2 &   0.9  &  3.3  &  1.8\\
  &        & B &         0.6 &   0.5 &   1.6 &   1.1  &  4.7  &  2.3\\[1mm]
  
17&081204  & A &         0.4 &   0.3 &   0.9 &   0.7  &  2.5  &  1.5\\
  &        & B &         0.4 &   0.4 &   1.1 &   0.8  &  2.9  &  1.7\\
  &        & C &         0.4 &   0.3 &   0.9 &   0.7  &  2.4  &  1.5\\
  &        & D &         0.6 &   0.5 &   1.5 &   1.0  &  4.3  &  2.2\\
  &        & E &         0.6 &   0.5 &   1.6 &   1.1  &  4.7  &  2.3\\ 
  &        & F &         1.1 &   0.8 &   3.1 &   1.7  &  --   &  3.7\\ 
\bottomrule
\end{tabular}
\label{tab:Color}
\end{center}
\tablefoot{
Columns (4) to (9) give the redshift
of the galaxy for different assumptions about its spectral slope $\beta$
and absolute magnitude $M_R$: 
= (0.0,$-$18), (1.0,$-$18), (0.0,$-$20), (1.0,$-$20), (0.0,$-$22), 
(1.0,$-$22). For greater detail, see Sect.~\ref{redshiftsEstimate}.
  GRB 050922B and GRB 080727A are not included here since they 
have no host galaxy candidates.}
\end{table}

\newpage

\begin{table}[htbpt!]
\begin{center}
\caption{Summary of the early-time upper limits 
based on observations with GROND.}
\begin{tabular}{rll lrr}
\toprule
\#& GRB     &  $t$ [h]   & Filter & UL \\
\midrule
10 & 080207  & 9.75   &  $g^\prime  $ & 24.0    \\
   &         & 9.75   &  $r^\prime  $ & 23.6    \\
   &         & 9.75   &  $i^\prime  $ & 23.1    \\
   &         & 9.75   &  $z^\prime  $ & 22.3    \\
   &         & 9.75   &  $J   $ & 21.0    \\
   &         & 9.75   &  $H   $ & 19.7    \\
   &         & 9.75   &  $K_s   $ & 19.0    \\[1mm]

11 & 080218B & 0.75   &  $g^\prime  $ & 21.4    \\
   &         & 0.75   &  $r^\prime  $ & 21.5    \\
   &         & 0.75   &  $i^\prime  $ & 20.6    \\
   &         & 0.75   &  $z^\prime  $ & 20.6    \\
   &         & 0.75   &  $J   $ & 20.8    \\
   &         & 0.75   &  $H   $ & 18.5    \\
   &         & 0.75   &  $K_s   $ & 17.8    \\

   &         & 3.2    &  $g^\prime  $ & 24.6    \\
   &         & 3.2    &  $r^\prime  $ & 24.7    \\
   &         & 3.2    &  $i^\prime  $ & 23.9    \\
   &         & 3.2    &  $z^\prime  $ & 24.7    \\
   &         & 3.2    &  $J   $ & 22.0    \\
   &         & 3.2    &  $H   $ & 20.8    \\
   &         & 3.2    &  $K_s   $ & 20.0    \\[1mm]

14 & 080915A  &  0.15  &  $g^\prime  $& 22.0    \\
   &          &  0.15  &  $r^\prime  $& 22.2    \\
   &          &  0.15  &  $i^\prime  $& 21.8    \\
   &          &  0.15  &  $z^\prime  $& 21.7    \\
   &          &  0.15  &  $J   $& 20.1    \\
   &          &  0.15  &  $H   $& 18.9    \\
   &          &  0.15  &  $K_s   $& 17.6    \\

   &          &  0.92  &  $g^\prime  $& 23.0    \\
   &          &  0.92  &  $r^\prime  $& 23.5    \\
   &          &  0.92  &  $i^\prime  $& 23.0    \\
   &          &  0.92  &  $z^\prime  $& 23.1    \\
   &          &  0.92  &  $J   $& 21.2    \\
   &          &  0.92  &  $H   $& 20.0    \\
   &          &  0.92  &  $K_s   $& 18.6    \\[1mm]

15 & 081012   &  19.35 &  $g^\prime  $&  23.2   \\
   &          &  19.35 &  $r^\prime  $&  23.5   \\
   &          &  19.35 &  $i^\prime  $&  22.8   \\
   &          &  19.35 &  $z^\prime  $&  22.8   \\
   &          &  19.35 &  $J   $&  21.5   \\
   &          &  19.35 &  $H   $&  20.4   \\
   &          &  19.35 &  $K_s   $&  19.4   \\[1mm]

16 & 081105   & 12.84  &  $g^\prime  $& 24.0    \\
   &          & 12.84  &  $r^\prime  $& 23.0    \\
   &          & 12.84  &  $i^\prime  $& 22.1    \\
   &          & 12.84  &  $z^\prime  $& 21.8    \\
   &          & 12.84  &  $J   $& 20.7    \\
   &          & 12.84  &  $H   $& 19.6    \\
   &          & 12.84  &  $K_s   $& 18.2    \\[1mm]

17 & 081204   & 9.60   &  $g^\prime  $& 24.2    \\ 
   &          & 9.60   &  $r^\prime  $& 24.1    \\ 
   &          & 9.60   &  $i^\prime  $& 23.2    \\ 
   &          & 9.60   &  $z^\prime  $& 22.4    \\ 
   &          & 9.60   &  $J   $& 20.7    \\ 
   &          & 9.60   &  $H   $& 19.4    \\ 
   &          & 9.60   &  $K_s   $& 18.3    \\ 
\bottomrule
\end{tabular}
\label{tab:darkULs}
\end{center} 
\tablefoot{For early-time observations by other groups  see,
e.g., the web page of J. Greiner at {\tt
www.mpe.mpg.de/$\sim$jcg/grbgen.html} or GRBlog at {\tt http://grblog.org/grblog.php}. 
In all cases, 
the data given here supersede the values given in the 
corresponding GRB circulars:
GCN 7279, GRB 080207  \citep{Yoldas2008GCN7279}; 
GCN 7319, GRB 080218B \citep{Rossi2008GCN7319};
GCN 8268, GRB 080915A \citep{Rossi2008GCN8268};
GCN 8373, GRB 081012  \citep{Filgas2008};
GCN 8492, GRB 081105  \citep{Clemens2008};
GCN 8627, GRB 081204  \citep{Updike2008}.}
\end{table}

\newpage

\begin{table*}[htbp!]
\enlargethispage{2cm}
\caption[]{Log of the late-time optical/NIR observations to search for
a GRB host candidate.}
\renewcommand{\tabcolsep}{8.5pt}
\begin{center}
\begin{tabular}{rlcccccc}
\toprule
\# & GRB   &  Instrument & Filter &Date obs       & Calib      & FWHM & Exp. (s) \\        
\midrule                                            
1 & 050717 &  GROND      &$g^\prime r^\prime i^\prime z^\prime $     &2007/07/24-26  & SA114-750  & 1\farcs0    & 8880 \\
  &        &  GROND      &$JHK_s$        &2007/07/24-26  & 2MASS      & 1\farcs4    & 7200 \\[2.3mm]

2 & 050922B&  FORS2      &$R_C$            & 2009/08/15    & ESO ZP     & 0\farcs7    & 2930 \\
  &        &  ISAAC      &$K_s$          & 2009/07/06    & 2MASS      & 0\farcs7    & 1920 \\
  &        &  NEWFIRM    &$K_s$          & 2008/11/08    & 2MASS      & 1\farcs2    & 1800 \\[2.3mm]

3 & 060211A&  GROND      &$g^\prime r^\prime i^\prime z^\prime $     &2007/10/20-22  & SA95-190   & 1\farcs0    & 10360 \\
  &        &  GROND      &$JHK_s$        &2007/10/20-22  & 2MASS      & 1\farcs6    & 8400 \\
  &        &  NEWFIRM    &$J$            &2009/01/17     & 2MASS      & 1\farcs1    & 10200 \\
  &        &  NEWFIRM    &$K_s$          &2009/01/17     & 2MASS      & 1\farcs2    & 3600 \\[2.3mm]

4 & 060805A&  GROND      &$g^\prime r^\prime i^\prime z^\prime $     &2008/05/05-07  & SDSS       & 0\farcs9    & 4440 \\
  &        &  GROND      &$JHK_s$        &2008/05/05-07  & 2MASS      & 1\farcs6    & 3600 \\
  [2.3mm]

5 & 060919 &  FORS1      &$R_C$            & 2008/04/10    & SA110-362  & 0\farcs8    & 2930 \\
  &        &  ISAAC      &$K_s$          & 2008/05/18    & 2MASS      & 0\farcs6    & 1920 \\[2.3mm]

6 & 060923B&  FORS1      &$R_C$            & 2008/04/05    & NGC2437     & 0\farcs8    & 2930 \\
  &        &  ISAAC      &$K_s$          & 2008/04/15    & 2MASS      & 0\farcs5    & 1920 \\[2.3mm]

7 & 061102 &  FORS1      &$R_C$            & 2008/04/06    & NGC2437     & 0\farcs7    & 2930 \\
  &        &  ISAAC      &$K_s$          & 2008/04/18    & 2MASS      & 0\farcs7    & 1920 \\[2.3mm]

8 & 070429A&  FORS1      &$R_C$            & 2008/04/08    & SA110-362  & 1\farcs0    & 2930 \\
  &        &  ISAAC      &$K_s$          & 2008/05/18    & 2MASS      & 0\farcs6    & 2400 \\[2.3mm]

9 & 070517&  FORS1      &$R_C$            & 2008/04/10    & SA110-362  & 0\farcs6    & 2930 \\
  &        &  ISAAC      &$K_s$          & 2008/08/05    & 2MASS      & 0\farcs5    & 1920 \\[2.3mm]

10& 080207 &  VIMOS      &$R_C$            & 2010/02/10    & PG1047+3   & 0\farcs8    & 2930 \\
  &        &  ISAAC      &$K_s$          & 2010/02/07    & 2MASS      & 0\farcs6    & 1920 \\[2.3mm]

11& 080218B&  FORS2      &$R_C$            & 2009/05/26    & PG1047     & 0\farcs5    & 2930 \\
  &        &  ISAAC      &$K$            & 2009/03/20    & 2MASS      & 0\farcs5    & 1920 \\[2.3mm]

12& 080602 &  GROND      &$g^\prime r^\prime i^\prime z^\prime $     & 2009/11/24    & SDSS       & 1\farcs1    & 4440  \\
  &        &  GROND      &$JHK_s$        & 2009/11/24    & 2MASS      & 1\farcs3    & 3600  \\ 
  &        &  FORS2      &$R_C$            & 2009/06/05    & NGC7006   & 0\farcs8    & 2930 \\
  &        &  ISAAC      &$K_s$          & 2009/07/06    & 2MASS      & 0\farcs7    & 1920 \\[2.3mm]

13& 080727A &  FORS1      &$R_C$            & 2008/07/27    & E5-Stetson & 0\farcs8    & 2930 \\
  &        &  ISAAC      &$K_s$          & 2010/02/10    & 2MASS      & 0\farcs6    & 1920 \\[2.3mm]
14& 080915A &  FORS1      &$R_C$            & 2008/09/27    & E7         & 1\farcs4    & 968  \\
  &        &  HAWK-I     &$K_s$          & 2008/09/16    & 2MASS      & 0\farcs6    &  840 \\[2.3mm]

15& 081012 &  VIMOS      &$R_C$            & 2009/10/21    & SA98       & 0\farcs8    & 2400 \\
  &        &  ISAAC      &$K_s$          & 2009/10/08    & 2MASS      & 0\farcs4    & 1920 \\[2.3mm]

16& 081105 &  VIMOS      &$R_C$            & 2009/10/21    & SA98       & 1\farcs0    & 2400 \\
  &        &  ISAAC      &$K_s$          & 2009/09/14    & 2MASS      & 0\farcs4    & 1920 \\[2.3mm]

17& 081204 &  VIMOS      &$R_C$            & 2009/10/21    & SA98       & 1\farcs0    & 2400 \\
  &        &  ISAAC      &$K_s$          & 2009/09/14    & 2MASS      & 0\farcs5    & 1920 \\
  &        &  SOFI       &$J$           & 2010/11/01    & 2MASS      & 0\farcs5    & 3600 \\
  &        &             &               &               &            &        &      \\
\bottomrule
\end{tabular}
\label{tab:obs_log}
\end{center}
\tablefoot{
{\it Notes for individual targets:}\  GRB 070517: a candidate optical afterglow
was found by \cite{Fox6420}, which we identify
as the GRB afterglow based on our data.
{\it Standard star fields:}\  The fields  PG1047+3,
E5, E7, NGC~2437, and NGC~7006 are from  the internet pages of
P. Stetson  {\tt
http://www3.cadc-ccda.hia-iha.nrc-cnrc.gc.ca/community/STETSON/}.  Landolt
equatorial standards stars (SA) for the $R_C$ band were obtained from the
internet page of the Canada-France-Hawaii Telescope  {\tt
http://www.cfht.hawaii.edu/ObsInfo/Standards/Landolt/}.  SA standard star
fields for GROND optical calibrations are downloaded from the SDSS archive
server at {\tt http://www.sdss.org/}. ZP stands for 
photometric zero-point calibration. {\it
Filters:}\ Observations with FORS2 were performed using the  $R_{\rm
 special+76}$ filter. Both, FORS1 and VIMOS used the $R_{\rm Bessel+36}$ filter. 
 The FWHM column refers to the FWHM of the average stellar PSF.}
\enlargethispage{0cm}
\end{table*}

\end{appendix}
 
\end{document}